%% file: main.tex
\documentclass[preprint,12pt]{elsarticle}

\usepackage{dcolumn}
\usepackage{bm}
\usepackage[utf8]{inputenc}
\usepackage[T1]{fontenc}
\usepackage{mathptmx}
\usepackage{etoolbox}

\usepackage{graphicx}
\usepackage[english]{babel} 
\usepackage{url} 
\usepackage{todonotes} 
\usepackage{multirow} 
\usepackage{tabularx} 
\usepackage{booktabs} 
\usepackage[hidelinks]{hyperref}  
\usepackage{academicons} 
\definecolor{orcidlogocol}{HTML}{A6CE39} 
\newcommand{\orcid}[1]{\href{https://orcid.org/#1}{\textcolor{orcidlogocol}{\aiOrcid}}}

\usepackage{xcolor}
\usepackage{array}
\newcolumntype{C}[1]{>{\centering\arraybackslash}p{#1}}

\usepackage{amsmath}

\usepackage[margin=1in]{geometry}
\usepackage{graphicx}
\usepackage{subcaption}
\usepackage{booktabs}
\usepackage{amsmath}
\usepackage{siunitx}
\usepackage{float}
\usepackage{hyperref}
\usepackage{xcolor}
\usepackage{array}
\usepackage{grffile}  
\usepackage{xspace}   

\newcommand{\plotifexists}[1]{%
	\IfFileExists{#1}{\includegraphics[width=\linewidth]{#1}}{%
		\IfFileExists{../#1}{\includegraphics[width=\linewidth]{../#1}}{%
			\fbox{\parbox[c][4.5cm][c]{0.9\linewidth}{\centering Plot not found:\\ \texttt{#1}}}%
		}%
	}%
}

\newcommand{\mycomment}[1]{}

\newcommand{\tc}{$T_{\mathrm{C}}$\xspace}
\newcommand{\ms}{$\mu_{0}M_{\mathrm{s}}$\xspace}
\newcommand{\js}{$\mu_{0}M_{\mathrm{s}}$\xspace}
\newcommand{\ha}{$\mu_{0}H_{\mathrm{a}}$\xspace}
\newcommand{\hc}{$\mu_{0}H_{\mathrm{c}}$\xspace}
\newcommand{\epistemicunc}{$\sigma_{\mathrm{e}}^2$\xspace}

\newcommand{\ndfeb}{$\mathrm{Nd_{2}Fe_{14}B}$\xspace}

\journal{Computational Materials Science}

\begin{document}

\begin{frontmatter}

\title{Modelling magnetic material properties with uncertainty-aware neural networks\tnoteref{publication}}
\tnotetext[publication]{Published in Computational Materials Science (2026). DOI: \href{https://doi.org/10.1016/j.commatsci.2026.115073}{10.1016/j.commatsci.2026.115073}. Licensed under CC BY 4.0.}

\author[cdlab,diss,vdsp]{C.~Wager}
\author[diss]{H.~Moustafa}
\author[cdlab,diss]{A.~Kovacs}
\author[cdlab,diss]{Q.~Ali}
\author[diss]{H.~Oezelt}
\author[tmc]{H.~Yamano}
\author[tmc]{M.~Yano}
\author[tmc]{N.~Sakuma}
\author[tmc]{H.~Hosoi}
\author[tmc]{A.~Kinoshita}
\author[tmc]{T.~Shoji}
\author[tmc]{A.~Kato}
\author[cdlab,diss]{T.~Schrefl}

\affiliation[cdlab]{
    organization={Christian Doppler Laboratory for magnet design through physics informed machine learning, University for Continuing Education Krems},
    city={Wr. Neustadt},
    postcode={2700},
    country={Austria}
}
\affiliation[diss]{
    organization={Department for integrated sensor systems, University for Continuing Education Krems},
    city={Wr. Neustadt},
    postcode={2700},
    country={Austria}
}
\affiliation[vdsp]{
    organization={Vienna Doctoral School in Physics},
    city={Vienna},
    country={Austria}
}
\affiliation[tmc]{
    organization={Advanced Materials Engineering Division, Toyota Motor Corporation},
    city={Susono},
    country={Japan}
}

\begin{abstract}
    \input{0_abstract}
\end{abstract}


\end{frontmatter}


\input{1_introduction}
\input{2_uncertainty}
\input{2.1_bnn_model}
\input{3_model_evaluation}

\input{3.1_discussion_approach}
\input{3a_surrogate_TC_data_study}

\input{4_STUDY_intrinsic_prediction}

\input{4.2_model_comparison}
\input{4.3_models_summary}

\input{5_STUDY_coercivity_prediction}

\input{6_conclusion}

\input{888_appendix_study_details}
\input{999_post}

\bibliographystyle{elsarticle-num}
\bibliography{bibliography}

\end{document}

%% file: 0_abstract.tex

Machine learning is increasingly used to accelerate materials discovery across large compositional and structural design spaces, but limited and heterogeneous data make reliable uncertainty estimation essential. 
In this work, we investigate uncertainty quantification for permanent magnet modeling across three complementary studies. 
First, on a public Curie temperature surrogate dataset, we benchmark Gaussian process regression, random forest bagging, and dropout-based Bayesian neural networks and show that calibration, sharpness, and confidence curve diagnostics reveal differences in uncertainty quality that are not visible from point-prediction metrics alone. 
Second, we transfer this framework to the prediction of intrinsic magnetic properties in \ndfeb-based magnets. 
Third, we extend the same uncertainty-aware perspective to coercivity prediction from microstructural information using a graph neural network. 
Together, these studies show that uncertainty quantification improves the trustworthiness of magnetic material property predictions and can be transferred from composition-based surrogate models to more complex structure-sensitive learning tasks.

%% file: 1_introduction.tex
\section{Introduction}

    The discovery and design of novel magnetic materials increasingly rely on machine learning models that can navigate large compositional, processing, and microstructural design spaces
    \cite{moller2018, katsikas2021, kovacs2023}.
    Prior work has already shown that machine learning can predict a broad range of magnetic-material properties, including 
    magnetization \cite{katsikas2021}, 
    Curie temperature \cite{nelson2019a}, 
    composition-dependent hard-magnetic performance \cite{singh2023, moller2018}, 
    soft-magnetic saturation flux density and thermal stability \cite{lu2020}. 
    This trend is particularly attractive in permanent magnet research, where experimental synthesis, characterization, density-functional-theory screening, and multiscale micromagnetic simulation are costly,
    and where the relevant structure-property relations are strongly nonlinear and coupled across length scales
    \cite{katsikas2021, lu2020, moller2018, kovacs2023}.
    At the same time, the data regime is typically small, heterogeneous, and biased towards experimentally accessible regions of parameter space.
    As a result, predictive accuracy alone is not sufficient.
    Materials models are routinely used to rank candidates, guide follow-up experiments, and support optimization loops.
    For these tasks, the model must also indicate when its own prediction should not be trusted
    \cite{lookman2019, korolev2022, tran2020}.

    This requirement becomes especially important when machine learning is used for extrapolative discovery.
    In practice, models are often queried outside the training distribution in order to identify previously unexplored compounds or microstructures.
    This is common in magnetic materials because composition, crystal structure, processing history, and microstructure interact in a high-dimensional design space, while available labels are usually sparse and unevenly distributed
    \cite{katsikas2021, moller2018, singh2023}.
    Uncertainty quantification therefore plays a dual role.
    It supports risk-aware decision making for individual predictions, and it provides the acquisition signal required by active-learning and adaptive-sampling workflows in materials design
    \cite{lookman2019}.
    More broadly, recent reviews have emphasized that uncertainty-aware modeling is becoming a core methodological requirement in scientific machine learning.
    \cite{gawlikowski2023, tran2020}.

    Several recent data-driven studies in magnetic materials and related systems
    (i.e. \cite{srinithi2025, dengina2022, lai2022, demoraes2024, kovacs2023}),
    assemble datasets from mixed sources, including experiments, simulations, curated databases, and literature-derived values.
    Measurements originating from different laboratories, processing routes, and characterization protocols inherently carry different data noise.
    The corresponding data uncertainty, usually termed \textit{aleatoric uncertainty}, is therefore rarely well described by a homoscedastic assumption.
    Instead, the uncertainty is often heteroscedastic and depends on the local region of feature space, the measurement protocol, or the quality of the underlying source data
    \cite{kendall2017, gawlikowski2023, acar2021}.
    In parallel, limited coverage of the design space introduces \textit{epistemic uncertainty}, which reflects uncertainty in the model itself and is expected to increase in sparsely sampled or out-of-distribution regions
    \cite{kendall2017, gal2016b}.

    These distinctions matter because not every accurate regression model produces meaningful uncertainty estimates.
    A high coefficient of determination ($R^2$) does not guarantee calibrated predictive intervals, nor does it ensure that uncertainty correlates with actual prediction error
    \cite{tran2020, pernot2022}.
    Classical probabilistic surrogates such as Gaussian Process Regression provide a principled predictive distribution
    \cite{rasmussen2006},
    while bagged ensembles and related approaches can capture aspects of model uncertainty
    \cite{breiman2001, lakshminarayanan2017}.
    However, these methods do not necessarily expose uncertainty in the same way, and their practical usefulness depends on calibration, decomposition into epistemic and aleatoric contributions, and robustness
    under distribution shift
    \cite{gal2016b, gawlikowski2023}.
    This is precisely the gap in much of the existing magnetic-materials machine-learning literature.
    Recent studies demonstrate strong predictive performance and useful physical interpretation for magnetic-property prediction
    \cite{lu2020, katsikas2021, moller2018, singh2023},
    but they do not make uncertainty quality a central evaluation target.
    Incorporating uncertainty quantification would therefore improve the state of the art in two ways.
    First, it would allow magnetic-property predictions to be ranked together with calibrated confidence estimates, which is essential when selecting candidates for expensive synthesis or simulation.
    Second, it would reveal whether prediction errors arise from irreducible data noise or from insufficient data coverage, thereby supporting safer extrapolation and more efficient data acquisition
    \cite{lookman2019, acar2021, tavazza2021, palmer2022}.
    This motivates the present work, where uncertainty quality is treated as a primary evaluation target rather than a secondary diagnostic.

    To demonstrate the role of uncertainty quantification in magnetic materials research, we present two complementary studies.
    In the first study (Sec. \ref{sec:intrinsic_prediction}), we compare machine-learning models for predicting intrinsic magnetic properties and assess not only predictive accuracy but also the quality of the associated uncertainty estimates.
    This baseline setting allows us to examine Gaussian negative log-likelihood training and dropout-based Bayesian approximation in a controlled regression problem
    \cite{gal2016b, kendall2017}.

    Building on these insights, the second study (Sec. \ref{sec:gcnn4coercivity}) transfers the same uncertainty-aware perspective to a more challenging task, namely predicting coercivity from microstructural information with a graph neural network.
    The two studies therefore address complementary levels of description, from intrinsic property prediction to microstructure-sensitive coercivity modeling.
    Together, they illustrate both the reliability and the transferability of uncertainty-aware approaches across distinct prediction tasks in machine-learning-assisted permanent magnet design.

%% file: 2_uncertainty.tex
\section{Uncertainty for Materials Science}
\label{sec:uncertainty}
    
    In machine learning, especially when using neural networks, it is important to remember that model predictions are not guaranteed to be correct. Instead, they are educated guesses based on patterns the model has learned from past data. 
    Predictions carry inherent uncertainties, which can be categorized into two main types: \textit{aleatoric} and \textit{epistemic} uncertainty. 
    Usually, neural networks provide point estimates without clear indicators of confidence, which can be particularly problematic in scenarios requiring high reliability, such as medical diagnosis and autonomous driving.
    In materials science and design an estimate of confidence is necessary to guide exploration in under-sampled regions
    \cite{gal2016b, kendall2017, scalia2019}.
    Generally, predicted uncertainty is represented by the variance $\sigma^2$ or by the standard deviation $\sigma$. 
    The latter, has the advantage that its value has the same unit as the prediction.

\subsection{Aleatoric Uncertainty}
\label{sec:aleatoric_uncertainty}

    Aleatoric uncertainty $\sigma_\mathrm{a}^2$ arises from intrinsic randomness or noise in the data itself \cite{scalia2019}. It represents variability that cannot be reduced by collecting additional data, as it is inherent to the observation process or the underlying data-generating mechanism. 
    In materials science, such uncertainty often emerges when experimental measurements or computational results are obtained using different methods, instruments, or levels of approximation, each introducing distinct noise characteristics. 
    Aleatoric uncertainty can be classified as \textit{homoscedastic}, when the noise level remains constant across all inputs and is on the other side primarily data-dependent, or \textit{heteroscedastic}, when the noise magnitude varies with the input data \cite{kendall2018}.
    The latter is particularly relevant in materials science, where datasets frequently integrate measurements and simulations from diverse sources with varying reliability and accuracy.

    The aleatoric heteroscedastic uncertainty can be approximated directly from data by adapting the model and using a special loss function (Eq. \ref{eq:gnllLoss}), for example, by assuming an underlying Gaussian error and estimating both the mean and variance of the output distribution. 
    The model implicitly learns to increase the predicted variance for uncertain predictions, effectively reducing the impact of prediction errors on the overall loss \cite{kendall2017, scalia2019}.

\subsection{Epistemic Uncertainty}
\label{sec:epistemic_uncertainty}
    
    The epistemic uncertainty $\sigma_\mathrm{e}^2$ stems from a lack of knowledge about the true model parameters or the system being modeled. 
    It is often referred to as "model uncertainty" and can be reduced by acquiring more data or improving the model's understanding of the system. 
    Typically, the epistemic uncertainty is captured by means of ensemble methods, such as random forest \cite{breiman2001}. 
    It emerges from the slight differences in predictions obtained from sub-models that were trained independently from one another and thus have a different perception about the system. 
    
\begin{figure}
        \centering
        \includegraphics[width=1\linewidth]{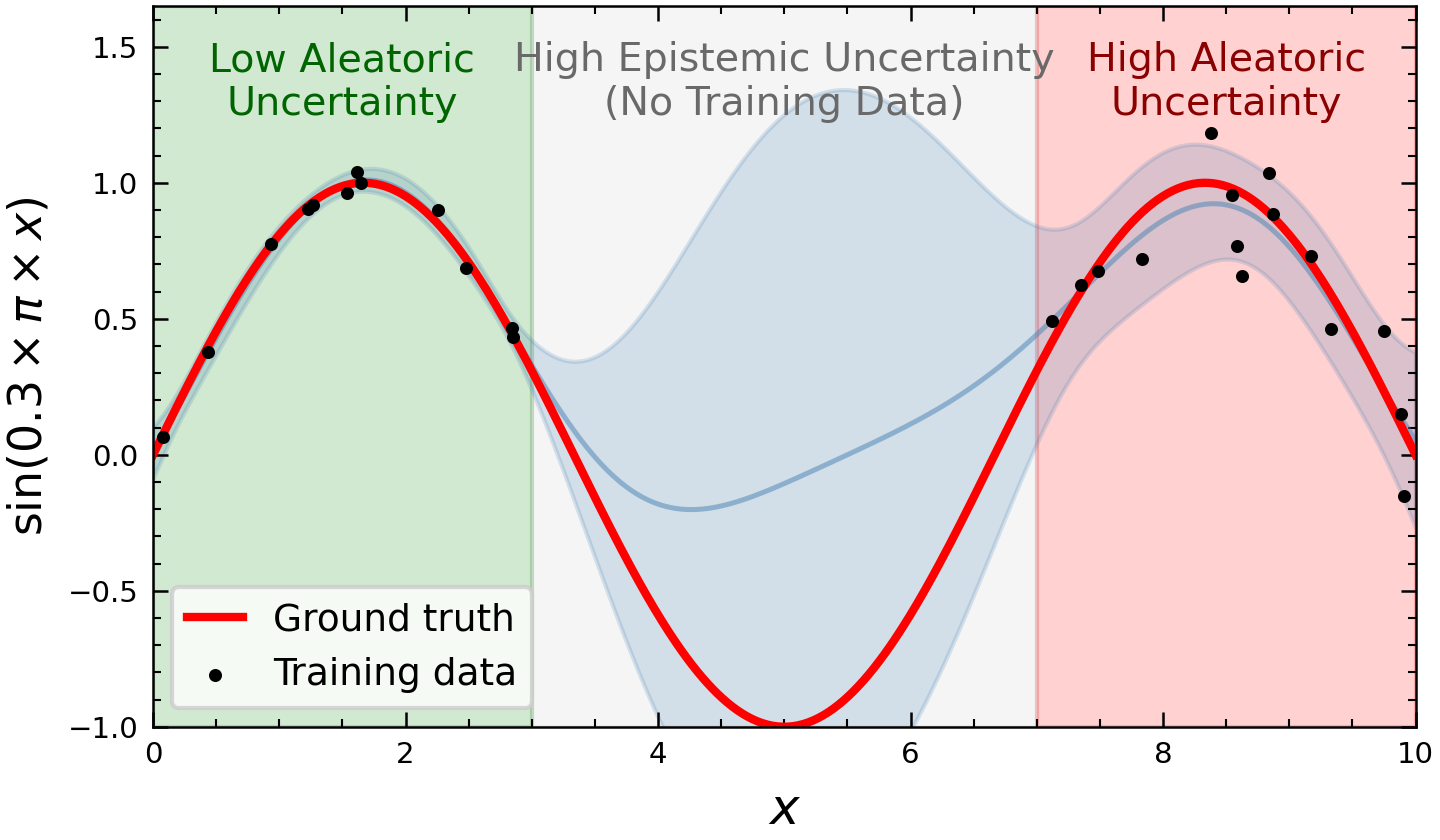}
        \caption{
            Illustration of aleatoric and epistemic uncertainty in a regression setting. 
            The ground truth function $\sin(0.3\pi x)$ (red curve) is observed only in two disjoint input regions. 
            In the left region ($x \in [0,3]$), training data samples are dense and low-noise, yielding \textbf{low aleatoric uncertainty}. 
            In the right region ($x \in [7,10]$), training data are sparse and high-noise, yielding \textbf{high aleatoric uncertainty}.
            In the central region ($x \in [3,7]$), no training data exist. Here, the model therefore exhibits \textbf{high epistemic uncertainty}, reflected in the widened predictive confidence band (blue shaded area). 
            Unlike aleatoric uncertainty, epistemic uncertainty is reducible: additional observations in the unobserved region would collapse the predictive variance towards the true function.
        }
        \label{fig:uncertainty_types}
    \end{figure}

    The combination of both aleatoric and epistemic uncertainties provides a more complete picture of the total uncertainty in a prediction \cite{kendall2017}.
    Figure \ref{fig:uncertainty_types} visualizes the sources that give rise to the two types of uncertainty. 
    Recognizing and quantifying these distinct uncertainties is essential for a robust assessment of model confidence.

%% file: 2.1_bnn_model.tex
\subsection{Dropout networks}
\label{sec:bnn}

    A class of neural network models that incorporates Bayesian inference principles is the Bayesian Neural Network (BNN). 
    In these models, probability distributions are assigned to the network weights, allowing the model to capture epistemic uncertainty.
    
    Kendall and Gal \cite{kendall2017} introduced a deep learning framework capable of modeling both aleatoric and epistemic uncertainty within a single neural network. 
    Their approach builds on the work of Gal and Ghahramani \cite{gal2016b}, who demonstrated that neural networks employing dropout layers, that are active during prediction time, can approximate Bayesian inference. 
    Such dropout networks thus mimic the behavior of BNNs. 
    In the following, we refer to these approximate Bayesian models interchangeably as Bayesian Neural Networks (BNNs) or dropout networks.
    By using a \textit{Gaussian negative log-likelihood} loss function (Eq. \ref{eq:gnllLoss}) this machine learning method also captures aleatoric uncertainty \cite{kendall2017}.

\begin{figure}[!b]
        \centering
        \includegraphics[width=1\linewidth]{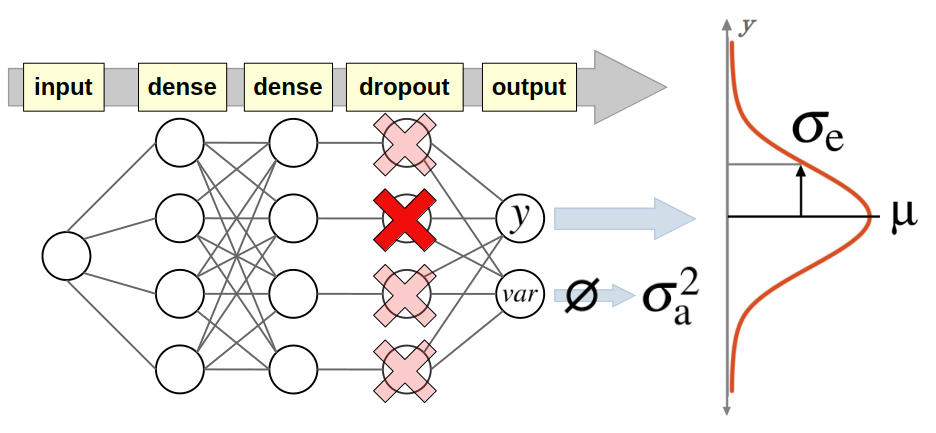}
        \caption{
            \textbf{Conceptual dropout network}: 
            This diagram illustrates an example dropout network with two dense layers followed by one stochastic dropout layer. 
            When predicting the label of a sample multiple forward passes are performed with the dropout layer active. 
            Each forward pass produces a slightly different prediction $y_i$, because the dropout layer randomly sets the output of some neurons in the previous layer to zero (shown as red crosses) \cite{chollet2015, geron2019}. 
            By repeating this process multiple times, a distribution over the predictions is obtained. 
            The mean $\mu$ of this distribution serves as the final prediction, while its first standard deviation $\sigma_\mathrm{e}$ quantifies the epistemic uncertainty of the model.
            With each forward pass the loss produces a variance $\mathrm{var}(\textbf{x})$. 
            The final aleatoric uncertainty $\sigma_\mathrm{a}^2$ is the average over all variances $\mathrm{var}(\textbf{x})$.
        }
        \label{fig:conceptual_bnn}
    \end{figure}

    Conceptually, these dropout networks resemble ensemble methods, since they generate multiple predictions for each sample that are ultimately averaged. 
    In practice, this is realized through dropout layers that are active during prediction time, which render the network slightly different at each forward pass. 
    By employing this so-called \textit{Monte Carlo dropout} (MC dropout)\cite{gal2016b} the output of some nodes of the previous layer is set to zero during prediction time. \cite{chollet2015, geron2019} 
    See Figure \ref{fig:conceptual_bnn} for a visual explanation of the model architecture. 
    A significant advantage of integrating MC dropout over ensemble methods is that only a single model needs to be trained, rather than a large number of separate models. This reduces training time and computational costs.
    This effect scales with increasing model complexity. 
    
    The mean $\mu$ of the $M$ stochastic predictions $y_i$ represents the final prediction value, 
    while the variance $\sigma_\mathrm{e}^2 =\mathrm{var}(\mathbf{y})$ of the distribution quantifies the epistemic uncertainty associated with this prediction.
    
    A natural way to capture data-dependent uncertainty in regression models is to assume that each target value is drawn from a probability distribution whose parameters are predicted by the network \cite{nix1994, kendall2017, lakshminarayanan2017}.
    In the case of Gaussian likelihood, the neural network outputs two quantities for each input $\textbf{x}$: the mean $\mu(\textbf{x})$, representing the central prediction, and the variance $\mathrm{var}(\textbf{x})$, representing the associated predictive uncertainty. 
    The training objective is then defined as the negative log-likelihood of the observed data sample $\textbf{x}$ under this Gaussian distribution. 
    Formally, for a target $y$, the loss function can be written as

    \begin{equation}
    \label{eq:gnllLoss}
        \mathcal{L}(y \,|\, \mu(\textbf{x}), \mathrm{var}(\textbf{x})) = \frac{1}{2} \left( \log \mathrm{var}(\textbf{x}) + \frac{(y - \mu(\textbf{x}))^2}{\mathrm{var}(\textbf{x})} \right)
    \end{equation}
    
    This expression covers two key aspects: the squared error between the prediction and the observation, scaled by the predicted variance, and a regularization term penalizing overly large variances. 
    Intuitively, if the network assigns low uncertainty (meaning high confidence) to its prediction, then errors are penalized more heavily. 
    Conversely, if it predicts high uncertainty, the penalty for prediction errors is reduced. 
    However, the logarithmic term prevents the model from inflating the variance arbitrarily 
    \cite{kendall2017}.
    %
    In practice, the loss is implemented slightly different to increase numerical stability. To make training more stable, we optimize a modified loss by having our model predict $log(\sigma^2)$ rather than $\sigma^2$. Thus, we ensure that the variance is positive \cite{geron2019, nguyen2020}.
    %
    Some other implementations also add a small constant, e.g. $10^{-6}$, to the variance improving numerical stability (see the PyTorch \cite{paszke2019} implementation for \href{https://docs.pytorch.org/docs/stable/generated/torch.nn.GaussianNLLLoss.html}{\textit{GaussianNLLLoss}}). 
    %
    This formulation was first suggested by Nix and Weigend (1994)\cite{nix1994} and later popularized in modern deep learning by Kendall and Gal (2017) \cite{kendall2017} and Lakshminarayanan et al. (2017) \cite{lakshminarayanan2017} for regression problems.

    When following Kendall and Gal's framework a single machine learning model can capture both uncertainty types at once.
    The final aleatoric uncertainty $\sigma_\mathrm{a}^2$ for a sample $x$ is the mean of the variances produced by the loss function over $M$ forward passes formally given by

    \begin{equation}
    \label{eq:aleatoric_uncertainty}
        \sigma_\mathrm{a}^2 = \frac{1}{M} \sum_i^M{{\mathrm{var}(\textbf{x})_i}} 
    \end{equation}

    In this work we consider a combination of the epistemic $\sigma_\mathrm{e}^2$ and aleatoric uncertainty $\sigma_\mathrm{a}^2$ to express and compare total uncertainty values of a model given by 
    
    \begin{equation} 
    \label{eq:total uncertainty}
        \sigma = \sqrt{\sigma_\mathrm{e}^2 + \sigma_\mathrm{a}^2}
    \end{equation}

    which has the advantage that $\sigma$ uses the same units as the target $y$.

    Concluding this section on the model architecture, our approach also draws inspiration from comparative studies such as Scalia et al. \cite{scalia2019} and Dewolf et al. \cite{dewolf2023}, 
    which provide systematic evaluations of ensemble methods, dropout networks, Monte Carlo integration, and dropout-based approaches.
    To implement the BNN architecture, we build custom classes in Python with the libraries Scikit-learn\cite{pedregosa2011} and Keras\cite{chollet2015} with a JAX\cite{bradbury2018} backend.  

%% file: 3_model_evaluation.tex

\subsection{Model evaluation methods} 
\label{sec:model_evaluation}

    The \textbf{predictive performance} of each machine learning model is assessed with a repeated train-test-split protocol.
    For each study, the available data are divided into training and test subsets, and the full train-fit-predict workflow is repeated for 10 different random seeds.
    All reported point-prediction metrics, including MAE, MSE, RMSE, and $R^2$, are computed on the held-out test split of each run and then averaged over these 10 randomly seeded runs in order to reduce dependence on a single favorable split and to obtain a statistically more robust evaluation.
    In addition, we visualize the model fit by plotting measured versus predicted values from the corresponding test sets of these repeated runs.
    This approach preserves evaluation on unseen data while revealing how stable the model behavior is across different train-test partitions.
    Similarly, the residuals are obtained from the test predictions of the repeated runs.
    By plotting them against the predicted values we can identify prediction biases more quickly.
    
\begin{figure}[bt]
        \centering
        \includegraphics[width=1.0\linewidth]{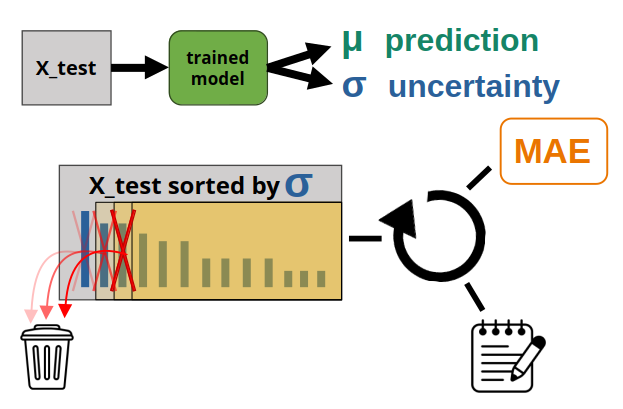}
        \caption{
            \textbf{Procedure to obtain a confidence curve}: 
            With this routine a plot can be created that visually quantifies how informed a model's uncertainty estimates are.
            First, an uncertainty-aware machine learning model is trained and predicts on an unseen test dataset.
            The machine learning model produces a prediction value $\mu$ and uncertainty estimates $\sigma$ for each predicted target. 
            Then we sort the predictions by the corresponding uncertainty estimates $\sigma$ in decreasing order. 
            The loop begins by calculating the error score (e.g. MAE) on the test dataset. 
            This value is stored.
            Then the sample with the most uncertain predicted label is removed. 
            This process is repeated until only one test sample remains.
            Finally, the error scores are plotted against the percentage of discarded samples. 
            }
        \label{fig:uncertainty evaluation protocol}
    \end{figure}

    The \textbf{uncertainty estimation} requires another method of evaluation.
    A central question in uncertainty-aware modeling is whether the predicted uncertainties can be trusted to effectively indicate prediction errors. 
    To evaluate this, we adopt the protocol proposed by Scalia et al. \cite{scalia2019, pernot2022}, which provides a systematic way to assess the quality of uncertainty estimates. 
    The underlying assumption is that, if the estimates are reliable, samples associated with high predicted uncertainty are more likely to correspond to large prediction errors. 
    Consequently, when such uncertain samples are removed from the evaluation dataset, the average prediction error on the remaining subset should decrease. 
    This relationship can be visualized through a confidence curve, which quantifies how the model’s prediction error evolves as uncertain predictions are excluded in decreasing order. 
    We construct the confidence curves as follows (Fig. \ref{fig:uncertainty evaluation protocol}): 
    \begin{enumerate}
    
        \item \textbf{Rank predictions by uncertainty.} First, the model's predictions on a test dataset are ranked in decreasing order based on their associated uncertainty estimates. This means that the predictions, which the model is most confident about, will be at the bottom of the ranking, and the least confident predictions at the top. 
        
        \item \textbf{Remove data points with highest uncertainty.} Next, samples are removed from the test set one by one from the top, starting with the data point carrying the highest uncertainty. 
        
        \item \textbf{Calculate prediction error on remaining test data.} As data points are removed, the average prediction error is calculated for the remaining test data points. The error metric can be the mean absolute error (MAE) or root mean squared error (RMSE). 
        The choice depends on whether it is important to detect large prediction errors (using RMSE) or if outliers should not dominate the metric (using MAE). 
        
        \item \textbf{Plot error vs. percent of discarded data.} Finally, the confidence curve is generated by plotting the error score for each subset of the test dataset against the percentage of test samples that have been discarded. 
        
    \end{enumerate}   

    In the confidence curve plot, we illustrate the ideal scenario by removing samples from the test dataset that exhibit the highest prediction error. The resulting curve, also called \textit{ideal curve}, represents the ideal behavior of the confidence curve if the uncertainty estimates perfectly rank the predictions from highest to lowest error. The resulting plot allows a visual evaluation of how meaningful the model’s uncertainty estimation is.

    The dataset is divided into a training and a test subset.
    The model is trained on the training set and subsequently used to generate predictions on the unseen test set.
    We evaluate the estimated uncertainty only for predictions on the test set.
    To reduce split-dependent variance, this test set evaluation is repeated over 10 randomly seeded train-test splits. Scalar uncertainty metrics such as MACE and Sharpness are computed on the held-out test split of each run and then averaged across these runs. When shown as plots, the calibration and sharpness diagnostics are constructed from the pooled test predictions across the same 10 train-test splits.
    
    Some uncertainty-aware machine learning models are capable of providing separate estimates for epistemic and aleatoric uncertainty. 
    In such cases, each type of uncertainty can be evaluated independently following the same procedure described above. Then we rank the prediction by the uncertainty estimate that we chose.  
    


    A practically useful uncertainty estimate should produce a confidence curve that slopes downwards like the ideal curve. 
    This indicates that as more data points with high uncertainty are discarded, moving towards the higher percentages on the x-axis, the average prediction error of the remaining data points decreases. 
    In other words, observing a downward slope means that the model's uncertainty estimates are effectively identifying the predictions that are more likely to be inaccurate. 
    Conversely, if the confidence curve is flat or non-decreasing, the plot suggests that the model's uncertainty estimates are not informative. 
    The steeper the decline of the confidence curve, i.e., the closer it follows the ideal curve, the more reliable and well-calibrated the uncertainty estimation is.    
    
%% file: 3.1_discussion_approach.tex
\subsection{Existing uncertainty-estimation strategies in materials science}
\label{sec:uq_literature}

    Existing literature in materials science and scientific machine learning broadly agrees that uncertainty quantification (UQ) must be treated as part of the predictive task rather than as a post hoc add-on.
    Acar's review of small-scale materials science organizes this problem explicitly by uncertainty source, algorithm, and length scale, and highlights that aleatoric and epistemic effects arise from different physical and modeling mechanisms \cite{acar2021}.
    Psaros et al. complement this view with a broader scientific machine learning comparison of uncertainty-aware methods and evaluation metrics, showing that method choice should be discussed together with how uncertainty quality is assessed \cite{psaros2023}.

    For materials-property prediction in particular, Tavazza et al. compare three routes to prediction intervals on 12 JARVIS-DFT properties: quantile-loss models, direct learning of the interval width, and Gaussian processes \cite{tavazza2021}.
    They report a slight advantage for direct learning of instance-wise uncertainty because it is easy to fit and often reduces over- and underestimation of the error.
    Our approximate Bayesian neural network is conceptually close to that objective because it also predicts sample-dependent uncertainty through the Gaussian negative log-likelihood loss.
    The important difference is that we do not stop at a single predictive spread.
    We further decompose the predictive variance into aleatoric and epistemic parts and test whether these components are informative for magnetic-property prediction.

    The literature also cautions that obtaining an uncertainty value is not the same as obtaining a trustworthy uncertainty estimate.
    Palmer et al. show that raw bootstrap ensemble variance is often miscalibrated and becomes much more useful only after calibration \cite{palmer2022}.

    For the evaluation protocol, the most relevant materials-specific reference is Tran et al., who note that materials screening lacks a standard procedure for comparing uncertainty estimates and therefore advocate reporting dedicated UQ metrics and figures rather than relying on prediction error alone \cite{tran2020}.
    This recommendation is implemented explicitly in the surrogate Curie temperature study of Section~\ref{sec:surrogate_tc_study}, where we compare the models with calibration curves, sharpness plots, and scalar calibration metrics generated with the Uncertainty Toolbox \cite{chung2021} before carrying the same workflow into the later magnetic-property studies.
    The confidence curve analysis follows that logic.
    By discarding the predictions with the largest estimated uncertainty and tracking the remaining error, the confidence curve tests whether uncertainty is operationally useful for selective prediction and screening.
    Pernot shows that confidence curves can be interpreted against a probabilistic reference, allowing them to assess whether larger predicted uncertainties are associated with larger errors and to test the calibration and tightness of prediction uncertainties.
    \cite{pernot2022}

    Our approach is therefore deliberately pragmatic.
    We use regression models that scale to the magnetic datasets studied here, but we judge them by a stricter criterion than point accuracy alone.
    A useful model must provide uncertainty estimates that are empirically checkable, and practically relevant for ranking, screening, and selective trust in predicted magnetic properties.

%% file: 3a_surrogate_TC_data_study.tex
\section{Curie Temperature surrogate study}
\label{sec:surrogate_tc_study}
This section compares three uncertainty-aware regression models on a publicly available Curie temperature (\tc) dataset.
Further, it defines the uncertainty quantification (UQ) workflow that is used later in the manuscript.

To quantify uncertainty we use not only point-prediction error but several methods. 
We analyse the prediction results with the confidence curve plot (Sec. \ref{sec:model_evaluation}) as it is a practical diagnostic tool. 
Following the recommendation of Tran et al., we also look at calibration and sharpness plots to assess the statistical reliability of the uncertainty estimates \cite{tran2020, chung2021}.

The Curie temperature dataset assembled by Nelson and Sanvito \cite{nelson2019a} and distributed through GitHub (\href[]{https://github.com/msg-byu/ML-for-CurieTemp-Predictions}{msg-byu/ML-for-CurieTemp-Predictions}).
It remains a useful reference point because it is public, based on chemical composition, and directly tied to magnetic-material screening.

\subsection{Unified workflow}
All three Curie temperature models are evaluated with the same pipeline to make them comparable.
The data split, preprocessing, and evaluation protocol are held fixed.

\begin{enumerate}
	\item The dataset is restricted to numerical descriptors, entries without target values are removed, and constant features are discarded.
	\item The data are split into train and test subsets with a test fraction of 0.3.
	\item The full split--fit--predict pipeline is repeated for 10 random seeds (123--132) to reduce dependence on a single favorable partition.
	\item A standard scaler is fitted on the training set and then applied to both train and test sets.
	\item Point-prediction performance is summarized with MAE, RMSE, and $R^2$, computed on the held-out test split of each run and then averaged over the 10 random seeds.
	\item Model fit is visualized as parity plots for one representative test set on a single seed.
	\item UQ is evaluated only on the unseen test set. Scalar UQ metrics such as MACE and Sharpness are computed on each held-out test split and then averaged over the 10 random seeds. The calibration and sharpness plots are constructed from the pooled test predictions across those 10 test splits, while the confidence curves retain the repeated-split evaluation.
\end{enumerate}

\subsection{Models}
Three representative uncertainty-aware regression models are compared: 
Gaussian process regression (GPR) \cite{rasmussen2006}, 
random forest regression with an additional bagging layer (RFR+bagging) \cite{breiman2001}, 
and a Bayesian neural network (BNN) implemented through Monte Carlo dropout \cite{gal2016b, kendall2017} (Sec. \ref{sec:bnn}).
Detailed hyperparameters are summarized in \hyperref[app:tc-study-details]{Appendix A}.

The key distinction between the models is the meaning of their uncertainty output.
The definitions of aleatoric uncertainty (Sec. \ref{sec:aleatoric_uncertainty}), epistemic uncertainty (Sec. \ref{sec:epistemic_uncertainty}), and their combination into total uncertainty (Eq. \ref{eq:total uncertainty}) were introduced earlier in sections
Here we only recall how each model exposes these quantities in practice.

\textbf{Gaussian process regression} \\
The GPR model uses a Mat\'ern kernel with an additive white-noise term.
It returns a predictive mean and predictive standard deviation.
In the present study, that standard deviation is interpreted as a total predictive uncertainty.
The white-noise term accounts for observation noise, but it should not be overinterpreted as a full heteroscedastic decomposition into separate aleatoric and epistemic channels. \cite{rasmussen2006}

\textbf{Random forest regression with bagging} \\
The RFR+bagging model uses a random forest regressor as the base learner inside an additional bagging ensemble.
The final prediction is the ensemble mean.
The spread across ensemble members is used as a heuristic epistemic uncertainty estimate.
This interpretation is common in ensemble methods. 
However, Palmer et al. suggest that  ensemble spread is not automatically calibrated, which is why explicit calibration diagnostics are necessary. \cite{palmer2022}

\textbf{Bayesian neural network} \\
The BNN uses Monte Carlo dropout as described in Sec. \ref{sec:bnn}, where dropout remains active during inference so that multiple stochastic forward passes approximate a predictive distribution \cite{gal2016b}.
Here the BNN is analyzed through three outputs: total uncertainty, aleatoric uncertainty, and epistemic uncertainty.
That decomposition is useful because if the newtork is well trained, it reveals whether it is uncertain because the training data is intrinsically noisy or because the model lacks knowledge in a region of the feature space \cite{kendall2017}.

\subsection{Surrogate data distribution}
Before comparing the models, it is useful to inspect the target distribution.
The Curie temperature target is not uniformly distributed across the dataset's full temperature range.
That matters because both point-prediction difficulty and uncertainty behavior depend on how densely different target regions are represented in the training data.

\begin{figure}[H]
	\centering
	\plotifexists{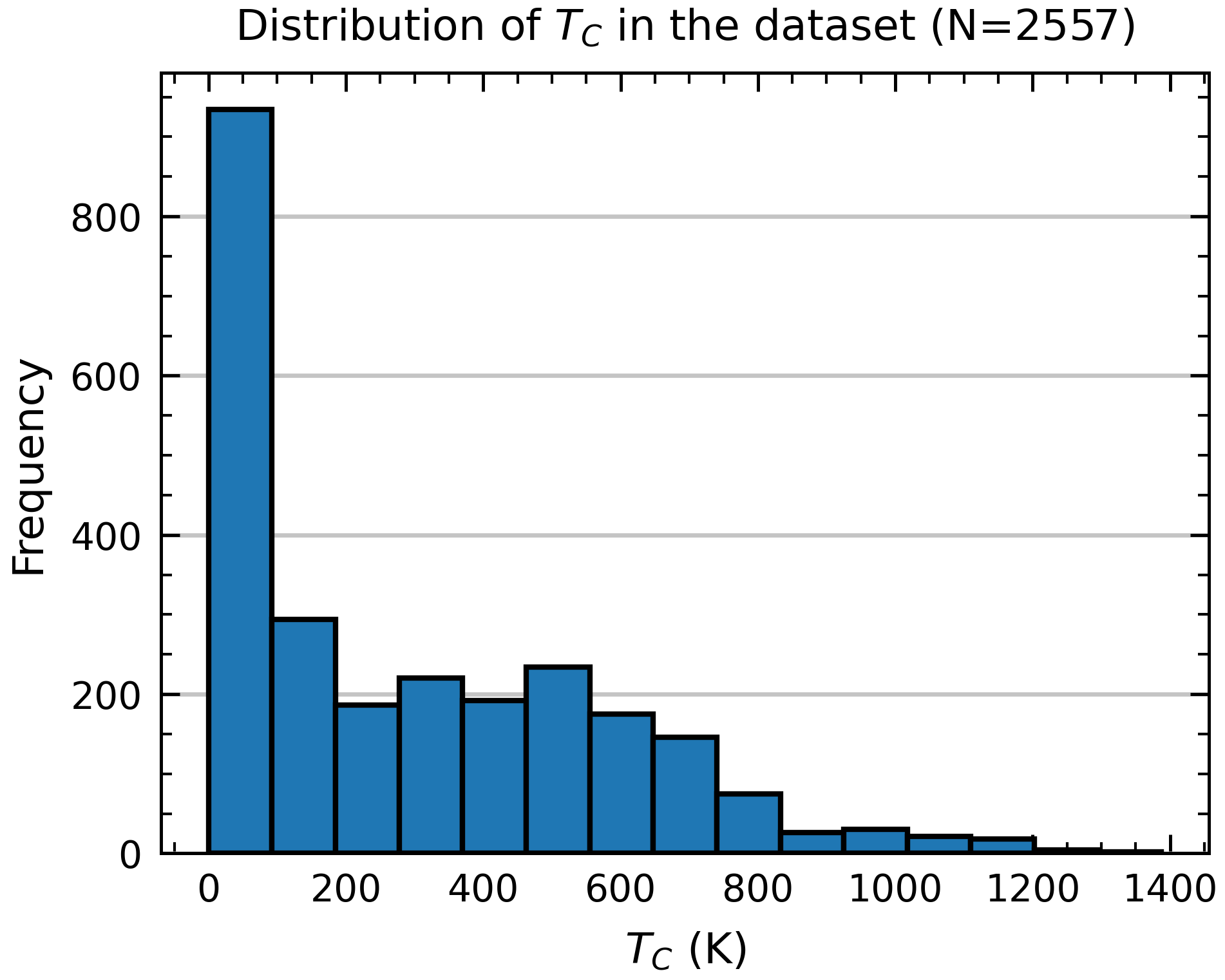}
	\caption{
		Distribution of Curie temperature values in the surrogate dataset.
		Dense regions correspond to well-represented parts of the target space, while a sparse tail is expected to be harder to predict.
		Thus, point-predictions and uncertainty estimates are expected to be more reliable in the dense than in the sparse region.
	}
	\label{fig:tc-distribution}
\end{figure}

Figure~\ref{fig:tc-distribution} shows a right-skewed target distribution with denser left regions and a sparse tail to the right.
A useful uncertainty model should reflect that imbalance by assigning larger uncertainty to underrepresented target regions.

\subsection{Point-prediction results}
Table~\ref{tab:tc-fit-summary} summarizes the test set predictive performance.
All reported metrics are computed on the held-out test split of each of the 10 randomly seeded train-test runs and then summarized as mean $\pm$ standard deviation.

\begin{table}[H]
	\centering
	\begin{tabularx}{\columnwidth}{>{\raggedright\arraybackslash}X c c c}
	\toprule
		Model & MAE (K) & RMSE (K) & $R^2$ \\
		\midrule
		GPR      & $78.52 \pm 3.47$ & $122.57 \pm 6.97$ & $0.809 \pm 0.023$ \\
		RFR+bag. & $76.15 \pm 3.30$ & $124.83 \pm 7.49$ & $0.801 \pm 0.025$ \\
		BNN      & $95.85 \pm 6.05$ & $156.49 \pm 9.52$ & $0.689 \pm 0.031$  \\
		\bottomrule
	\end{tabularx}
	\caption{
		Predictive performance on the Curie temperature test set averaged over ten repetitions with different system-wide random seeds. Values are reported as mean $\pm$ standard deviation.
	}
	\label{tab:tc-fit-summary}
\end{table}

RFR+bagging and GPR both have a good $R^2$-score of little over 80\% and their error scores are also quite close together.
The BNN reaches a reasonable predictive level with an $R^2$-score of just under 70\%. 

\begin{figure}[H]
	\centering
	\begin{subfigure}[t]{0.32\linewidth}
		\centering
		\plotifexists{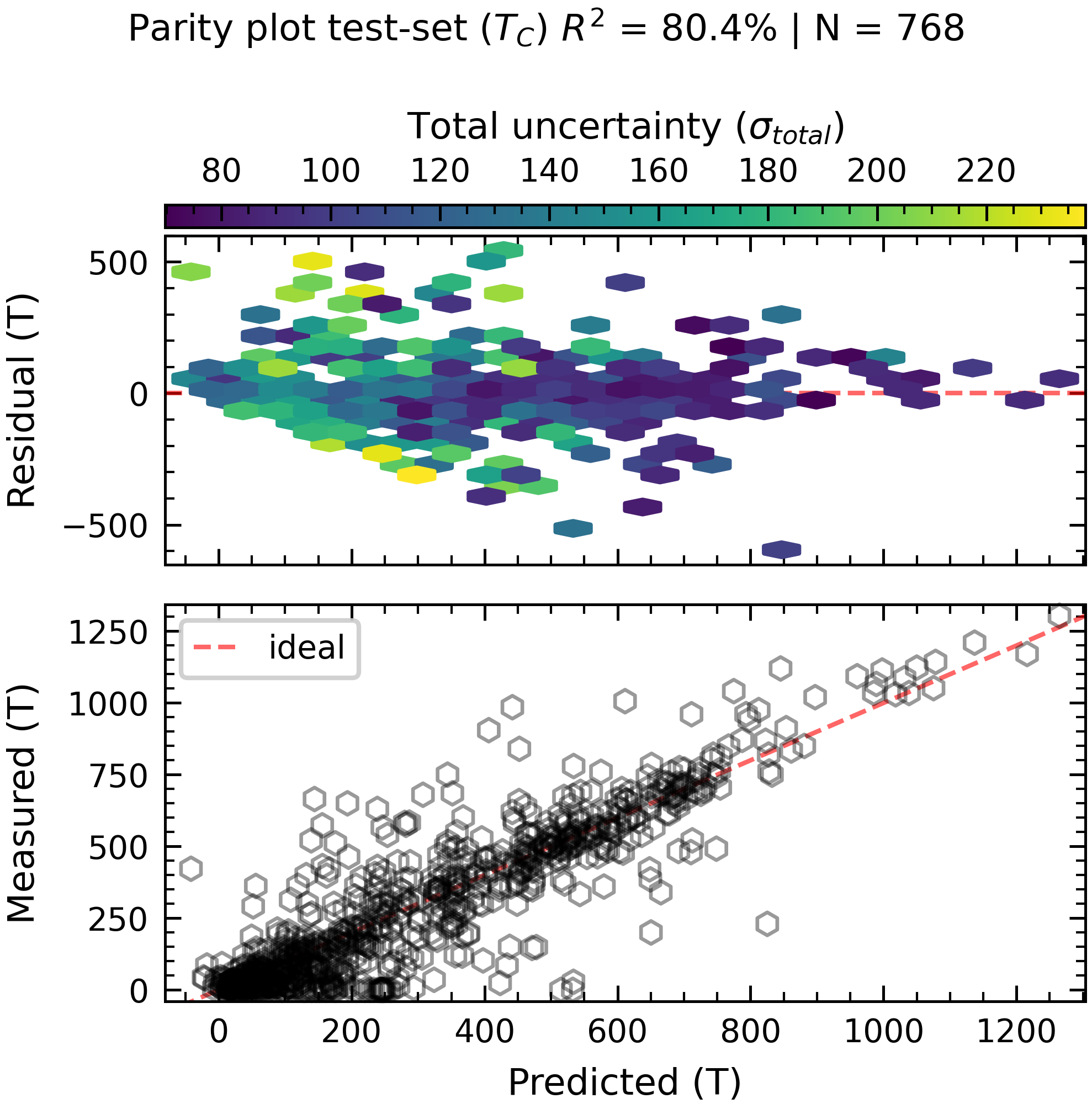}
		\caption{GPR.}
	\end{subfigure}
	\hfill
	\begin{subfigure}[t]{0.32\linewidth}
		\centering
		\plotifexists{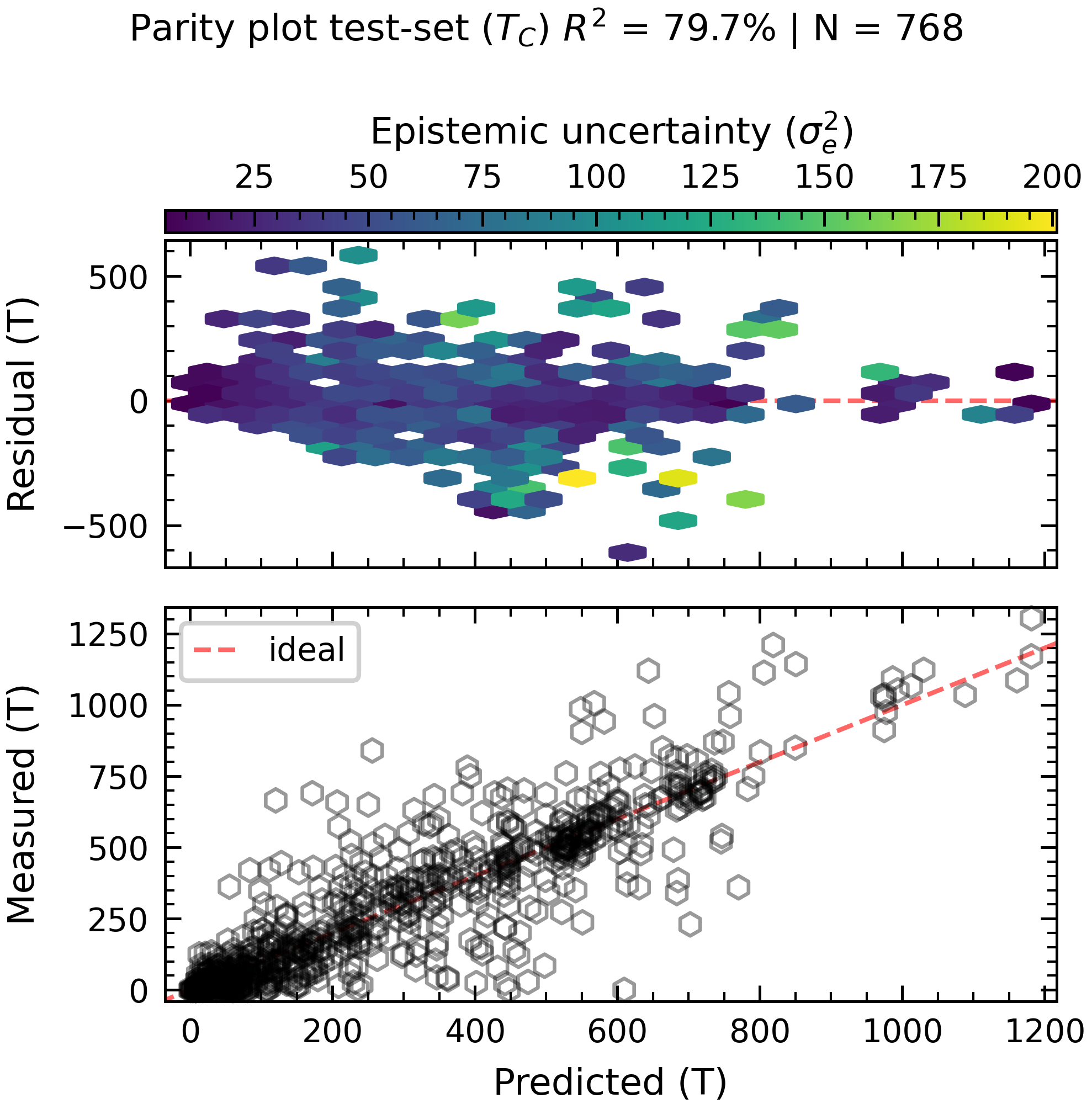}
		\caption{RFR+bagging.}
	\end{subfigure}
	\hfill
	\begin{subfigure}[t]{0.32\linewidth}
		\centering
		\plotifexists{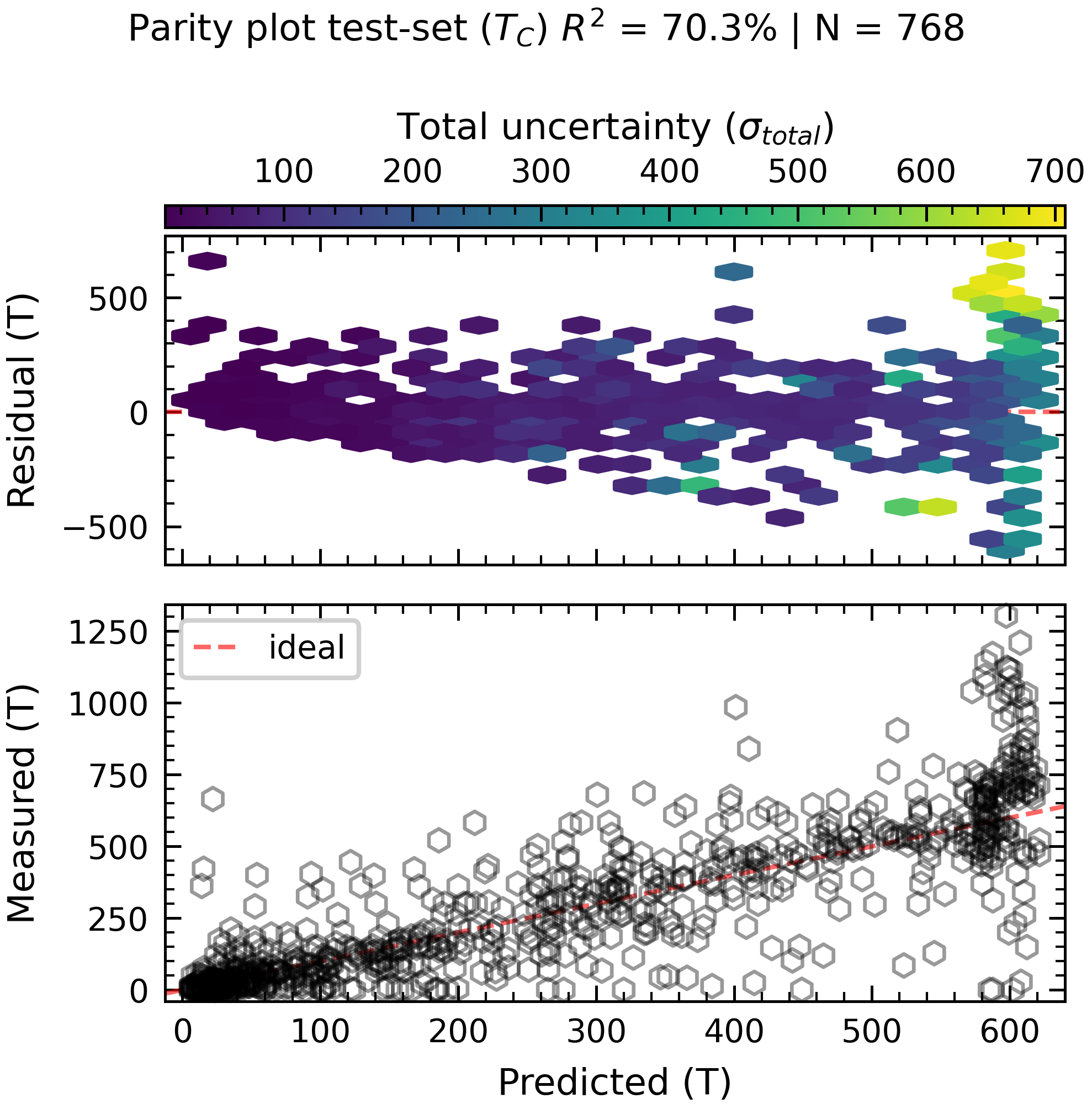}
		\caption{BNN.}
	\end{subfigure}
	\caption{
		Representative test set hexbin parity plots for the three Curie temperature models from one single random seed.
		For the lower panel, the ideal behavior is a narrow concentration along the diagonal.
		The upper panel plots residuals against predictions in hexbins.
		A lighter color indicates higher uncertainty.
		Hexbins farther from the red dashed line indicate larger predictive error.
	}
	\label{fig:tc-hexbins}
\end{figure}

\textbf{Hexbin Plots for model fit} \\
The parity hexbins in Figure~\ref{fig:tc-hexbins} complement Table~\ref{tab:tc-fit-summary} by showing both prediction quality and uncertainty estimates relative to residuals.
The upper panel shows the residuals against predictions in hexbins, where a lighter color indicates higher uncertainty.
Compared with ordinary scatter plots, hexbin plots remain readable when many points overlap.
Hexbins farther from the red dashed line indicate larger predictive error.
The lower panel shows the predicted versus true values, where a narrow concentration along the diagonal indicates good predictive performance.

Figure~\ref{fig:tc-hexbins} should be read together with Table~\ref{tab:tc-fit-summary}.
Both classical baselines have broader residuals in the low to middle \tc range, but capture the high \tc tail better than the BNN.
The BNN plot shows that the network captures the overall dependence, but with a wider residual spread, especially in the high \tc range.
The hexbin plots some correlation between predictive uncertainty and residuals, but a clear pattern is hard to spot with the naked eye.
The GPR has many highly uncertain predictions, the RFR+bagging has a few, and the BNN is mainly uncertain in the high \tc tail.
Also the BNN model's uncertainty estimates have a higher range from 0 to 700 K, while the GPR and RFR+bagging models have a range from 0 to 235 K or 200 K, respectively.

\subsection{Uncertainty quantification results}
\label{sec:tc-uq-results}
    The hexbin plots in Figure~\ref{fig:tc-hexbins} did not reveal a clear correlation between predictive uncertainty and residuals.
    So, the next step is to quantify uncertainty with more formal diagnostics.
    The UQ analysis is performed on the test set only.
    This is intentional because the relevant question is whether uncertainty estimates remain useful on unseen data.
    Following Tran et al. \cite{tran2020}, three complementary diagnostics are used.

\begin{itemize}
	\item \textbf{Calibration curve:} 
    	compares nominal coverage with empirical coverage.
    	A perfectly calibrated model lies on the diagonal.
    	Curves below the diagonal indicate overconfidence, while curves above the diagonal indicate underconfidence.
        \cite{chung2021, tran2020, chung2021}

    \item \textbf{Sharpness:} 
    	measures how concentrated the predictive distributions are.
    	In this context, smaller sharpness values correspond to narrower predictive intervals.
    	Sharpness is desirable only when calibration is also good.
        \cite{tran2020, chung2021}
     
	\item \textbf{Confidence curve:} 
    	measures whether uncertainty is useful for ranking predictions by trustworthiness.
    	If discarding the least confident predictions reduces error, then the uncertainty estimate is operationally informative. 
        \cite{scalia2019, pernot2022}
\end{itemize}

\textbf{Scalar UQ metrics} \\

Moreover, there are two scalar terms used below that act as summary statistics.
\textit{MACE} stands for mean absolute calibration error. 
It summarizes the average absolute gap between nominal coverage and observed coverage across confidence levels.
Lower MACE means that stated confidence levels are closer to the frequencies actually observed in the test set.
For example, a well-calibrated 80\% interval should contain roughly 80\% of the true targets.
\cite{chung2021}
\textit{Sharpness} is a summary of interval width.
A sharp model produces concentrated predictive intervals, but a sharp yet miscalibrated model is not trustworthy.
For that reason, MACE and sharpness are interpreted together rather than in isolation \cite{tran2020, chung2021}.

Table~\ref{tab:tc-uq-summary} summarizes these two uncertainty metrics of the three models.
The reported MACE and Sharpness values are computed on the held-out test split of each of the 10 randomly seeded train-test runs and then summarized as mean $\pm$ standard deviation.

\begin{table}[H]
	\centering
	\begin{tabularx}{\columnwidth}{>{\raggedright\arraybackslash}X c c c}
		\toprule
		Model / uncertainty & MACE & Sharpness \\
		\midrule
		GPR total & $0.099 \pm 0.010$ & $159.10  \pm 2.38$ \\
		RFR+bag.  & $0.254 \pm 0.008$ & $45.09   \pm 1.59$  \\
		BNN total & $0.039 \pm 0.014$ & $122.761 \pm 14.47$ \\
		BNN aleatoric & $0.055 \pm 0.028$ &  $117.36 \pm 15.41$ \\
		BNN epistemic & $0.256 \pm 0.030$ & $35.60 \pm 2.09$  \\
		\bottomrule
	\end{tabularx}
	\caption{
	    Test set uncertainty metrics for Curie temperature prediction averaged over ten randomly seeded repetitions. Values are reported as mean $\pm$ standard deviation. 
		Lower is better for MACE and sharpness.
	}
	\label{tab:tc-uq-summary}
\end{table}

Several conclusions follow directly from Table~\ref{tab:tc-uq-summary}.
First, the BNN total uncertainty provides the strongest predictive uncertainty estimate in this benchmark.
Its MACE is the lowest among all compared uncertainty outputs, while its sharpness remains substantially lower than that of GPR.
This combination indicates that the total BNN intervals are both comparatively well calibrated and not excessively broad.
Second, GPR provides a reliable uncertainty estimate in this benchmark, though worse than the BNN.
Its MACE is much lower than the RFR+bagging value, although its predictive intervals are broader and more conservative than those of the BNN total estimate.
Third, RFR+bagging does not have well calibrated uncertainty estimation even tough its point-prediction were good.
Its intervals are narrow, yet the high MACE shows that those intervals are not statistically trustworthy without further calibration.
That pattern is consistent with broader observations that raw ensemble spread should not be assumed to be calibrated automatically \cite{palmer2022}.
Fourth, the BNN is most informative when its uncertainty channels are combined.
The total and aleatoric outputs are both well calibrated, whereas the isolated epistemic term performs poorly.
Aleatoric uncertainty can therefore reliably capture the intrinsic noise in this dataset.  

\subsection{Calibration curves}
\label{sec:tc-calibration}
We can also plot the calibration curves to visualize the same information that the MACE metric summarizes.
Additionally, we can determine if a model systematically overestimates or underestimates its uncertainty.

\begin{figure}[H]
	\centering
	\begin{subfigure}[t]{0.32\linewidth}
		\centering
		\plotifexists{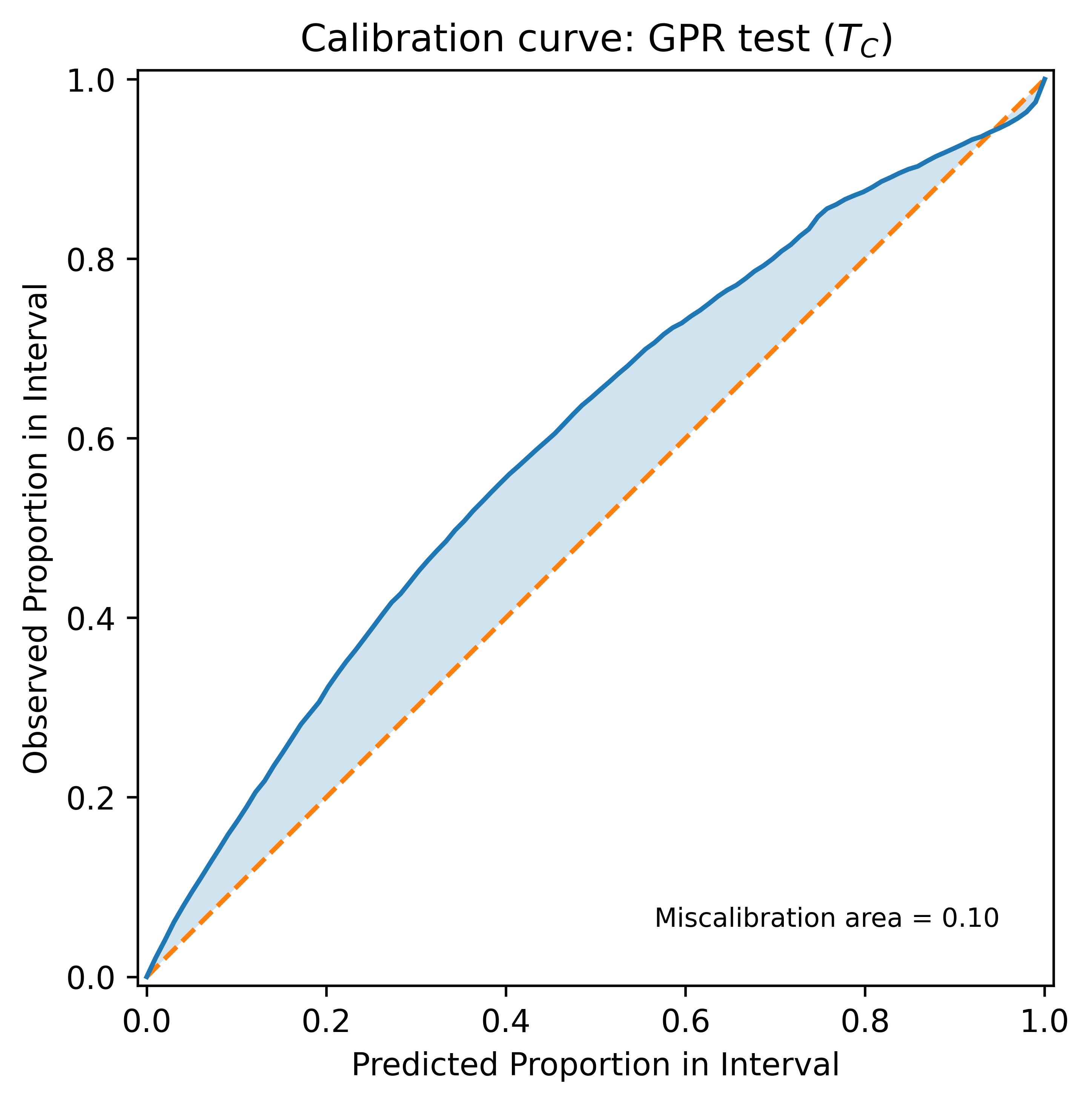}
		\caption{GPR total uncertainty.}
	\end{subfigure}
	\hfill
	\begin{subfigure}[t]{0.32\linewidth}
		\centering
		\plotifexists{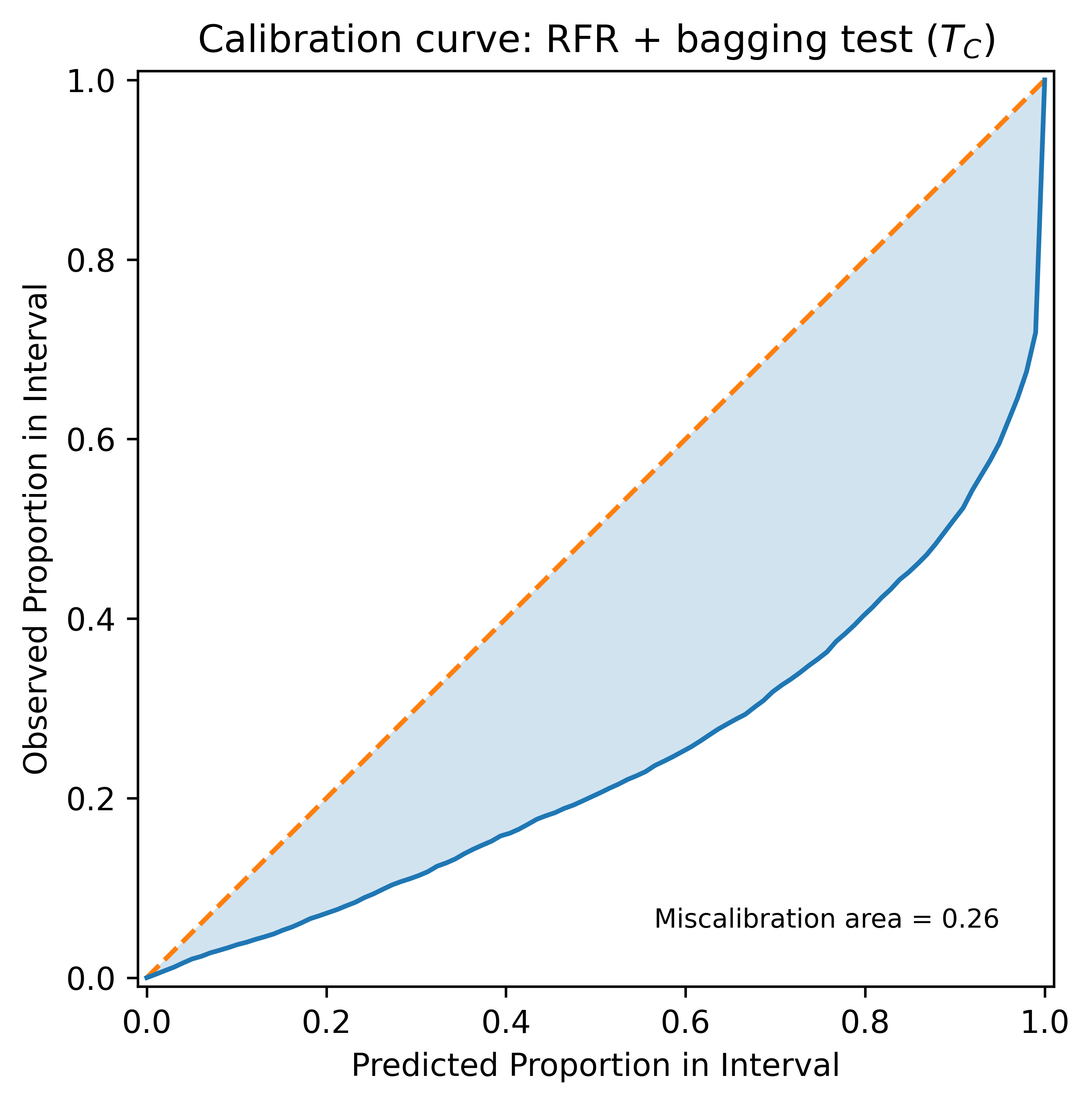}
		\caption{RFR+bagging epistemic uncertainty.}
	\end{subfigure}
	\hfill
	\begin{subfigure}[t]{0.32\linewidth}
		\centering
		\plotifexists{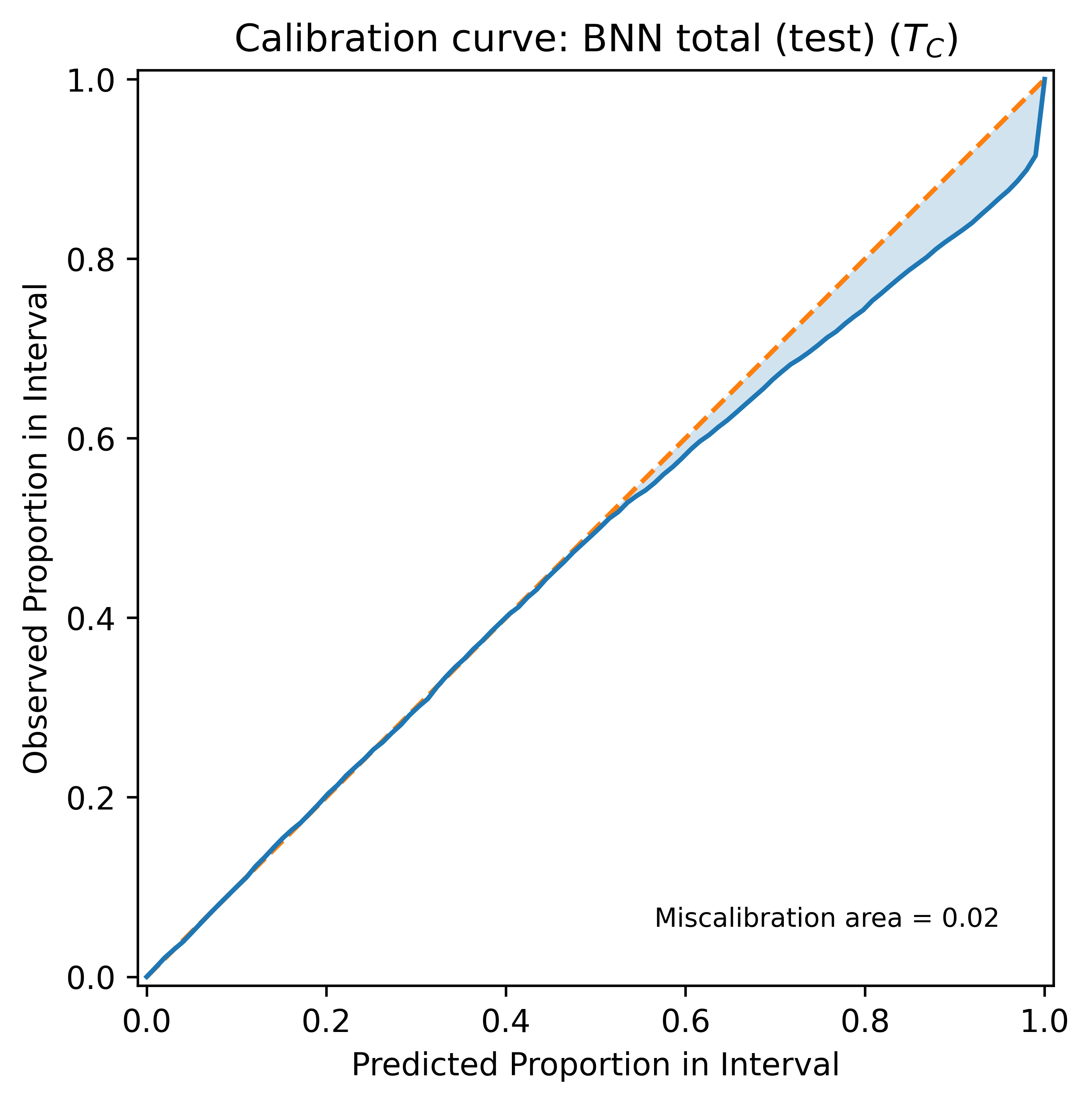}
		\caption{BNN total uncertainty.}
	\end{subfigure}
	\caption{
		Test set calibration curves for the three Curie temperature models constructed from the pooled test predictions across the 10 randomly seeded train-test splits.
		The diagonal corresponds to perfect calibration.
		Curves below the diagonal indicate systematic overconfidence, while curves above the diagonal indicate systematic underconfidence.
	}
	\label{fig:tc-calibration}
\end{figure}

Figure~\ref{fig:tc-calibration} confirms the trend already visible in Table~\ref{tab:tc-uq-summary}.
The BNN total uncertainty remains closest to the diagonal over most of the region, which is consistent with its lowest MACE.
GPR also stays close to the diagonal, but is clearly underconfident.
RFR+bagging deviates strongly and exposed the model's overconfidence in its predictions.
It is therefore indicated that the BNN total uncertainty offers the best calibrated predictive intervals among the compared models for this task. 

As the BNN exposes multiple uncertainty channels, it is useful to inspect its calibration behavior in more detail.

\begin{figure}[H]
	\centering
	\begin{subfigure}[t]{0.32\linewidth}
		\centering
		\plotifexists{eu_TC_bnn__TC_test_total_calibration_curve.png}
		\caption{Total.}
	\end{subfigure}
	\hfill
	\begin{subfigure}[t]{0.32\linewidth}
		\centering
		\plotifexists{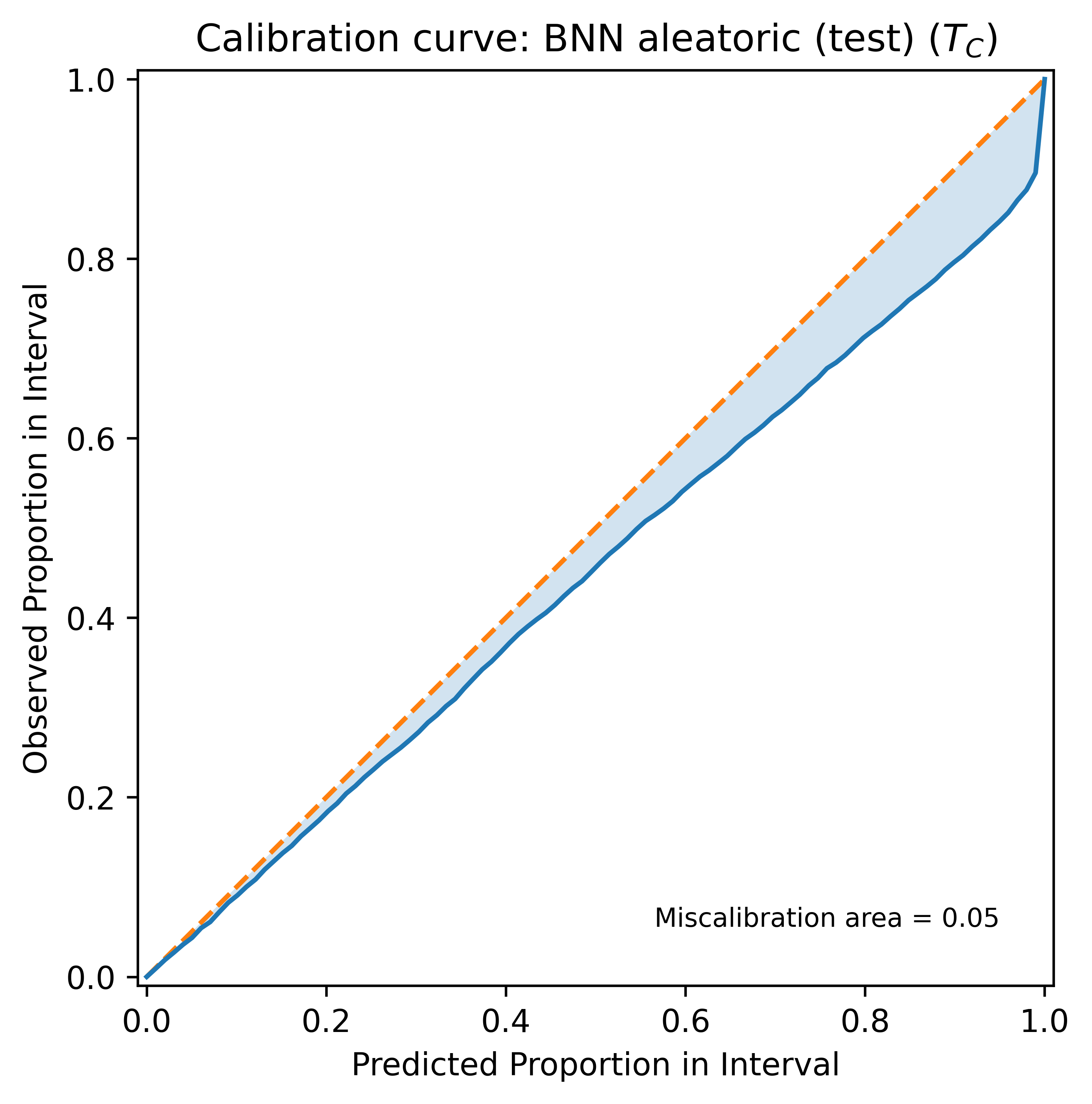}
		\caption{Aleatoric.}
	\end{subfigure}
	\hfill
	\begin{subfigure}[t]{0.32\linewidth}
		\centering
		\plotifexists{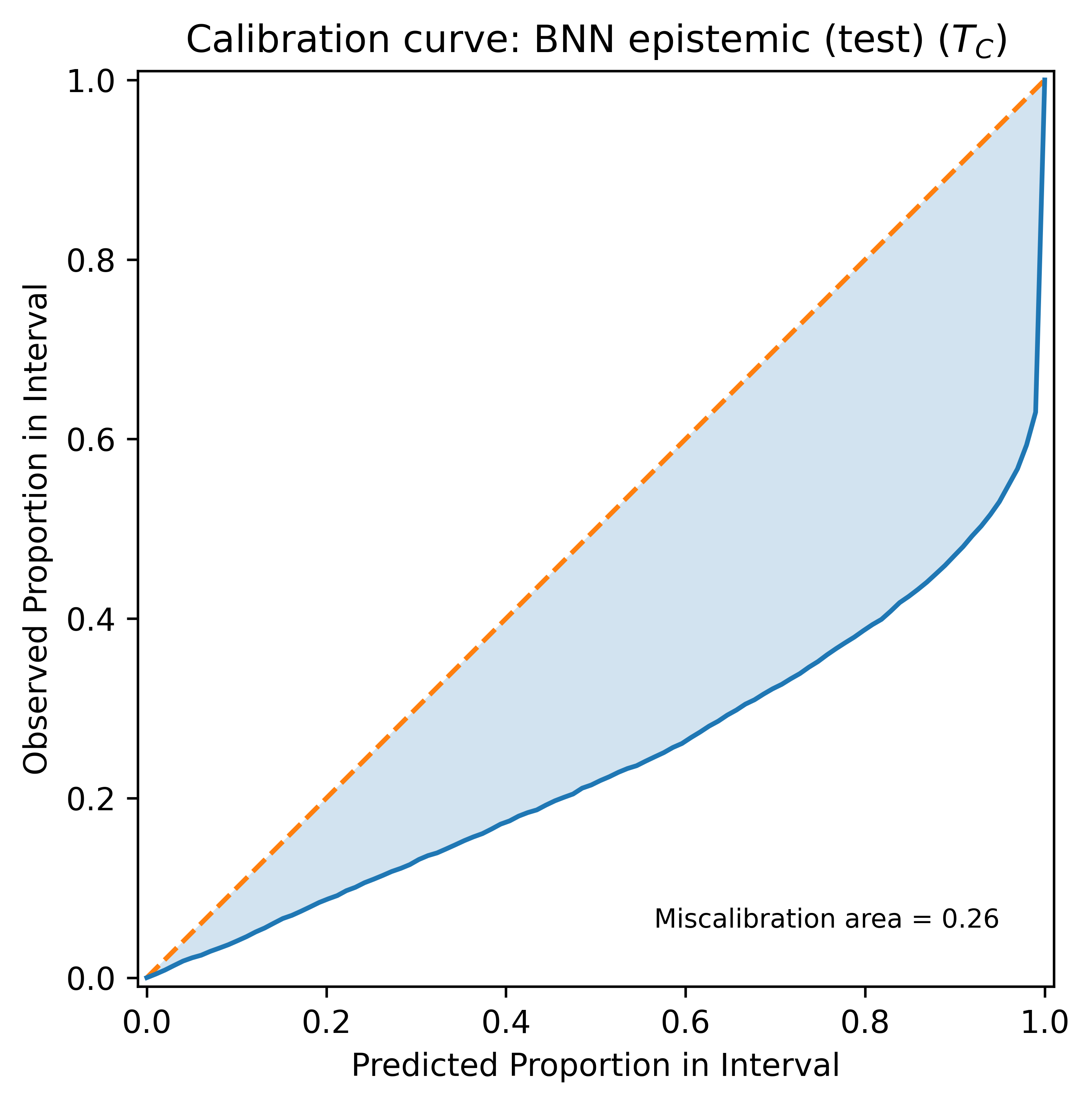}
		\caption{Epistemic.}
	\end{subfigure}
	\caption{
		Calibration curves for the three BNN uncertainty measures constructed from the pooled test predictions across the 10 randomly seeded train-test splits.
		The total and aleatoric uncertainties are substantially better calibrated than the isolated epistemic uncertainty in this model.
	}
	\label{fig:tc-bnn-calibration}
\end{figure}

Figure~\ref{fig:tc-bnn-calibration} shows that not every uncertainty channel of the BNN model should be interpreted with equal confidence.
The total and aleatoric curves remain close to the diagonal, whereas the epistemic curve departs much more strongly.
By contrast, the epistemic curve is too confident, meaning its predictive intervals are too narrow and do not contain the true targets as often as they should.
Thse plots match the results in Table~\ref{tab:tc-uq-summary} and additionally expose the BNN's systematic overconfidence. 
In practical terms, the decomposition is useful to analyse the strenghts of the models uncertainty estimates. 

\subsection{Sharpness plots}
\label{sec:tc-sharpness}
Sharpness plots complement calibration plots by visualizing the concentration of predictive intervals.
Putting it simply, calibration tests honesty, while sharpness tests concentration.
A model with broad but calibrated intervals may be statistically reliable but operationally conservative.
A model with very narrow intervals may look attractive until calibration reveals that it is overconfident.
The histogram bins of the sharpness plots show the distribution of uncertainty estimates. 
The dashed line indicates the mean sharpness value, which is also reported in Table~\ref{tab:tc-uq-summary}.
The further left this dashed line is (i.e. the more concentrated the predictive intervals are), the better.
\cite{chung2021, tran2020}

\begin{figure}[H]
	\centering
	\begin{subfigure}[t]{0.32\linewidth}
		\centering
		\plotifexists{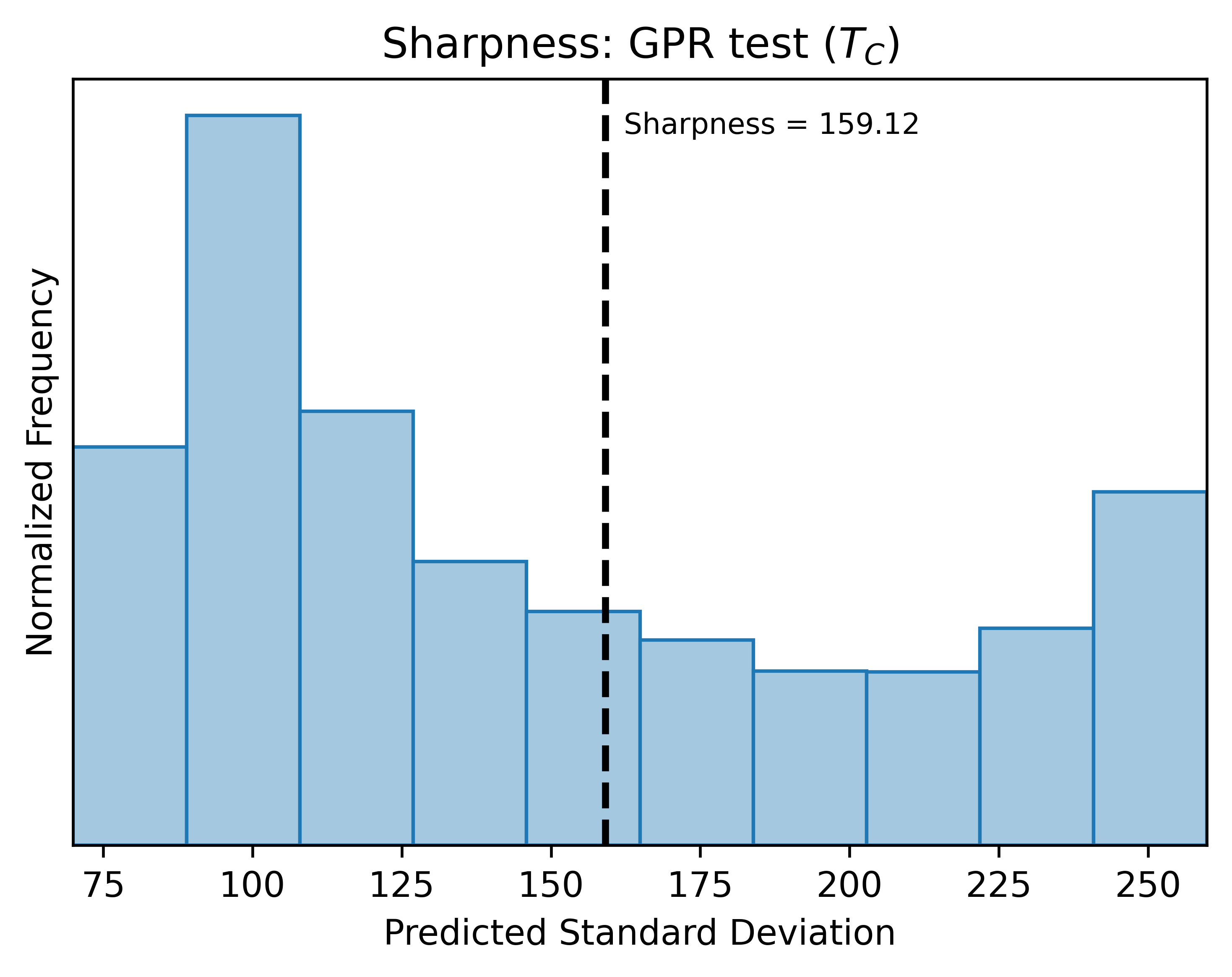}
		\caption{GPR total uncertainty.}
	\end{subfigure}
	\hfill
	\begin{subfigure}[t]{0.32\linewidth}
		\centering
		\plotifexists{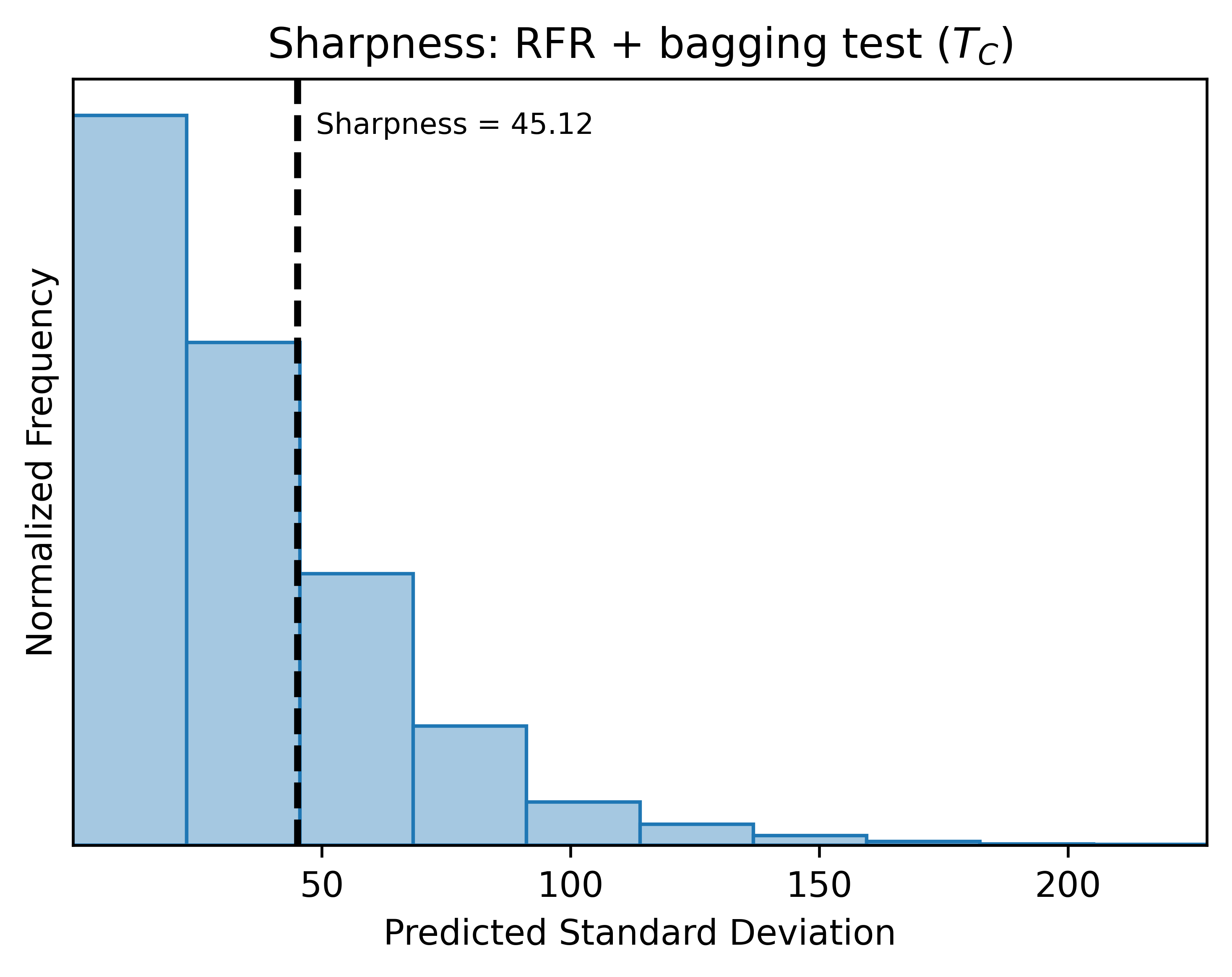}
		\caption{RFR+bagging epistemic uncertainty.}
	\end{subfigure}
	\hfill
	\begin{subfigure}[t]{0.32\linewidth}
		\centering
		\plotifexists{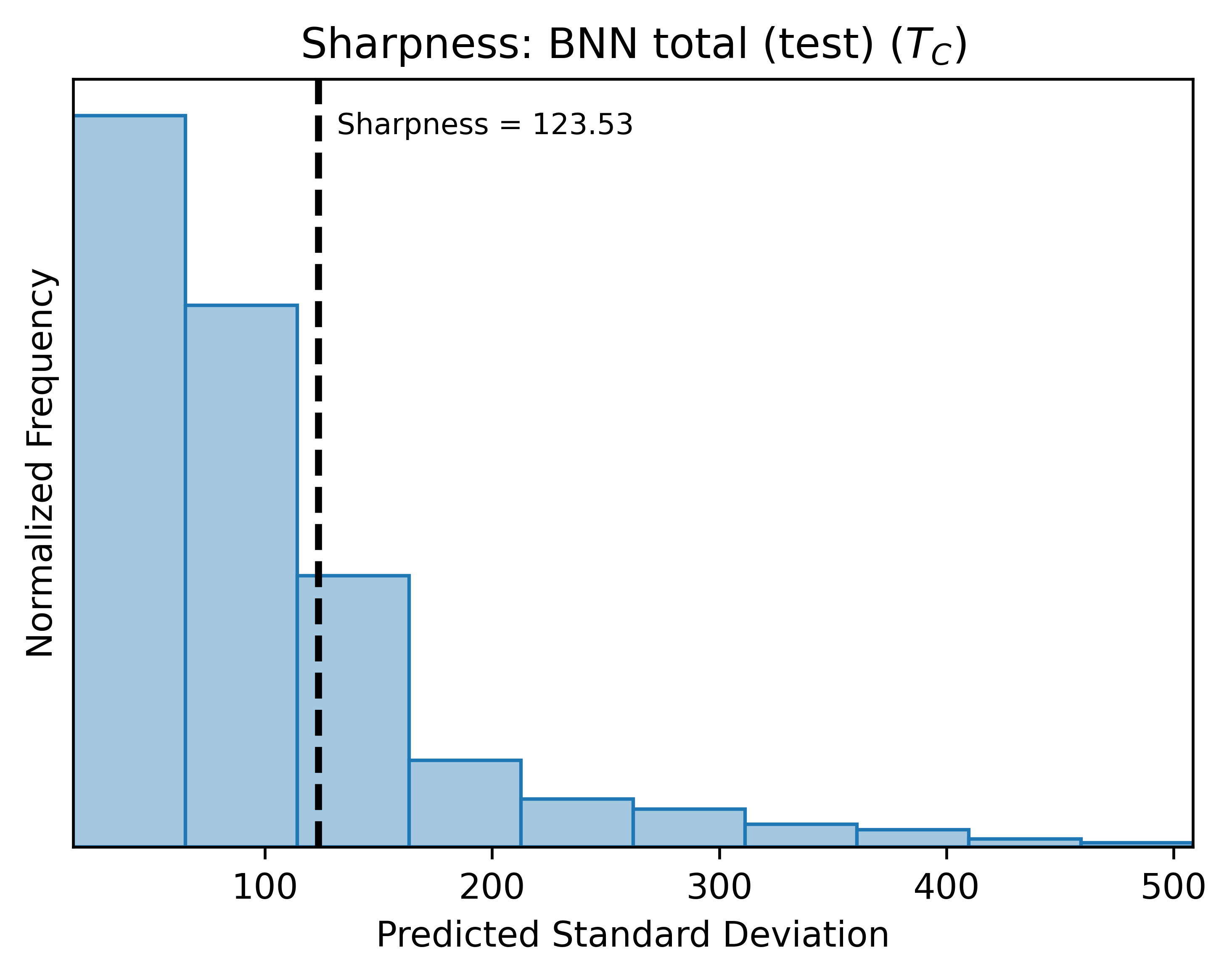}
		\caption{BNN total uncertainty.}
	\end{subfigure}
	\caption{
		Test set sharpness plots for the three Curie temperature models constructed from the pooled test predictions across the 10 randomly seeded train-test splits.
		Sharpness should always be interpreted together with calibration because narrow uncertainty is only desirable if it is calibrated.
	}
	\label{fig:tc-sharpness}
\end{figure}

The sharpness comparison (see Figure~\ref{fig:tc-sharpness}) highlights the main trade-off.
RFR+bagging is the sharpest model, but that apparent advantage is undermined by poor calibration.
The isolated BNN epistemic uncertainty is similarly sharp (see Table \ref{tab:tc-uq-summary}), yet likewise too poorly calibrated to serve as a reliable predictive interval by itself.
GPR model's uncertainty estimates are not sharp and also quite uniformly distributed across the target range. 
So, there the model is not very confident in any of its predictions. 
The BNN model scores visibly well: its total uncertainty is quite sharp and has a long tail in this plot.

\subsection{Confidence curves as an applied UQ diagnostic}
\label{sec:tc-confidence}
The confidence curve is the most application-oriented diagnostic in this section.
Instead of asking only whether uncertainty is statistically calibrated and precise, it asks whether uncertainty can improve decisions.
If the model retains only its most confident predictions, the prediction error should decrease under the assumption that the model can rank bad predictions with its uncertainty. 

In many practical workflows (i.e. surrogate-assisted design), the goal is not to trust every prediction equally.
The goal is to identify a subset of predictions that can be acted upon with higher confidence.
A strong confidence curve means that uncertainty is not only descriptive but also operational.
\cite{scalia2019, pernot2022}

\begin{figure}[H]
	\centering
	\begin{subfigure}[t]{0.32\linewidth}
		\centering
		\plotifexists{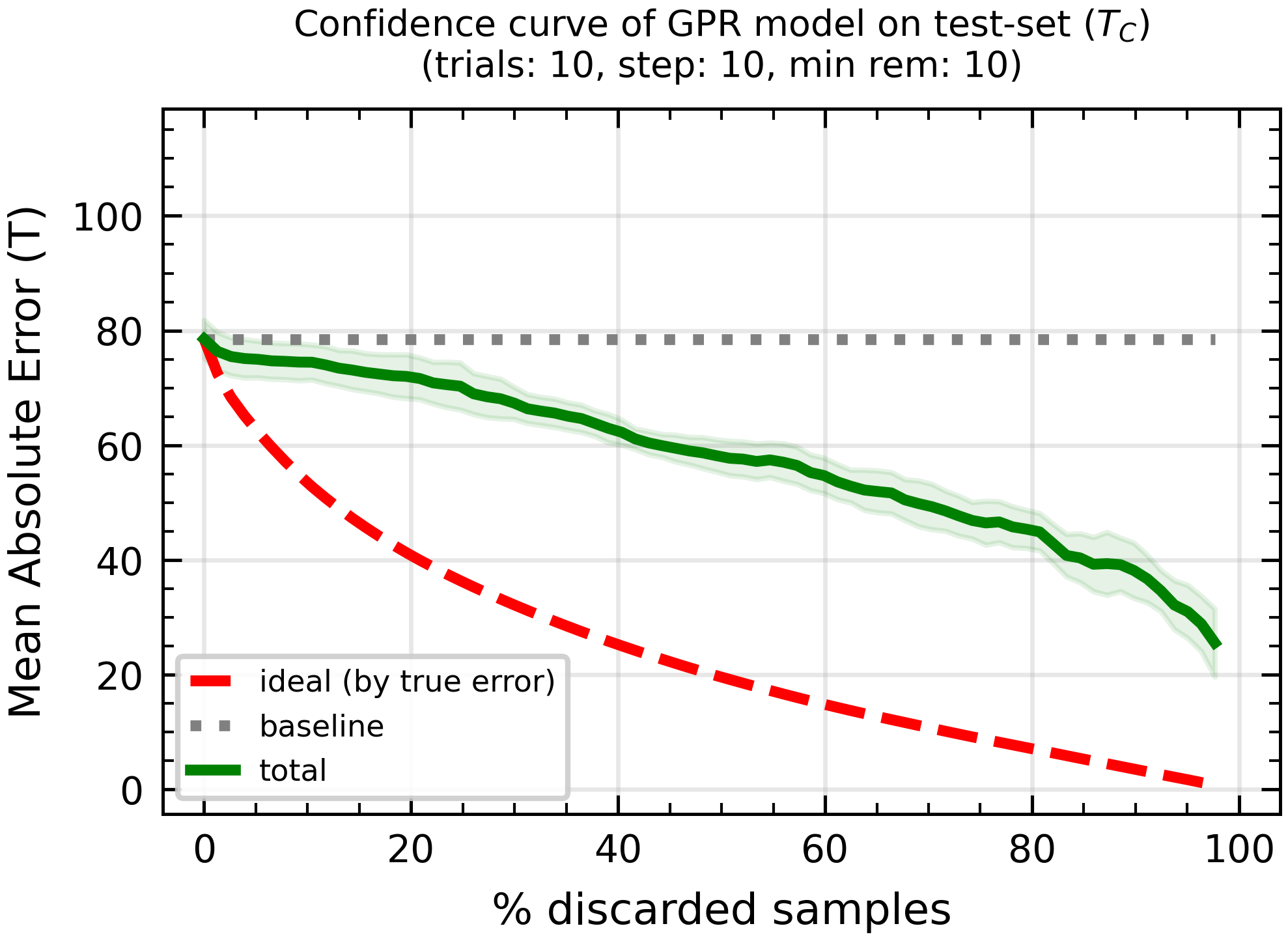}
		\caption{GPR.}
	\end{subfigure}
	\hfill
	\begin{subfigure}[t]{0.32\linewidth}
		\centering
		\plotifexists{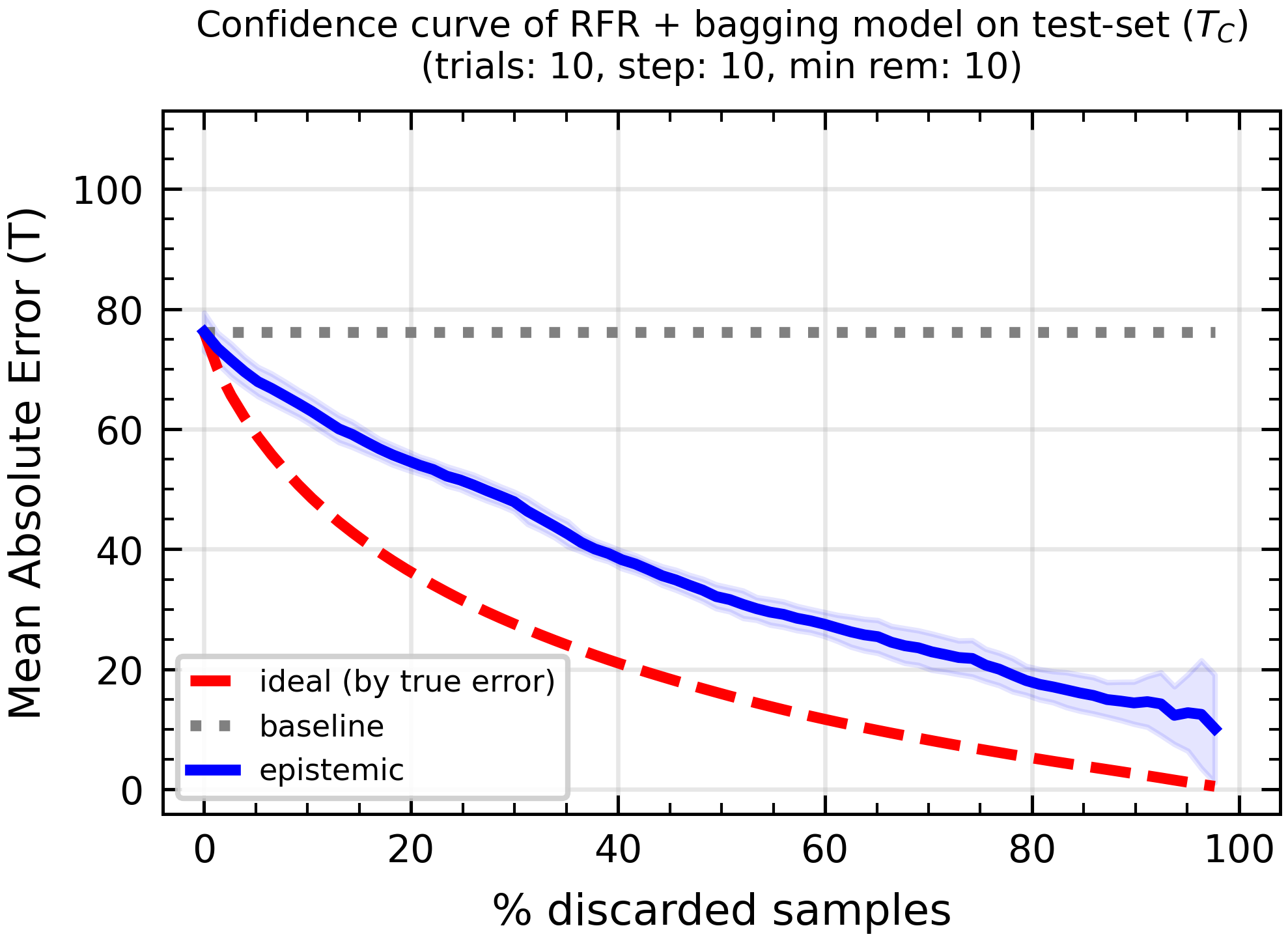}
		\caption{RFR+bagging.}
	\end{subfigure}
	\hfill
	\begin{subfigure}[t]{0.32\linewidth}
		\centering
		\plotifexists{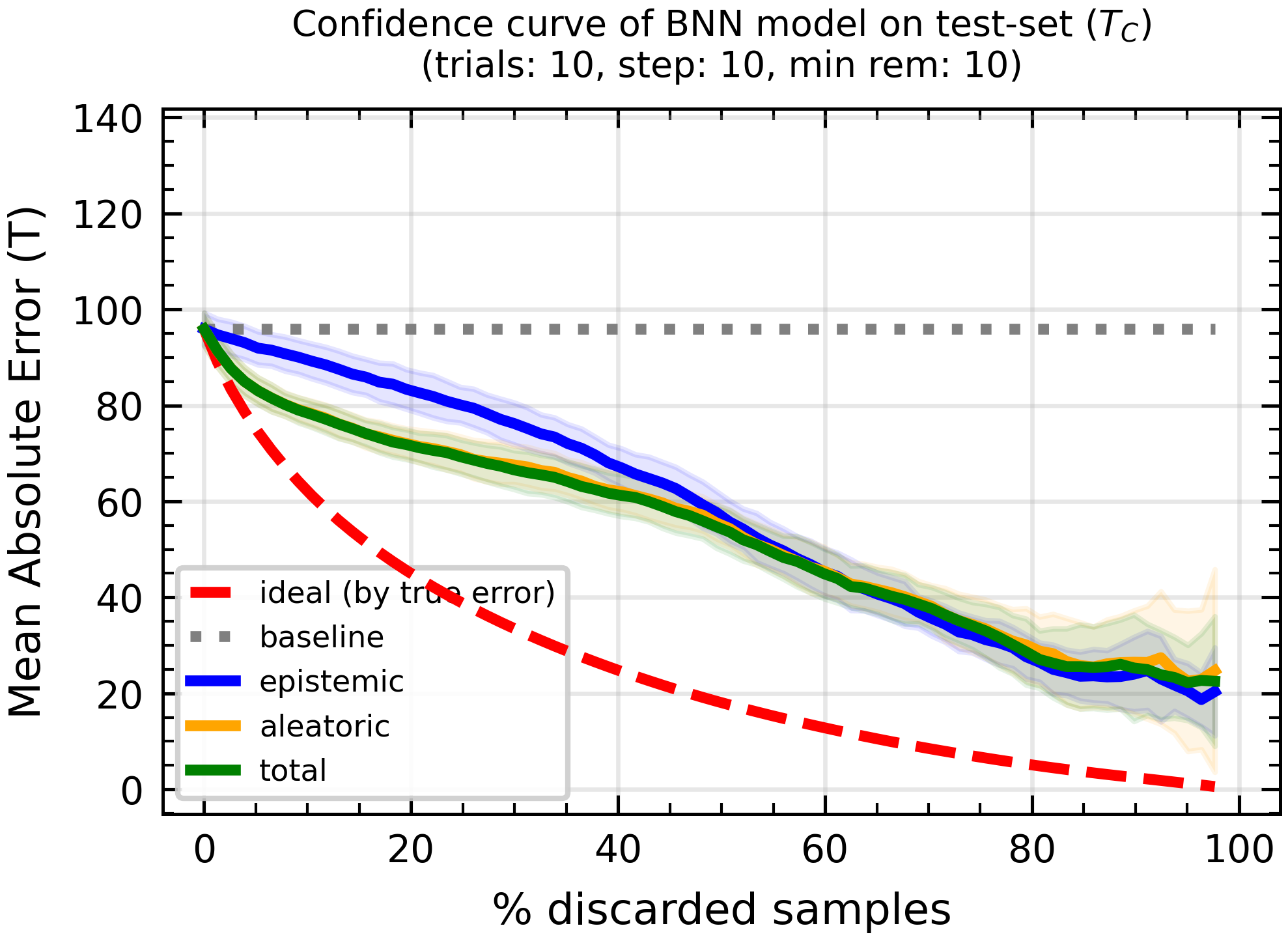}
		\caption{BNN.}
	\end{subfigure}
	\caption{
		Confidence curves on the repeated test-split evaluation across ten randomly seeded train-test repetitions.
		The shadow indicates the standard deviation across the ten repetitions.
		If the confidence curve does not fall below the grey dashed baseline, it indicates that the model's uncertainty estimates do not correlate with predictive error meaningfully.
		A practically useful uncertainty estimate should produce a downward-sloping curve similar to the red dashed ideal reference curve.
		A useful uncertainty estimate should improve predictive accuracy when the least confident predictions are removed.
	}
	\label{fig:tc-confidence}
\end{figure}

Figure~\ref{fig:tc-confidence} shows the confidence curves for the three Curie temperature models.
The GPR confidence curve is the least steep, which is consistent with its moderate calibration and overall high sharpness value.
Surprisingly, the RFR+bagging confidence curve is the steepest, which is inconsistent with its poor calibration. 
We assume its sharpness is good enough for the ranking task of the confidence curve plot, despite a systematic overconfidence.
All three curves of the BNN's uncertainty channels are downward sloping. 
The epistemic curve is the least steep of the three channels, which is consistent with its poor calibration.
The total uncertainty follows the aleatoric uncertainty closely. 
Both are steeper than the GPR curve, which is consistent with their better calibration and sharper predictive intervals.

\subsection{Discussion of the three Curie temperature models}
\label{sec:tc-discussion}
The Curie temperature study places the three models at different points in the accuracy--uncertainty landscape.

\textbf{RFR+bagging} \\
The random forest bagging regressor is a strong point predictor, but its raw ensemble-based uncertainty is not well calibrated.
Further, calibration might improve its predictive uncertainty according to Palmer et al. \cite{palmer2022}. 
Nevertheless, the RFR+bagging model has sharp uncertainty estimates and the bad overconfidence did not prevent it from ranking predictions by trustworthiness in the confidence curve plot.

\textbf{GPR} \\
The GPR is a strong predictor for this task. 
Its calibration is better than that of the RFR+bagging model, but this reliability is obtained with the broadest predictive standard deviations in the comparison.
This model ranked worst on the confidence curve plot, which is consistent with its moderate calibration and overall high sharpness value.

\textbf{BNN} \\
The BNN is the richest model from a UQ-analysis perspective because it exposes total, aleatoric, and epistemic uncertainty separately.
In this benchmark study, its total and aleatoric outputs are informative and well calibrated, whereas the isolated epistemic channel is not.
The total BNN uncertainty is the strongest overall predictive uncertainty estimate because it combines the lowest calibration error with a moderate sharpness level.
The BNN is therefore not the best pure regressor here, 
but offers great insight into uncertainty decomposition and is the strongest option for reliable predictive uncertainty reporting.

%% file: 4_STUDY_intrinsic_prediction.tex
\section{Intrinsic property prediction with uncertainty quantification} 
\label{sec:intrinsic_prediction}

    This section extends the uncertainty-aware model comparison from the Curie temperature benchmark to the prediction of intrinsic magnetic properties in \ndfeb-based permanent magnets. 
    The focus is placed on spontaneous magnetization \js and anisotropy field \ha.
    In this work, the anisotropy field $H_{\mathrm{A}}$ is treated as an effective scalar material parameter, representing the magnitude of the anisotropy field relevant for the considered magnetization direction, rather than as a spatially or directionally resolved vector field. 
    The aim of the study is not only to compare point-prediction accuracy, but also to assess whether the associated uncertainty estimates remain informative across these two physically distinct targets.
    We will use the same uncertainty-aware models as in the Curie temperature benchmark, but their hyperparameters are adapted for the new prediction task.

\begin{figure}[!htb]
        \centering
        \includegraphics[width=1.0\linewidth]{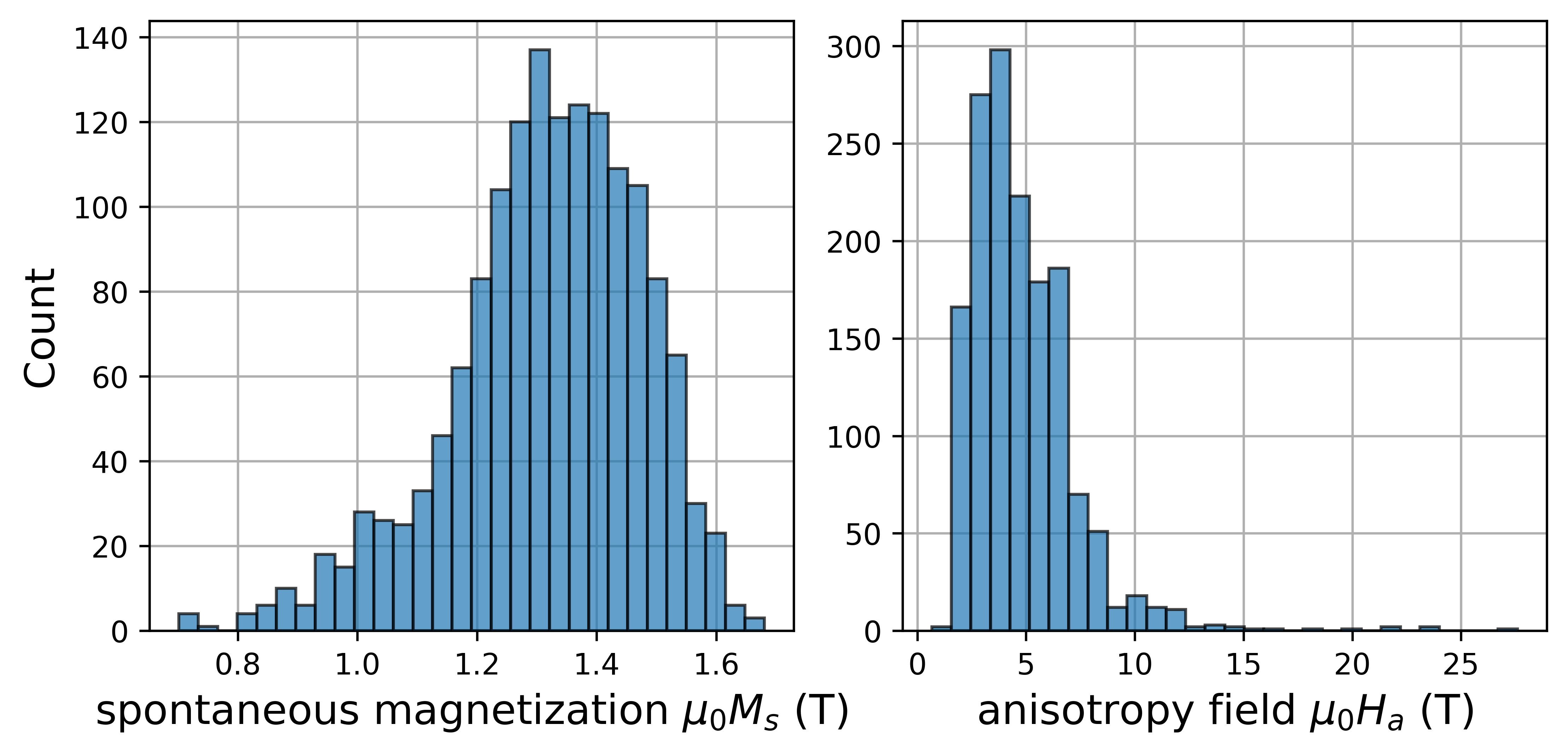}
        \caption{
            Distribution of experimental measurement labels of all 1519 datapoints. 
            The plot shows the intrinsic property labels spontaneous magnetization and anisotropy field at all five temperatures in the dataset.
        }
        \label{fig:label_distribution}
    \end{figure}

    The basis of this study is a dataset of experimental measurements of about 300 chemical compositions of Nd\textsubscript{2}Fe\textsubscript{14}B-based permanent magnets, their corresponding \js and \ha, along with their temperature during measurement. 
    Each sample's intrinsic properties were measured at five different temperatures ranging between 300 and 453 K. 
    The label distribution is plotted in Figure \ref{fig:label_distribution}.
    
\begin{figure}[htb]
    	\centering
    	\includegraphics[width=1\linewidth]{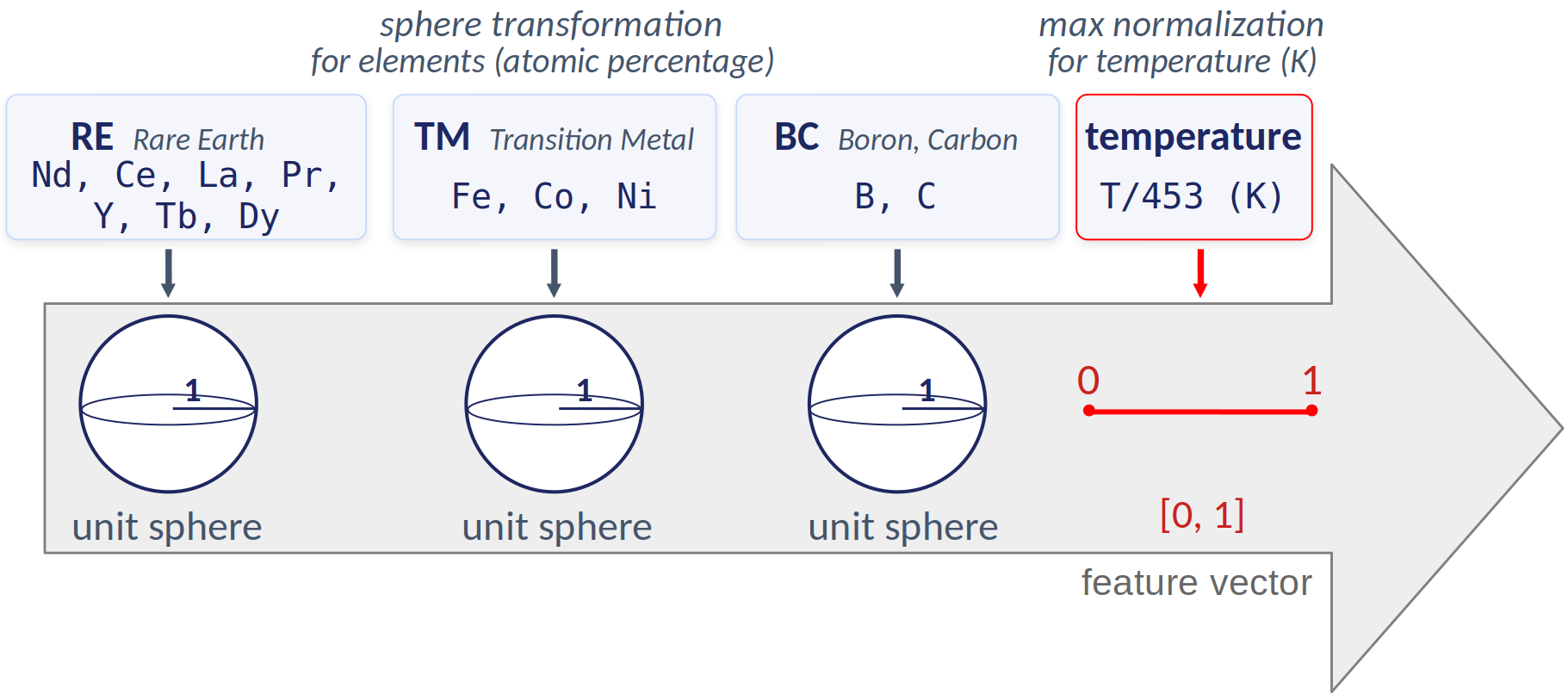}
    	\caption[Feature vector construction]{
    		\textbf{Feature vector construction}: The elemental groups are projected onto unit spheres. The temperature is scaled to $[0,1]$.
    		}
    	\label{fig:feature_vector}
    \end{figure}

    A 13-dimensional feature vector encodes both chemical composition and temperature in a form suitable for machine learning (Fig. \ref{fig:feature_vector}).
    The compositional features form three groups: rare-earth elements (Nd, Ce, La, Pr, Y, Tb, Dy), transition metals (Fe, Co, Ni), and light elements (B, C). The chemical composition is expressed as atomic fractions per group, e.g., (Nd$_{0.06}$La$_{0.94}$)$_{2}$Fe$_{14}$B$_{0.46}$C$_{0.54}$. 
    The compositional features are \textit{sphere-transformed}, meaning each group is mapped onto the unit sphere by means of normalization with the L2 norm \cite{park2022}. 
    Lastly, temperature is encoded in the feature vector scaled by division with the maximum temperature in the dataset (453 K). 
    
%% file: 4.2_model_comparison.tex

For the \ha and \js studies, the evaluation protocol follows the same repeated-split strategy as the surrogate \tc data study. 
A 70/30 train-test split was used, and the full train-evaluate workflow was repeated 10 times with different random seeds. 
All reported scalar metrics, including $R^2$, MAE, and RMSE, are computed on the held-out test split of each run and then averaged across these 10 repetitions. 
The parity plots shown below are representative test set plots from one single seed, whereas the confidence curve shadows summarize variability across ten splits. 
Both the GPR and RFR models also share the same preprocessing pipeline before fitting. 
Data is loaded just like for the \tc study, and the same train-test split is applied.
A group-wise sphere transformation of the rare-earth, transition-metal, and B/C composition blocks is applied. 
Also, the temperature is rescaled by the maximum temperature in the dataset. 
Detailed hyperparameters are summarized in \hyperref[app:ipp-study-details]{Appendix B}.

\subsection{Gaussian process model}
\label{sec:comparing_gp}

\begin{figure}[!htb]
        \centering
        \includegraphics[width=0.49\linewidth]{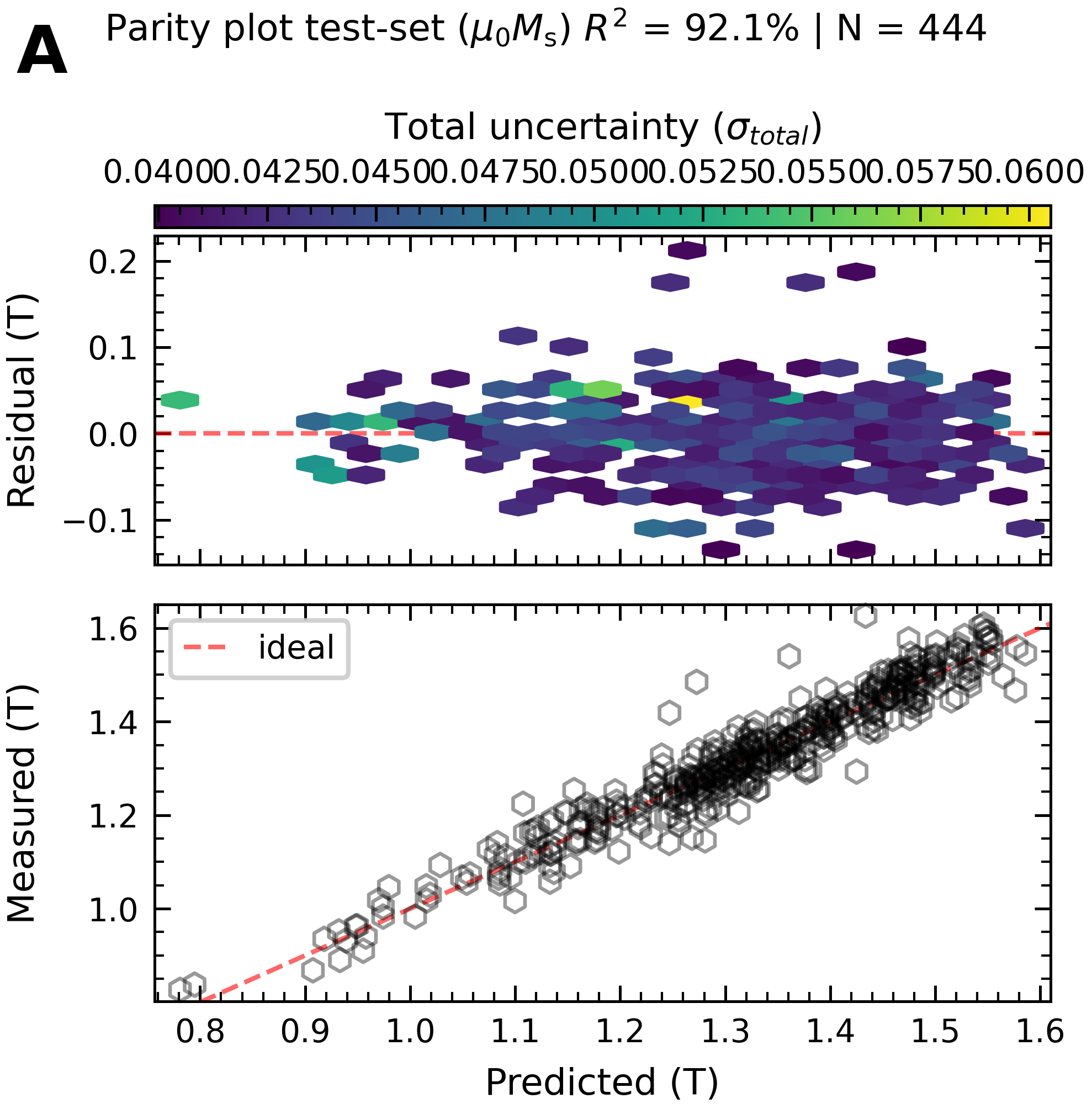}
        \hfill
        \includegraphics[width=0.49\linewidth]{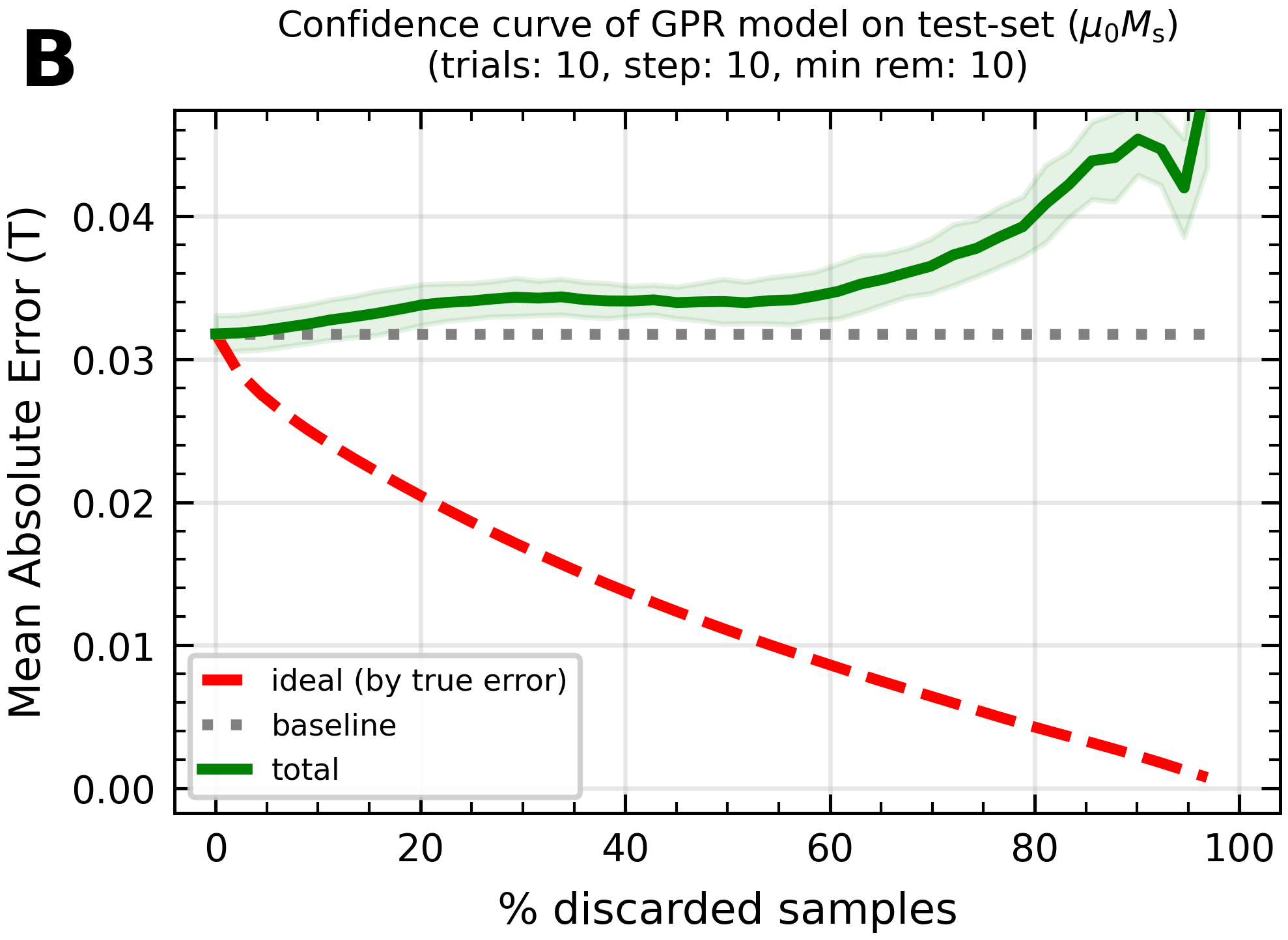}
        \caption{
            \textbf{A} Representative test set parity and residual plots of the Gaussian process model predicting spontaneous magnetization \js from one single random seed.  
            \textbf{B} The increasing confidence curve indicates that the model's uncertainty estimation is meaningless. 
            The light green shadow indicates the standard deviation across the 10 repeated random-seed runs. 
        }
        \label{fig:gpr_ms}
    \end{figure}

\begin{figure}[!htb]
        \centering
        \includegraphics[width=0.49\linewidth]{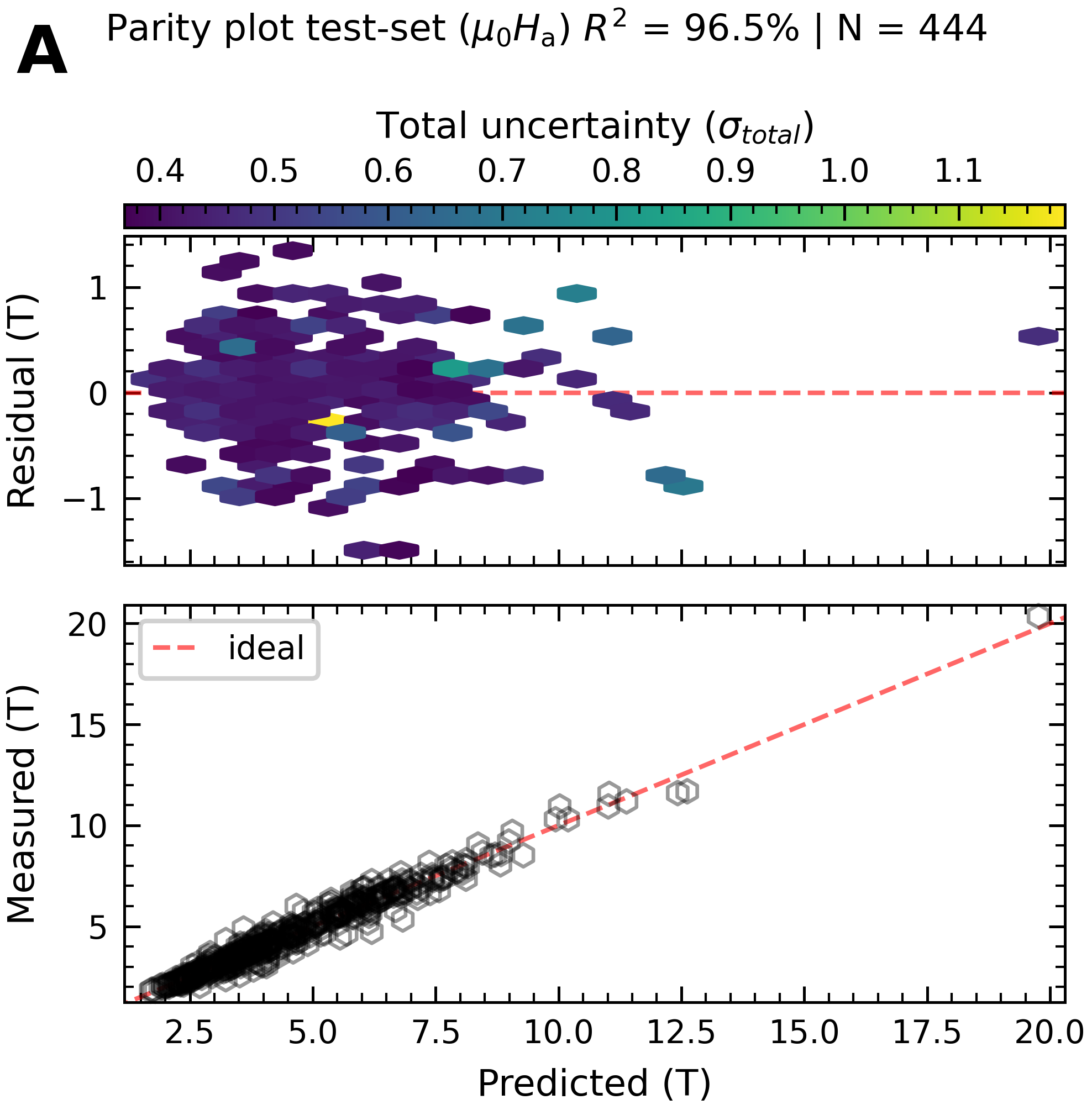}
        \hfill
        \includegraphics[width=0.49\linewidth]{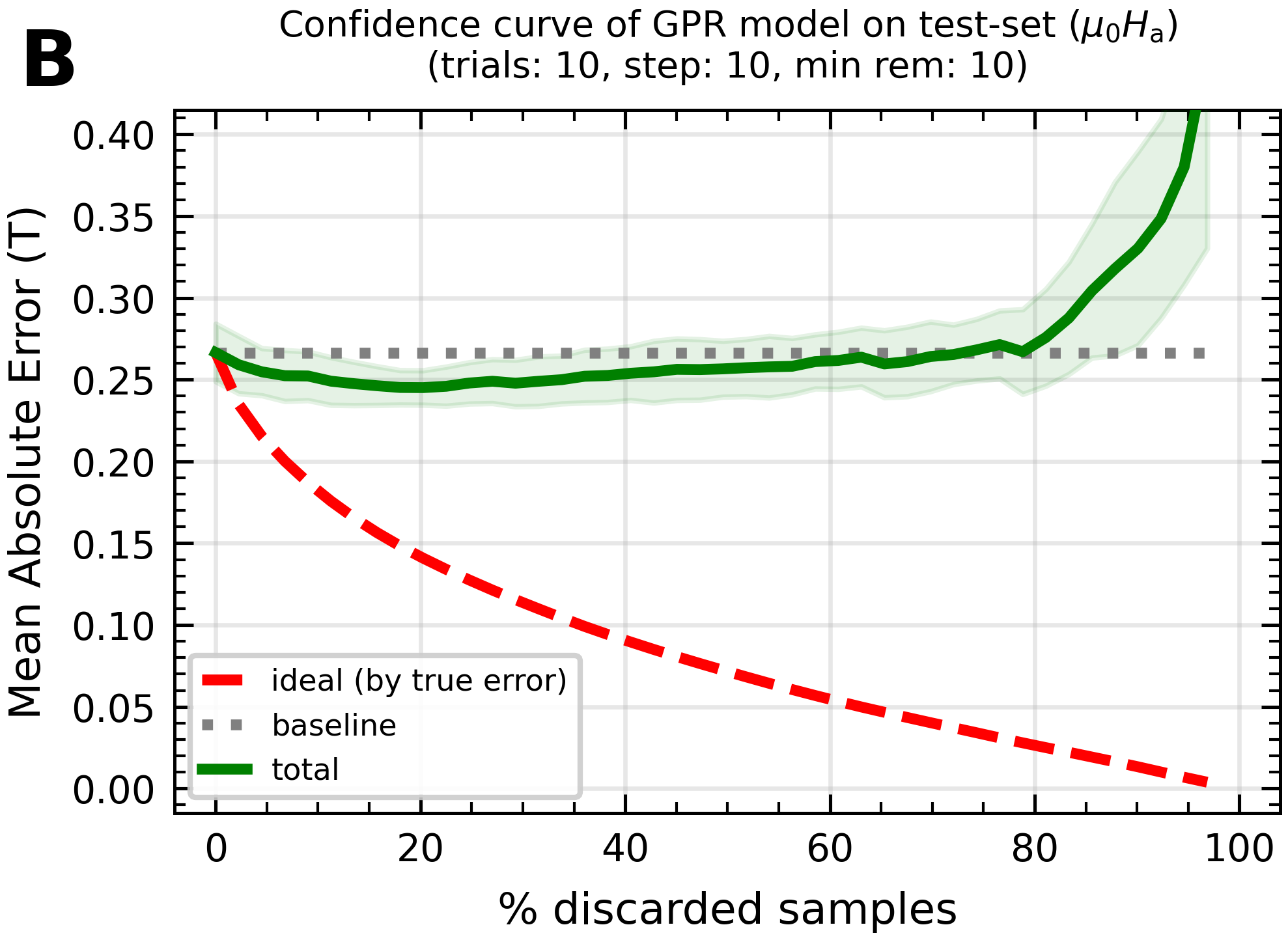}
        \caption{
            \textbf{A} Representative test set parity and residual plots of the Gaussian process model predicting anisotropy field \ha from one single random seed.  
            \textbf{B} The increasing confidence curve indicates that the model's uncertainty estimation is meaningless. 
            The light blue shadow indicates the standard deviation across the 10 repeated random-seed runs. 
    }
        \label{fig:gpr_ha}
    \end{figure}

    The first model architecture in our study was a Gaussian process regression model. Two models were trained and achieved a high $R^2$-score of 92.6\% on the magnetization model (Fig. \ref{fig:gpr_ms}\textbf{A}) and 96.9\% on the anisotropy model (Fig. \ref{fig:gpr_ha}\textbf{A}). The models for the two labels were trained with the same hyperparameters and the same preprocessing pipeline, including the standard-scaling step applied after the train-test split.
    The Gaussian process model returns a combined predictive uncertainty. In this study, we interpret it as a total uncertainty estimate because the White Noise kernel contributes a data-noise component in addition to model uncertainty.

    As an initial configuration, a kernel composed of a squared Matern kernel with smoothness parameter $\text{nu = 1.5}$ and $\text{length\_scale = 1}$, combined with a white noise kernel of $\text{noise\_level = 1}$ was used. 
    The kernel hyperparameter optimization was carried out with the limited-memory BFGS algorithm ($\text{fmin\_l\_bfgs\_b}$) using two restarts to avoid local minima. The target values were not normalized ($\text{normalize\_y}=False$). 
        
    Both Gaussian process regression models exhibit a rising slope of the confidence curve (Fig. \ref{fig:gpr_ms}\textbf{B} and Fig. \ref{fig:gpr_ha}\textbf{B}), which means that their uncertainty estimates for this task are incorrect. 
    In the evaluation plots (upper panel of Fig. \ref{fig:gpr_ms}\textbf{A} and Fig. \ref{fig:gpr_ha}\textbf{A}) datapoints of the residual plot are colored according to their estimated combined uncertainty. 
    We observe that the their color coding exhibits no pattern related to the prediction error. 
    If the uncertainty estimation was meaningful, the darker points with higher uncertainty would be located further away from the red line in this plot, while the more certain predictions colored in yellow would sit close to the red line.
    The estimated uncertainty values are very small, so it is assumed the two models could not capture sufficient information about the model uncertainty to distinguish effectively between confident and non-confident predictions. 
    Thus, the Gaussian process model's uncertainty estimation is uninformed and meaningless for this task.

%
\subsection{Random forest model}
\label{sec:comparing_rfr}

    As a second approach, an ensemble of tree-based models was considered: a bagging framework with 200 independent random forest model. 
    The base random forest was configured with 100 trees and unrestricted maximum depth. 
    The same hyperparamters were used for both models predicting \js and \ha, respectively. As for the GPR baseline, the random forest pipeline used the shared preprocessing described above, including the standard scaler fitted on the training split and applied to both train and test features. 
    Note that only the epistemic uncertainty $\sigma^2_e$ was evaluated for these models in the present study. More generally, Gaussian process and random forest based methods can represent both epistemic and aleatoric uncertainty depending on the specific formulation and implementation \cite{hermkes2014,shaker2020}. Here, however, the bagged random forest model reports the spread across ensemble members, which we interpret as an epistemic uncertainty. Since this ensemble does not learn an explicit conditional noise model or heteroscedastic variance term, a separate aleatoric uncertainty estimate is not available in the present work.
    
\begin{figure}[!htb]
        \centering
        \includegraphics[width=0.49\linewidth]{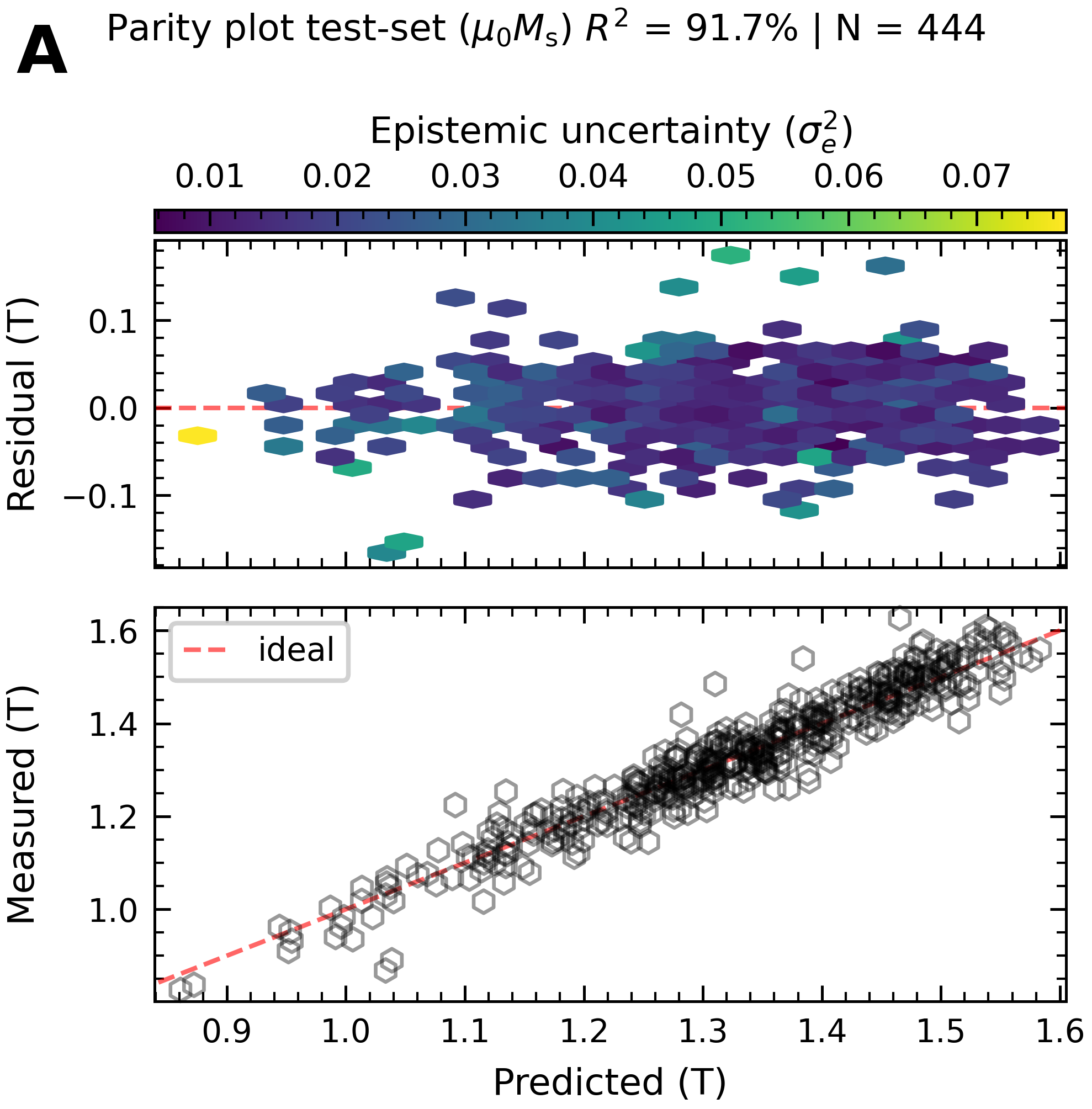}
        \hfill
        \includegraphics[width=0.49\linewidth]{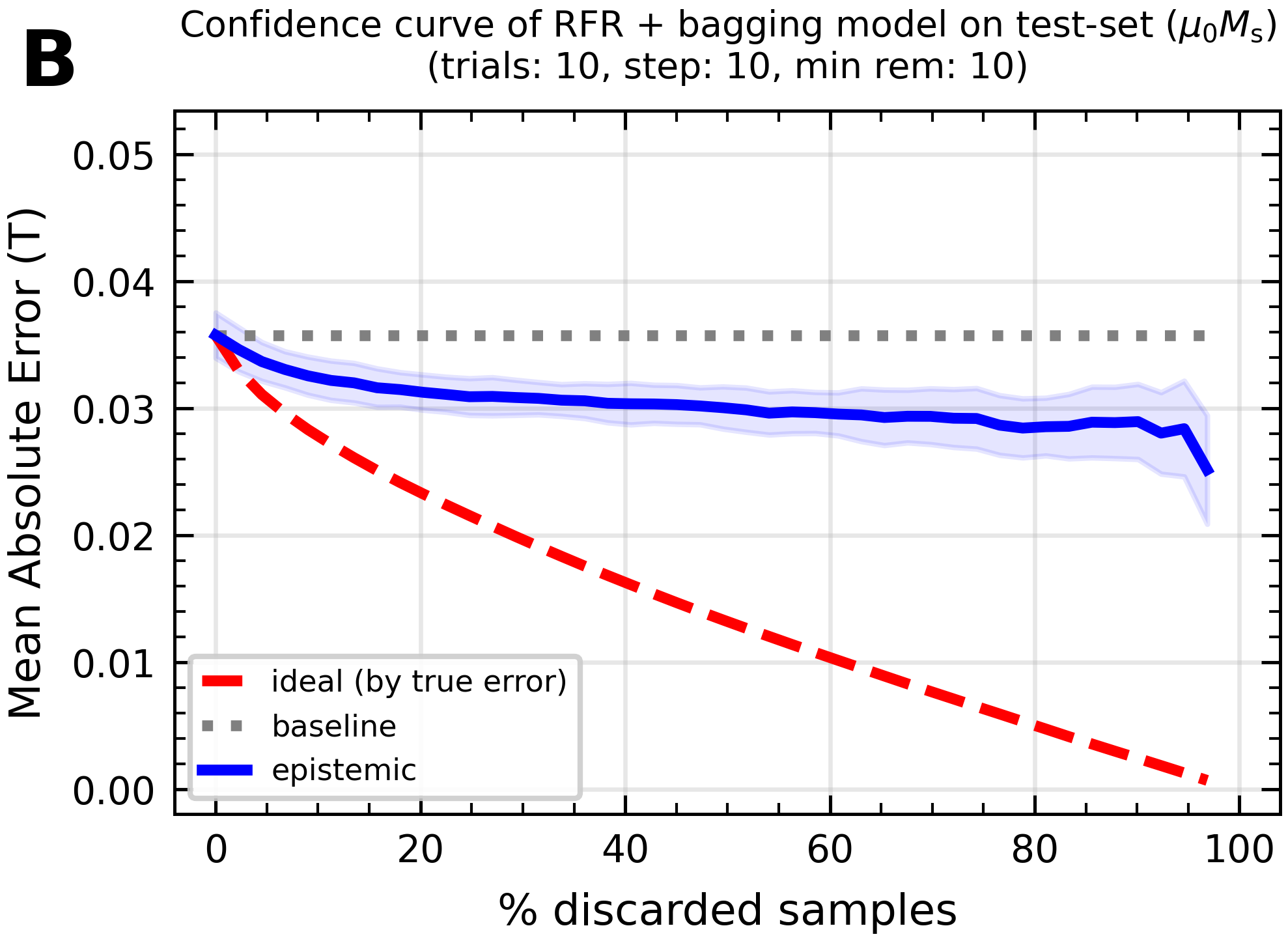}
        \caption{
            Only the epistemic uncertainty \epistemicunc was evaluated for this plot, because the random forest bagging regressor based on random forest is the only model in this comparison that estimates epistemic uncertainty alone.
            \textbf{A} Representative test set parity and residual plots of the random forest model predicting spontaneous magnetization \js from one single random seed.  
            \textbf{B} The confidence curve shows a slight negative slope.
            The light blue shadow indicates the standard deviation across the 10 repeated random-seed runs. 
        }    
        \label{fig:rfr_ms}
    \end{figure}

\begin{figure}[h]
        \centering
        \includegraphics[width=0.49\linewidth]{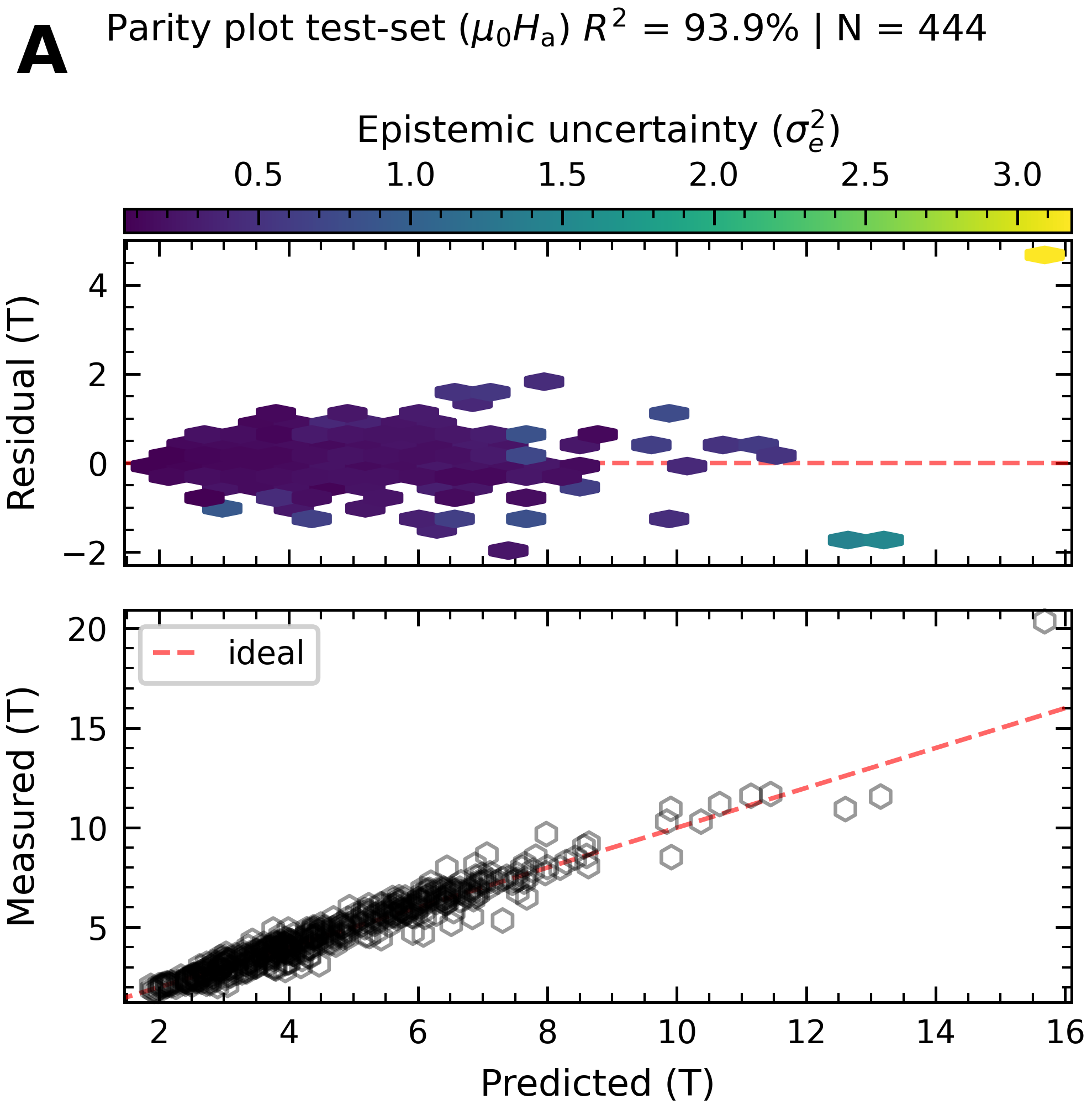}
        \hfill
        \includegraphics[width=0.49\linewidth]{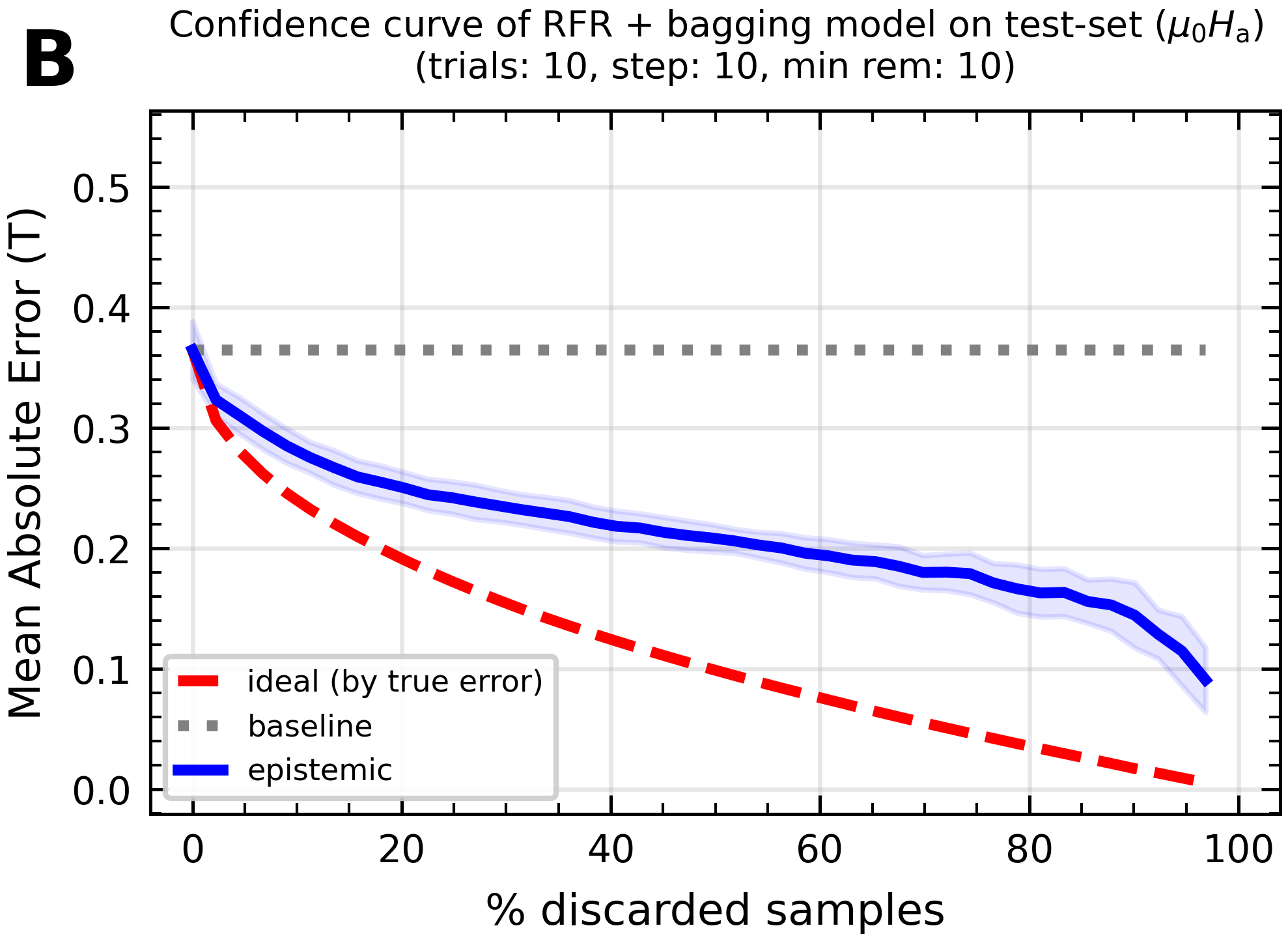}
        \caption{
            Only the epistemic uncertainty $\sigma_\mathrm{e}^2$ was evaluated for this plot, because the random forest bagging regressor based on random forests is the only model in this comparison that estimates epistemic uncertainty alone.
            \textbf{A} Representative test set parity and residual plots of the random forest model predicting anisotropy field \ha from one single random seed. 
            \textbf{B} The declining confidence curve indicates good uncertainty estimation. 
            We observe an S-shape that indicates that highly uncertain predictions correlate well with the prediction error (in the first 10\% of discarded samples).  
            The light blue shadow indicates the standard deviation across the 10 repeated random-seed runs. 
        }
        \label{fig:rfr_ha}
    \end{figure}

    The model fits the data very well.
    The magnetization model achieved a 90.9\% $R^2$-score and the anisotropy model 92.2\%.     
    The confidence curves decrease (Fig. \ref{fig:rfr_ms}\textbf{B} and Fig. \ref{fig:rfr_ha}\textbf{B}) and thus indicate a link between the uncertainty estimation and the prediction error for both models.
    Additionally, we can see in the residual plots how the darker more uncertain points tend to appear at the far ends of the point cloud, with some exceptions.

    The distribution of less confident predictions (upper panel of Fig. \ref{fig:rfr_ms}\textbf{A} and Fig. \ref{fig:rfr_ha}\textbf{A}) correlates with the distribution of labels in the training data (Fig. \ref{fig:label_distribution}).
    In other words, in regions were less data is available, the uncertainty is higher.
    Furthermore, we observe that the confidence curve of the \ha model exhibits an S-shape which means that highly uncertain predictions correlate well with the prediction error (in the first 10\% of discarded samples).  
    The S-shape indicates the presence of a region where the curve flattens, suggesting that within this range of the dataset, the model can less accurately distinguish between high- and low-confidence predictions. Overall we still see a negative slope.    
    
\subsection{Bayesian neural network model}
\label{sec:comparing_bnn}
    Lastly, we developed two distinct Bayesian neural network (BNN) models with the Keras \cite{chollet2015} library for the two target properties.
    The reason for using two architectures is pragmatic rather than conceptual: we tested several dense-layer sequences and dropout ratios for each target separately and found empirically that one shared architecture did not give the best joint compromise between predictive accuracy and informative uncertainty estimates for both labels.
    We therefore retained, for each target, the architecture that performed best in our repeated train-test experiments, using both a high $R^2$ score and the steepest possible declining confidence curve as selection criteria.
    The two networks should thus be interpreted as task-specific empirical optima within the same dropout-based BNN framework, not as two different theoretical model classes. 
    Note that in contrast to the random forest model presented before, both the BNN and the GPR model provide a combined predictive uncertainty estimate, while only the BNN additionally allows an explicit decomposition into epistemic and aleatoric parts in the present implementation. For the GPR model, the predictive variance is treated as a combined uncertainty because it contains both function uncertainty and an observation-noise contribution, but it is not decomposed further here \cite{hermkes2014}. 

\begin{figure}[!htb]
        \centering
        \includegraphics[width=0.49\linewidth]{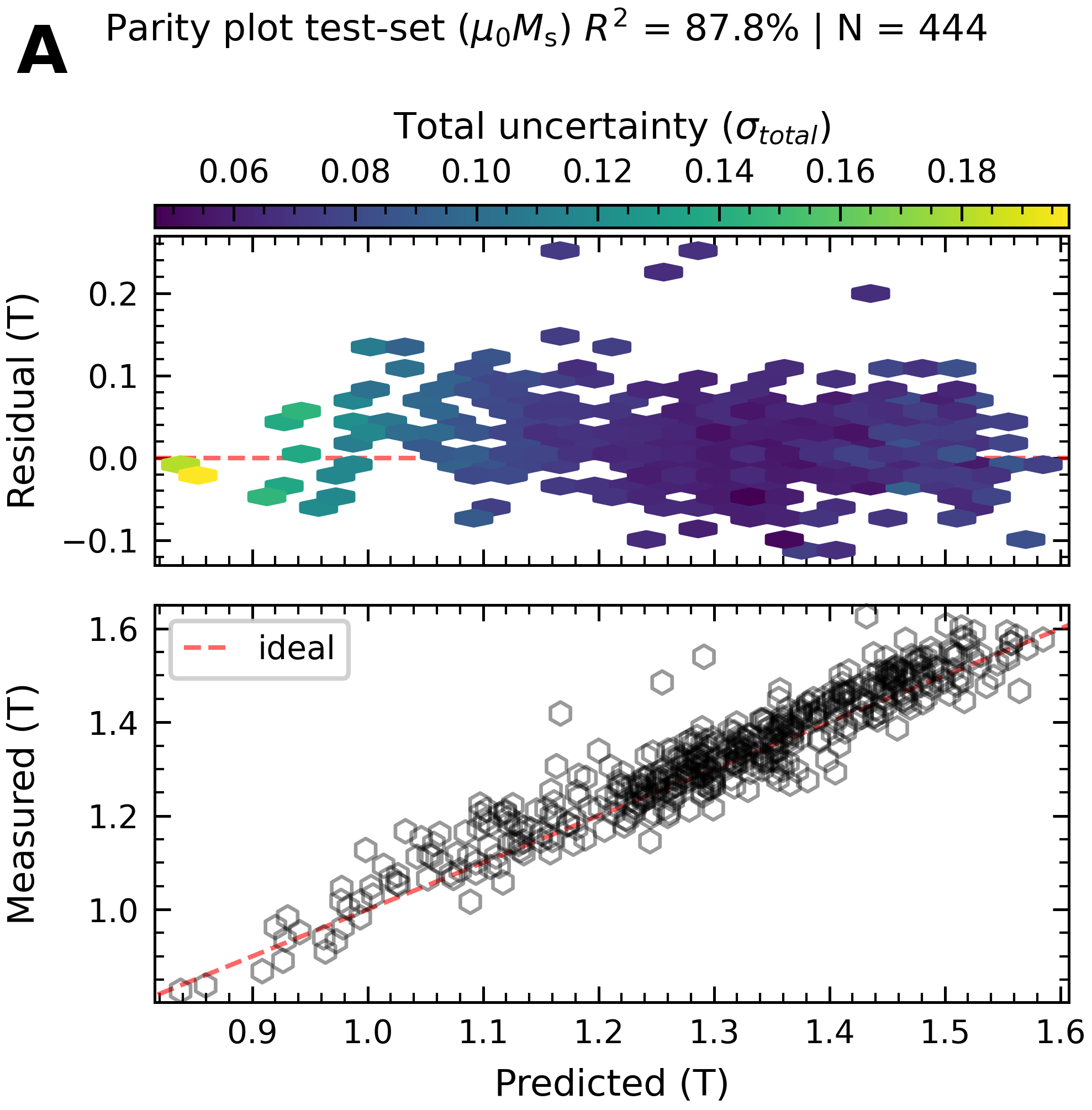}
        \hfill
        \includegraphics[width=0.49\linewidth]{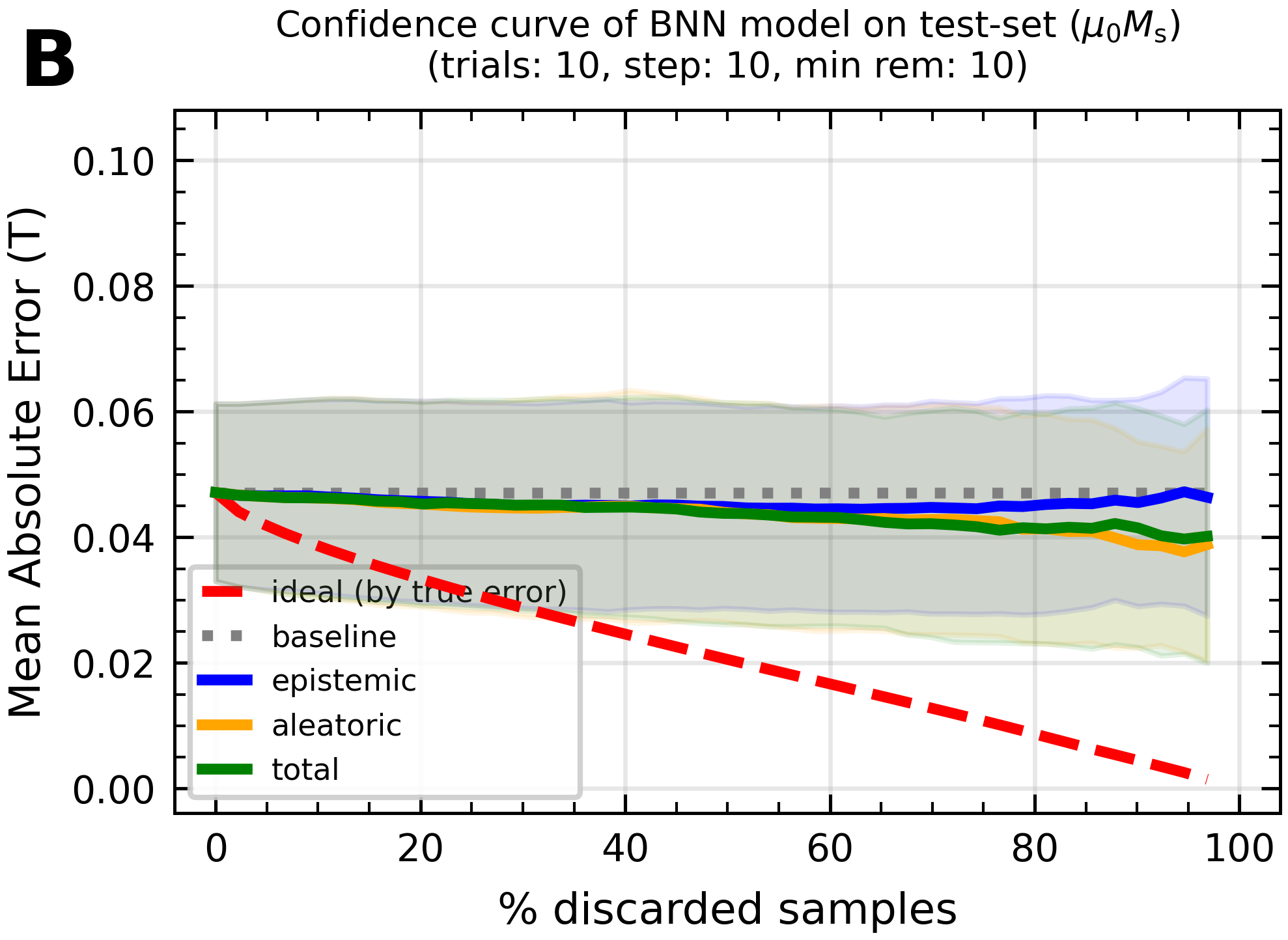}
        \caption{
            \textbf{A} Representative test set parity and residual plots of the Bayesian neural network model predicting spontaneous magnetization \js from one single random seed. 
            The uncertainty estimation colored in the residual plot (upper panel) does not coincide with the residuals.
            \textbf{B} The aleatoric and the total uncertainty's confidence curves decline slowly but steadily. 
            The epistemic uncertainty's confidence curve declines at first, but rises for that last 40\% of discarded data slowly then for the last 3\% sharply .
            The shadow indicates the standard deviation across the 10 repeated random-seed runs. 
        }    
        \label{fig:bnn_ms}
    \end{figure}

    For the prediction of the spontaneous magnetization \js, we implemented a neural network that contains three blocks.
    The first and the last block consist of 
    one hidden layer of 32 units and ReLU activation, followed by 
    a second hidden layer of 32 units with hyperbolic tangent activation, and 
    lastly a dropout layer with a dropout-rate of 0.5.
    The middle block consists of the same layers but has 64 instead of 32 nodes in the hidden layers. 
    The model was trained using the RMSprop optimizer \cite{chollet2015} over 800 epochs, and for each prediction 50 forward passes were sampled to estimate the model uncertainty.
    Among the tested magnetization architectures, this deeper three-block configuration gave the best empirical compromise between predictive fit and useful uncertainty estimates.
    It achieved an $R^2$-score of 88.4\%. 
    Looking, at the upper panel of Figure \ref{fig:bnn_ms}\textbf{A} we observe the color-coded uncertainty
    does not coincide with the residuals. 
    While we would expect larger residuals in absolute terms to align with higher uncertainty, the 
    observed color pattern appears instead
    to be dominated by the epistemic uncertainty arising from the label distributions (see Fig. \ref{fig:label_distribution}). 
    In Figure \ref{fig:bnn_ms}\textbf{B} we observe that 
    the aleatoric and the total uncertainty’s confidence curves decline slowly but steadily. The epistemic uncertainty’s confidence curve declines at first, but rises for that last 40\% of discarded data slowly then for the last 3\% sharply. 

\begin{figure}[!htb]
        \centering
        \includegraphics[width=0.49\linewidth]{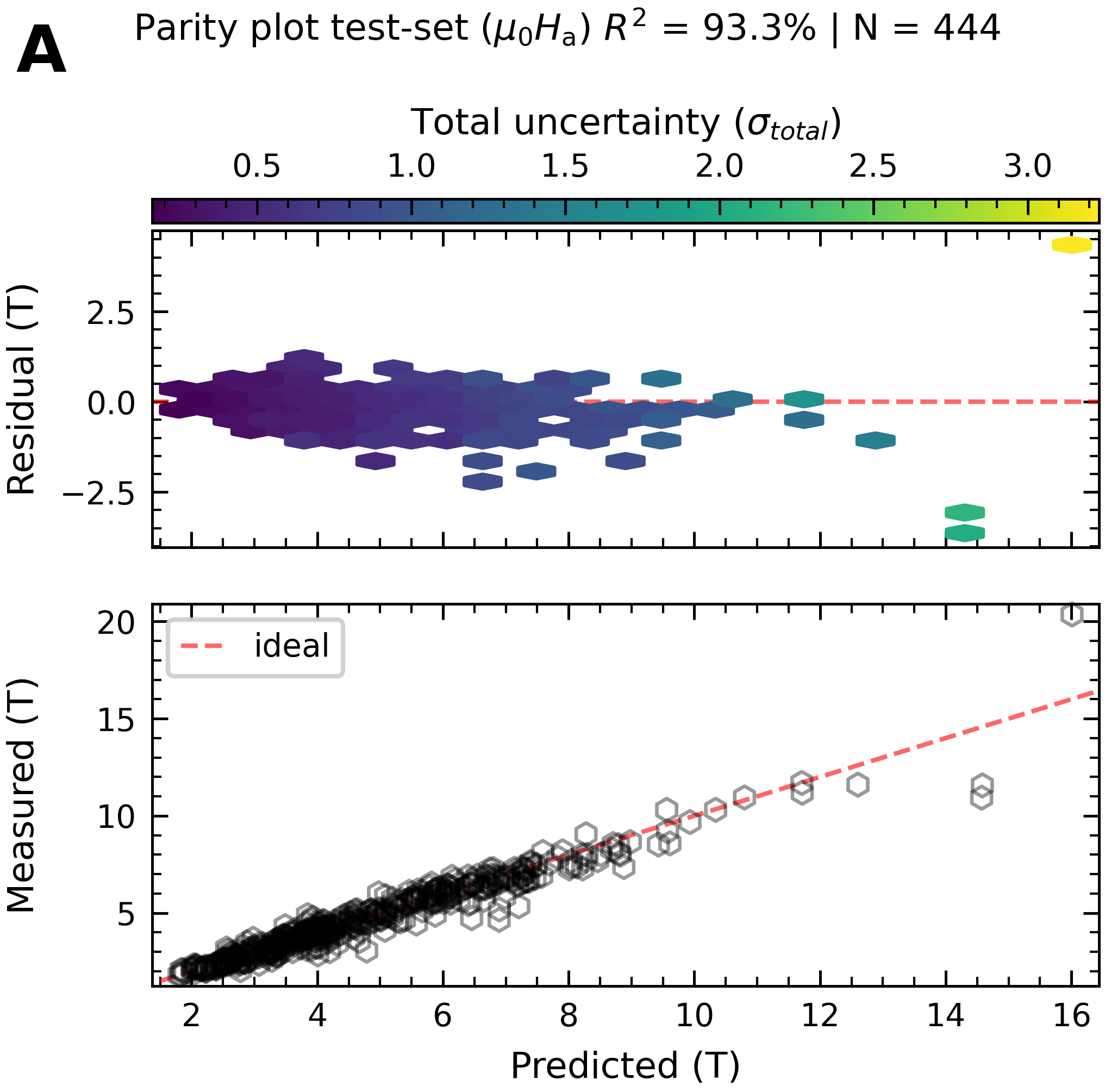}
        \hfill
        \includegraphics[width=0.49\linewidth]{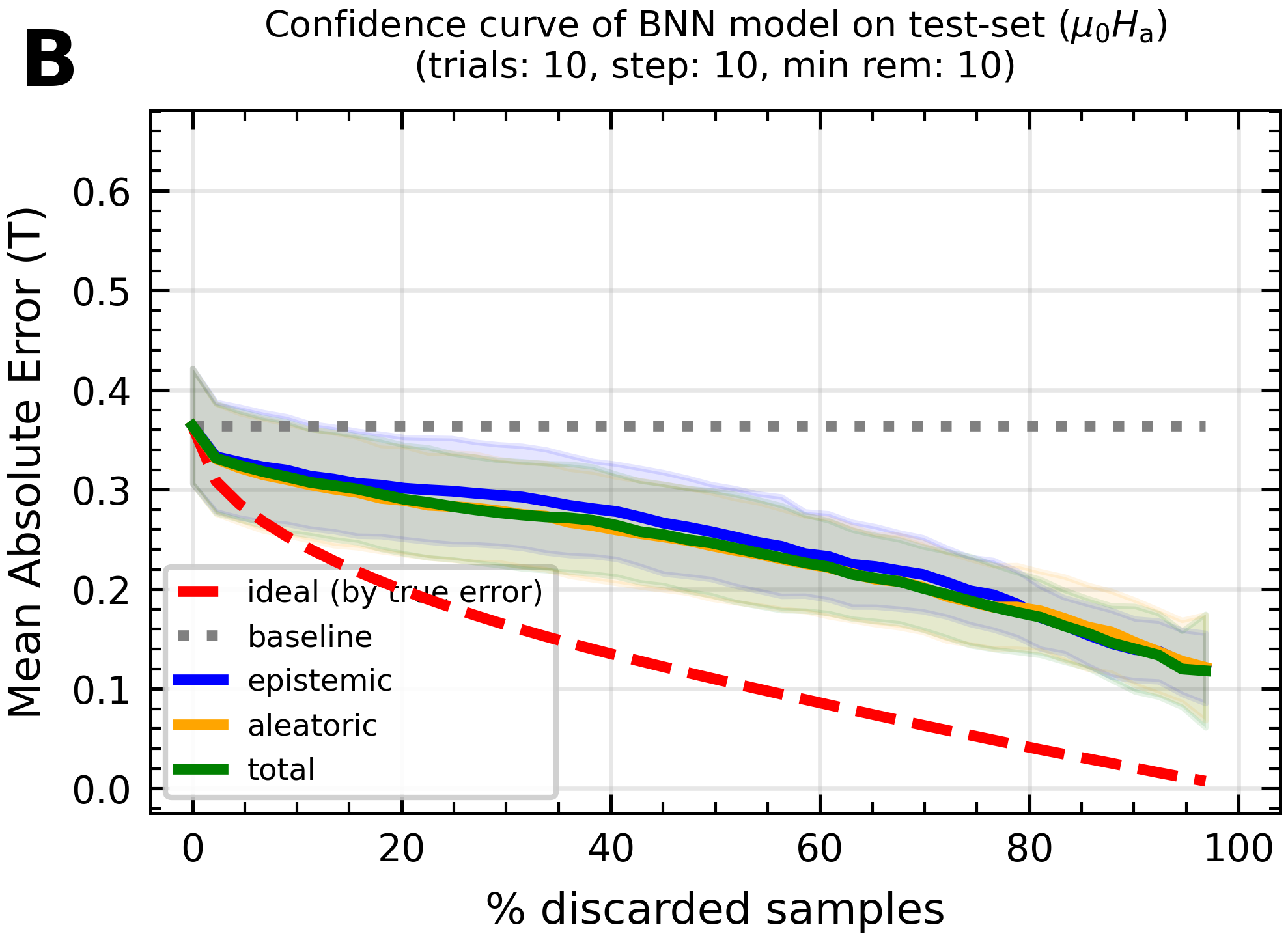}
        \caption{
            \textbf{A} Representative test set parity and residual plots of the Bayesian neural network model predicting the anisotropy field \ha from one single random seed.  
            \textbf{B} The confidence curves corresponding to all three uncertainty estimations (epistemic, aleatoric and their combination) show a consistent and steep decline, indicating that higher predicted uncertainty reliably corresponds to larger prediction errors demonstrating well-calibrated uncertainty estimates.
            The shadow indicates the standard deviation across the 10 repeated random-seed runs. 
        }
        \label{fig:bnn_ha}
    \end{figure}

    The second BNN model designed to predict the anisotropy field followed a hidden layer structure with 
    two dense layers of 128 units each, activated by ReLU and hyperbolic tangent functions, respectively. Subsequently, 
    a dropout layer with dropout-rate of 0.5 was employed. 
    The training strategy is very similar to the magnetization BNN model (RMSprop optimization for 400 epochs with 50 forward passes per sample prediction).
    In our empirical tests, the anisotropy target did not benefit from the deeper three-block layout used for \js prediction. A simpler two-layer architecture was sufficient and produced the best overall result for \ha, both in predictive accuracy and in uncertainty quality.
    This model achieved a high $R^2$-score of 93.1\%.
    The uncertainty estimation was very well-calibrated, as all three confidence curves drop quickly (Fig. \ref{fig:bnn_ha}\textbf{B}) and the colored residual points become gradually darker towards the region with few training samples and high prediction errors (upper panel of Fig. \ref{fig:bnn_ha}\textbf{A}). 
    
%% file: 4.3_models_summary.tex
\subsection{Discussion of the Model Performance}
\label{sec:study1_insights}

    For all three model classes, the approaches achieved a high coefficient of determination ($R^2$) for both spontaneous magnetization \js and anisotropy field \ha. 
    However, their ability to provide meaningful uncertainty estimates varied considerably. 
    The Gaussian process regression produced the highest $R^2$-scores (92.6\% for \js and 96.9\% for \ha), but its uncertainty estimates were meaningless and showed no correlation with prediction error. 
    The random forest models achieved slightly lower $R^2$-scores (90.9\% and 92.2\%, respectively) and offered good uncertainty estimates (declining confidence curve), but they remain limited to epistemic uncertainty only. 
    The Bayesian neural networks not only have high $R^2$-scores but also capture epistemic and aleatoric uncertainty explicitly. 
    The model predicting the anisotropy field (93.1\% $R^2$-score) estimates both uncertainties very well. 
    However, the model predicting spontaneous magnetization (88.4\% $R^2$-score) was less effective in estimating uncertainty. 
    The \js model's confidence curves declined for the aleatoric and the total uncertainty, but not for the epistemic uncertainty. 
    Neither the random forest nor the BNN model were particularly effective at capturing uncertainty when predicting spontaneous magnetization, still they provided a better uncertainty estimate than the GPR model.
    The model evaluation metrics are listed in Table \ref{tab:study1_insights}. 

     \begin{table}[!hb]
		\centering
		\begin{tabularx}{\columnwidth}{>{\raggedright\arraybackslash}X c c c}

        \hline
            \textbf{Model} & \textbf{$R^2$} & \textbf{MAE} & \textbf{RMSE} \\
            \hline            
                \js GPR & $0.926 \pm 0.007$ & $0.032 \pm 0.001$ & $0.043 \pm 0.002$ \\ 
                \js RFR & $0.909 \pm 0.007$ & $0.036 \pm 0.002$ & $0.048 \pm 0.003$ \\
                \js BNN & $0.884 \pm 0.030$ & $0.041 \pm 0.008$ & $0.054 \pm 0.007$ \\ 
            \hline
                \ha GPR & $0.969 \pm 0.004$ & $0.265 \pm 0.019$ & $0.418 \pm 0.044$ \\ 
                \ha RFR & $0.922 \pm 0.027$ & $0.366 \pm 0.028$ & $0.667 \pm 0.161$ \\ 
                \ha BNN & $0.931 \pm 0.020$ & $0.360 \pm 0.045$ & $0.627 \pm 0.140$ \\ 
    \hline
         \end{tabularx}
         \caption{
            Overview of the model evaluations on the held-out test splits. Reported values are mean $\pm$ standard deviation across 10 randomly seeded train-test runs. For the GPR model, the reported uncertainty is a combined predictive uncertainty of epistemic and aleatoric. The random forest model outputs epistemic uncertainty only. The BNN provides a combined uncertainty estimate and can also be decomposed into epistemic and aleatoric parts.
         }
         \label{tab:study1_insights}
     \end{table}

%% file: 5_STUDY_coercivity_prediction.tex
\section{Graph Neural Network for Coercivity Prediction with Uncertainty Quantification}
\label{sec:gcnn4coercivity}

The performance of magnetic materials depends not only on their chemical composition but also on their microstructural characteristics. 
To account for these effects, we developed a graph neural network (GNN) that predicts the coercive field of Nd\textsubscript{2}Fe\textsubscript{14}B-based permanent magnets using microstructural features and the distribution of easy-axis orientations \cite{moustafa_graph_2025}. 
Each microstructure is represented as a graph, where nodes correspond to individual grains and edges represent grain boundaries, enabling the model to capture complex grain connectivity and orientation relationships.

The dataset comprises over 1,000 synthetic microstructures, generated via the software Neper \cite{quey_large-scale_2011} and labeled with the coercive field computed by our reduced-order micromagnetic model \cite{moustafa_reduced_2024}. 
The node-level features include grain size, aspect ratio, sphericity, crystallographic orientation, and Stoner–Wohlfarth switching field, while the edge features encode the grain boundary thickness.
All features were scaled to ensure numerical stability.
Details on the feature vector construction can be found in the paper by Moustafa and his colleagues \cite{moustafa_reduced_2024}. 

The graph neural network was implemented using PyTorch Geometric \cite{fey_fast_2019}. The architecture consists of a customized convolutional layer, inspired by Dai et al. \cite{dai_graph_2021}, followed by a fully connected layer, global pooling, and fully connected layers. To enable uncertainty quantification, the model was trained with a Gaussian negative log-likelihood loss (Eq. \ref{eq:gnllLoss}) and employed Monte-Carlo dropout during prediction time, allowing the decomposition of predictive uncertainty into epistemic and aleatoric components.
Detailed hyperparameters are summarized in \hyperref[app:hc-study-details]{Appendix C}.

\begin{figure}[!ht]
    \centering
    \includegraphics[width=0.49\linewidth]{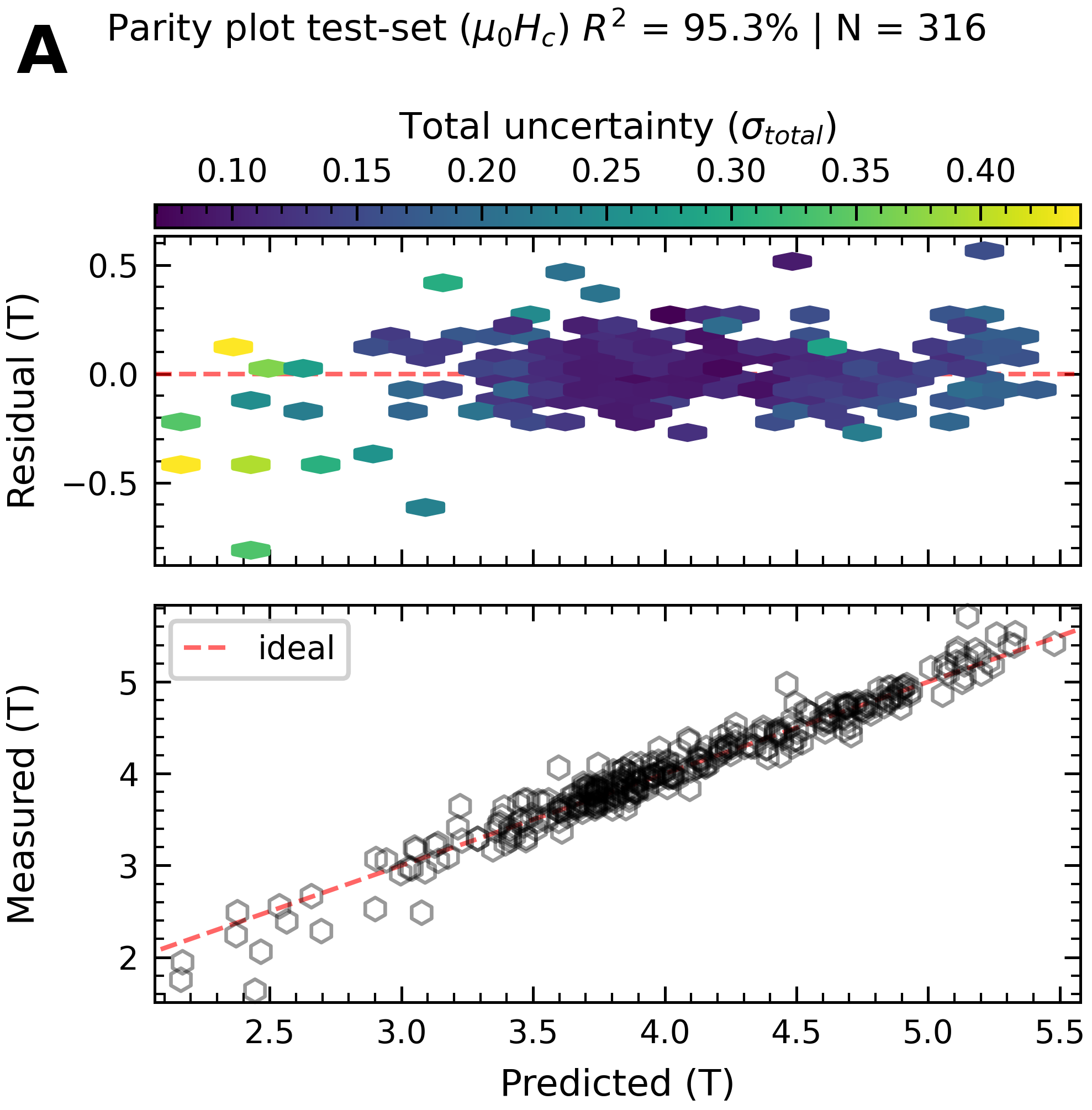}
    \hfill
    \includegraphics[width=0.49\linewidth]{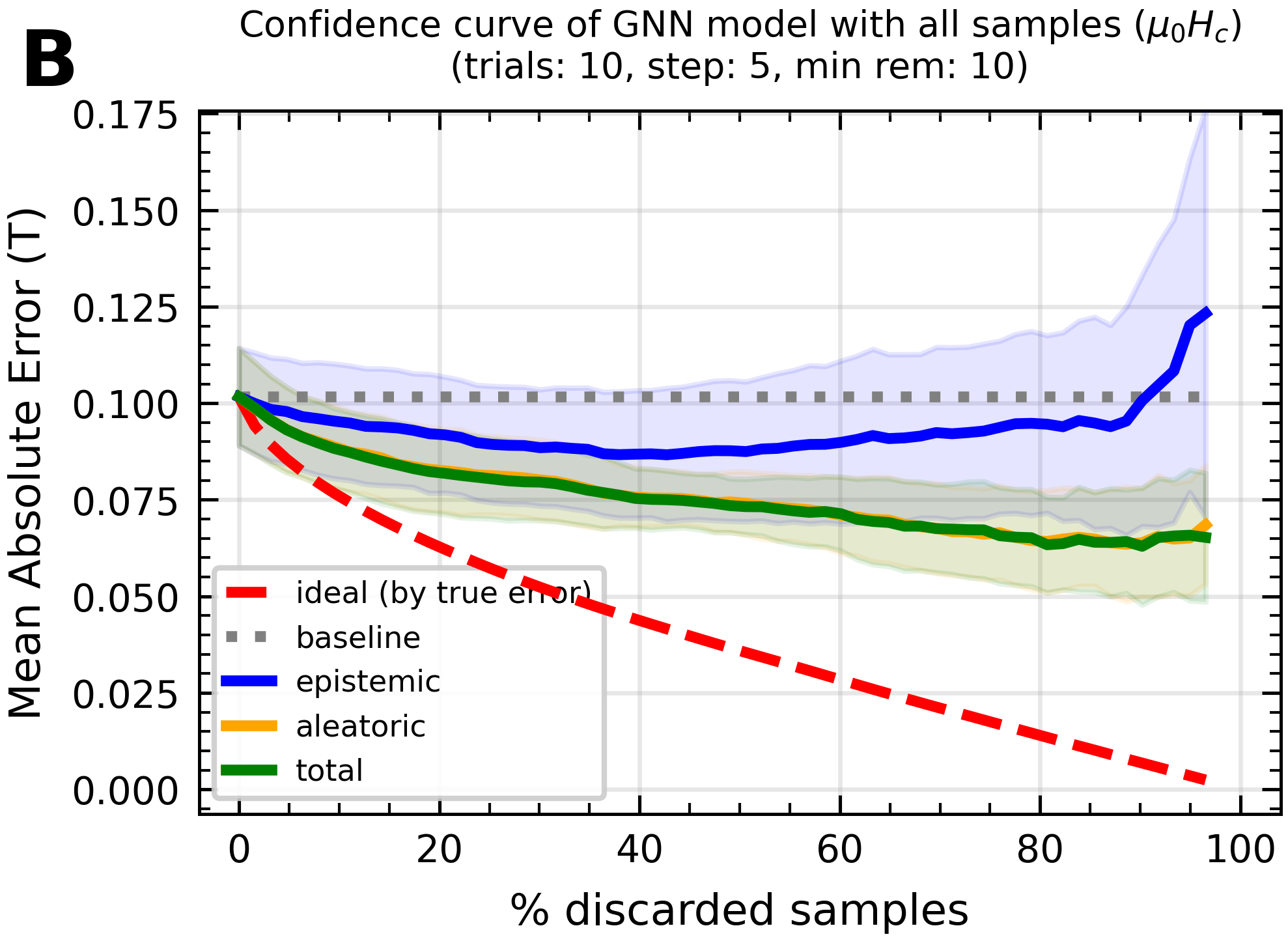}
    \caption{
        \textbf{A} Representative test set parity and residual plots of the graph neural network model predicting coercivity $\mu_0H_\mathrm{c}$ from one single random seed.  
        \textbf{B} The confidence curve, which is evaluated on the test set, demonstrates the effectiveness of the model’s uncertainty estimates, especially for the aleatoric and total uncertainty.
        The model was trained using 70\% of the total dataset and predicted on a test set comprising 15\% of the unseen dataset. 
    }
    \label{fig:gcn_uq}
\end{figure}

The GNN achieved high predictive accuracy when evaluated on the held-out test split of each of 10 randomly seeded train-test runs and then averaged across those runs: a coefficient of determination $R^2$ of 96.1\% $\pm$ 0.0172, a MAE of 0.0963 $\pm$ 0.0171 and MSE of 0.0184 $\pm$ 0.0071. 
The residual analysis indicates an increasing residual error for the lower coercivity range (upper panel of Fig. \ref{fig:gcn_uq}\textbf{A}). 

For the evaluation of the uncertainty estimation the model was trained using 70\% of the total dataset and predicted on a test set comprising 15\% of the unseen dataset. 
The confidence curve (Sec. \ref{sec:model_evaluation}) illustrates that the predicted variance of the aleatoric uncertainty correlates strongly with the mean absolute error. 
The predicted epistemic uncertainty is slightly less informative, as can be seen from the flatter blue confidence curve.
Also, the model becomes less effective at distinguishing between predictions with low estimated epistemic uncertainty. 
This is why the blue confidence curve rises towards the end of the plot, but still shows a negative slope for the first 90\% of discarded samples.
The green confidence curve for the total uncertainty estimation created a curve with a slope of -0.04\%. 
Overall, this confirms the informativeness of the uncertainty estimates, especially for predicted variance of the aleatoric and total uncertainty (Fig. \ref{fig:gcn_uq}\textbf{B}).

Uncertainty quantification in the present graph neural network provides a reliability measure for the predicted coercivity values.
Predictions with high estimated uncertainty can be flagged for follow-up simulation or experimental verification, while low-uncertainty predictions can be used more confidently for screening.
In that sense, the proposed workflow improves the usability and risk awareness of the model.

Since coercivity depends on the orientation of the easy axes relative to the applied field, it is inherently a rotation-dependent quantity. 
While our current architecture encodes crystallographic orientation as node features, future work could employ rotation-equivariant GNNs to enforce this symmetry directly and thereby improve the physical consistency of the predictions. 

%% file: 6_conclusion.tex
\section{Conclusion}
\label{sec:conclusion}
    
    Across three complementary studies, we systematically assessed how different model architectures estimate predictive uncertainty.
    In the Curie temperature surrogate study, we established the uncertainty quantification workflow used throughout the manuscript on a public benchmark dataset.
    The Gaussian process regression and random forest bagging models delivered the strongest point-prediction accuracy on this task, both reaching $R^2$-scores of about 0.8.
    However, the scalar calibration and sharpness diagnostics showed that point-prediction quality alone is not sufficient for judging predictive uncertainty.
    In the present implementation, the RF bagging regressor's uncertainty estimate is derived from ensemble spread and is therefore interpreted as an epistemic uncertainty.
    The dropout networks, trained with Gaussian negative log-likelihood loss (GNLL) and dropout-based Bayesian approximation (BNN) achieved the lowest MACE among the compared uncertainty outputs. 
    Therefore the BNN provided the most reliable predictive intervals in this benchmark, even though its point-prediction accuracy was lower than that of the classical baselines.
    This surrogate data study therefore provided the first clear evidence that calibration, sharpness, and confidence curve analysis reveal differences between models that are not visible from $R^2$, MAE, or RMSE alone.
    The BNN separately captures aleatoric and epistemic uncertainty, which can help decide whether a prediction is untrustworthy because of insufficient training data or because of inherent noise in the measurement.
    
    In the intrinsic-property study, focused on magnetic properties, we showed again that a very good model fit does not necessarily imply trustworthy uncertainty estimation. 
    While the Gaussian process regression models achieved high $R^2$-scores and produced a combined predictive uncertainty, this uncertainty estimate was not sufficiently informative for the present task.
    The random forest bagging regressor based on random forests also exhibit high $R^2$-scores and worked well for the confidence curve analysis.
    It even outperformed the BNN in predicting \ms.  
    The BNN models had high $R^2$-scores their uncertainty estimation correlated meaningfully with prediction error for predicting \ha but not for \ms. 
    This highlights that dropout networks can maintain both high $R^2$-score and informative uncertainty estimates.
    
    The coercivity study then extended this uncertainty-aware framework to the prediction of coercivity from microstructural information using a graph neural network. 
    Despite the increased complexity of the task, the same uncertainty quantification approach proved effective and interpretable. 
    Considering the total uncertainty in model predictions allowed us to effectively identify untrustworthy predictions.
    
    Together, these studies illustrate that uncertainty quantification is not merely an auxiliary feature but a core requirement for machine learning in materials science.
    The demonstrated framework, which combines Gaussian likelihood loss with dropout-based Bayesian inference, offers a unified approach that can be applied to both composition- and structure-based prediction tasks.
    Further, it was demonstrated that these two modeling techniques, capturing model and data uncertainty separately, can be transferred from simple to complex neural networks. 
    Future work may benefit from integrating uncertainty-aware models into autonomous materials discovery pipelines.
    The uncertainty estimates could be used as constraints in optimization tasks to rule out less-promising candidates. 
    Another application is to use uncertainty estimates to guide active learning, where the model can identify which new data points would be most informative to acquire next.

%% file: 888_appendix_study_details.tex
\clearpage
\phantomsection

\section{Appendix A. Reproducibility details for the surrogate study (\tc)}
\label{app:tc-study-details}
    
    This appendix collects implementation details that are useful for reproducing the Curie temperature surrogate study.
    The same machine was used to train and evaluate the models for all three studies in this manuscript, namely the Curie temperature surrogate study, the intrinsic-property study, and the coercivity graph-neural-network study.
    
    \subsection{Shared software environment and system information}
    \label{app:shared-env}
    The study environment was based on Python 3.10 with the key package versions listed in the project environment specification: NumPy 1.26.4, pandas 2.2.3, matplotlib 3.10.3, Keras 3.9.2, scikit-learn 1.6.1, SciPy 1.15.3, joblib 1.5.0, plotly 6.1.0, JAX 0.6.0, uncertainty-toolbox 0.1.1, and seaborn 0.13.2.
    The current workspace reports a Linux operating system, kernel 6.8.0-59-generic, x86\_64 architecture, and a 12th Gen Intel(R) Core(TM) i5-1245U CPU with 12 logical processors running Ubuntu 24.04.4 LTS.
    No GPU information was available from the current workspace.
    
    \subsection{Data handling and repeated evaluation}
    \label{app:tc-data-handling}
    All three models were trained on the Curie temperature surrogate dataset compiled by Sanvito and Nelson \cite{nelson2019a}.
    Only numerical descriptors were retained.
    Samples without target values were removed, and constant input columns were discarded before training.
    The train-test split was fixed to 70/30.
    A standard scaler was fitted on the training set and applied to both train and test sets.
    To reduce dependence on a single favorable split, the full split-fit-predict workflow was repeated for ten random seeds, 123 through 132.
    Reported aggregate metrics in the main text are based on these repeated runs.
    
    \subsection{Gaussian process regression}
    \label{app:tc-gpr}
    The Gaussian process regression (GPR) model used the implementation from scikit-learn with the kernel expression $kernel = ConstantKernel * Matern + WhiteKernel$.
    The kernel was initialized as follows:
    \begin{itemize}
        \item \texttt{ConstantKernel(1.0, (1e-3, 1e3))}
        \item \texttt{Matern(length\_scale=1.0, length\_scale\_bounds=(1e-2, 1e2), nu=1.5)}
        \item \texttt{WhiteKernel(noise\_level=1.0, noise\_level\_bounds=(1e-10, 1e2))}
    \end{itemize}
    Additional settings were \texttt{normalize\_y=True}, \texttt{n\_restarts\_optimizer=0}, and \texttt{random\_state=seed}.
    The train/test standard scaling described above was applied before fitting.
    
    \subsection{Random forest with bagging}
    \label{app:tc-rfr}
    The random forest baseline was implemented as a bagged ensemble.
    The base learner was \texttt{RandomForestRegressor} with
    \begin{itemize}
        \item \texttt{n\_estimators=50}
        \item \texttt{random\_state=seed}
        \item \texttt{n\_jobs=8}
    \end{itemize}
    This learner was wrapped in \texttt{BaggingRegressor} with
    \begin{itemize}
        \item \texttt{n\_estimators=100}
        \item \texttt{random\_state=seed}
        \item \texttt{n\_jobs=8}
    \end{itemize}
    All other arguments followed the defaults of scikit-learn 1.6.1.
    The predictive mean was computed from the bagged ensemble mean.
    The uncertainty estimate was taken as the standard deviation across the bagging estimators and interpreted as an epistemic uncertainty.
    
    \subsection{Bayesian neural network}
    \label{app:tc-bnn}
    The Bayesian neural network (BNN) was implemented in Keras 3 with the JAX backend.
    The architecture consisted of
    \begin{itemize}
        \item an input layer with \texttt{n\_features} nodes
        \item a dense layer with 128 units and \texttt{relu} activation
        \item a dense layer with 128 units and \texttt{tanh} activation
        \item dropout with rate 0.2
        \item a dense layer with 128 units and \texttt{relu} activation
        \item a dense layer with 128 units and \texttt{tanh} activation
        \item dropout with rate 0.2
        \item a linear output layer with two outputs
    \end{itemize}
    The two outputs represent the predictive mean and a variance-related quantity used in the Gaussian negative log-likelihood loss.
    Training used Adam with learning rate 0.005, batch size 32, and 1000 epochs.
    No early stopping and no validation split were used in the main training script.
    
    Predictive uncertainty was estimated by Monte Carlo dropout at inference time using 50 stochastic forward passes with dropout active.
    Aleatoric variance $\sigma_\mathrm{a}^2$ was computed as the mean predicted variance across passes, epistemic variance $\sigma_\mathrm{e}^2$ as the variance of the predicted means, and total variance as the square root of the sum of both contributions $\sigma = \sqrt{\sigma_\mathrm{e}^2 + \sigma_\mathrm{a}^2}$.
    
    \subsection{Evaluation settings}
    \label{app:tc-eval}
    Uncertainty quantification was evaluated on the held-out test set only.
    The main diagnostics were calibration curves, sharpness plots, and confidence curves. The calibration and sharpness plots were constructed from pooled test predictions across the 10 randomly seeded train-test splits.
    Confidence curves were reported using MAE as the ranking metric with retained-sample increments of 10.


\section{Appendix B. Reproducibility details for the intrinsic-property study (\js and \ha)}
\label{app:ipp-study-details}
    This appendix collects implementation details that are useful for reproducing intrinsic-property study on spontaneous magnetization \js and anisotropy field \ha.
    Again, the same machine was used to train and evaluate the models as in \hyperref[app:tc-study-details]{Appendix A}. 
    
    \subsection{Shared data handling and repeated evaluation}
    \label{app:ipp-data-handling}
    The intrinsic-property study used the same data-handling and repeated-evaluation protocol as the Curie temperature surrogate study (\hyperref[app:tc-study-details]{Appendix A}, Sec.~\ref{app:tc-data-handling}). Thus, all reported $R^2$, MAE, and RMSE values are computed on the held-out test split of each of the 10 randomly seeded train-test runs and then averaged, while the exported parity plots are representative test set plots from one single seed.
    
    For this study we only had 13 numerical input features available, which were used for both targets.
    All intrinsic-property scripts use the following elemental features in atomic percentage, plus a temperature feature in Kelvin:
    \begin{equation}
        \{\texttt{Nd}, \texttt{Ce}, \texttt{La}, \texttt{Pr}, \texttt{Y}, \texttt{Tb}, \texttt{Dy}, \texttt{Fe}, \texttt{Co}, \texttt{Ni}, \texttt{B}, \texttt{C}, \texttt{temp}\}, 
    \end{equation}
    Entries with missing target values are removed, and constant input columns are discarded before training.
    The intrinsic-property GPR and random forest scripts apply a \texttt{StandardScaler} after the train-test split.
    The scaler is fitted on the training subset and then applied to both train and test data.
    By contrast, the BNN wrapper scripts apply the sphere transformation to the numerical feature matrix. 
    
    \subsection{Shared Gaussian process settings}
    \label{app:ipp-gpr}
    The \js and \ha GPR scripts use the same model definition as in the Curie temperature surrogate study (\hyperref[app:tc-study-details]{Appendix A}, Sec.~\ref{app:tc-gpr}).

    \subsection{Shared random forest-with-bagging settings}
    \label{app:ipp-rfr}
    The \js and \ha random forest scripts also use identical hyperparameters as in the Curie temperature surrogate study (\hyperref[app:tc-study-details]{Appendix A}, Sec.~\ref{app:tc-rfr}).
    
    \subsection{Bayesian-neural-network model settings}
    \label{app:ipp-bnn}
    The \js and \ha BNN scripts share the same wrapper-level evaluation protocol.
    Both set \texttt{KERAS\_BACKEND=jax}, synchronize NumPy, Python, and Keras random seeds for each repetition, and instantiate a target-specific regressor class through
    \begin{itemize}
        \item \texttt{BnnMs(n\_features=n\_features, random\_seed=seed)} for spontaneous magnetization
        \item \texttt{BnnHa(n\_features=n\_features, random\_seed=seed)} for anisotropy field
    \end{itemize}
    We evaluate uncertainty on the test set by calling a predictor that returns four arrays per sample: predictive mean, total uncertainty, aleatoric uncertainty, and epistemic uncertainty.
    Thus, the uncertainty decomposition used in the intrinsic-property study is produced directly during inference by the target-specific BNN classes.
    
    The internal layer configuration, optimizer choice, and training-epoch settings of these target-specific helper classes are as follows.
    For the \ms BNN, the network consists of an input layer with \texttt{n\_features} nodes, followed by the dense-block sequence
    \begin{itemize}
        \item \texttt{Dense(32, activation='relu')}
        \item \texttt{Dense(32, activation='tanh')}
        \item \texttt{Dropout(0.5)}
        \item \texttt{Dense(64, activation='relu')}
        \item \texttt{Dense(64, activation='tanh')}
        \item \texttt{Dropout(0.5)}
        \item \texttt{Dense(32, activation='relu')}
        \item \texttt{Dense(32, activation='tanh')}
        \item \texttt{Dropout(0.5)}
        \item \texttt{Dense(2, activation=None)}
    \end{itemize}
    For the \ha BNN, the network consists of an input layer with \texttt{n\_features} nodes, followed by
    \begin{itemize}
        \item \texttt{Dense(128, activation='relu')}
        \item \texttt{Dense(128, activation='tanh')}
        \item \texttt{Dropout(0.5)}
        \item \texttt{Dense(2, activation=None)}
    \end{itemize}
    In both cases, the two linear output values represent the predictive mean and a variance-related quantity used by the Gaussian negative log-likelihood loss.
    Both target-specific models are compiled with \texttt{keras.optimizers.RMSprop()} and the custom loss \texttt{gaussian\_nll}.
    The class-level training durations are \texttt{epochs=800} for the \ms BNN and \texttt{epochs=400} for the \ha BNN.
    The shared base-class fit routine uses \texttt{batch\_size=32} and \texttt{verbose=0}.
    Training is performed directly with \texttt{fit()} on the training split, so no validation split and no early stopping are used in the reported repeated-seed runs.
    At inference time, both helper classes use Monte Carlo dropout with \texttt{n\_samples=50} to obtain predictive mean, total uncertainty, aleatoric uncertainty, and epistemic uncertainty.


\section{Appendix C. Reproducibility details for GNN study (\hc)}
\label{app:hc-study-details}
    This appendix collects implementation details that are useful for reproducing the coercivity graph-neural-network study.

    \subsection{Graph-neural-network data handling and repeated evaluation}
    \label{app:hc-data-handling}
    For the \hc study, an additional validation split is used only once to determine a reusable training length: after the seed-123 train-test split, the training portion is split again with \texttt{VAL\_SIZE=0.1}, and early stopping on validation loss selects a best epoch count.
    That epoch count is then reused for all ten train-test repetitions, so the reported seed-to-seed comparison remains aligned.
    
    The graph dataset is assembled in from the pickled files. 
    Graphs with fewer than two nodes are removed.
    For each node, the workflow first reconstructs physical quantities from stored normalized values, appends one additional radial-distance feature, and then retains the 13 configured node-feature indices.
    The scalar edge attribute is then constructed from the feature at local index 7 before that feature is removed from the node-feature matrix, so the final graph representation uses 12 node features per grain and one edge feature per edge.
    
    As in the tabular studies, scaling is fitted on the training subset only.
    Separate \texttt{StandardScaler} instances are fitted for node features, labels, and edge attributes, and the fitted transforms are then applied to the corresponding train, validation, and test graphs for that split.
    Mini-batches are created with \texttt{torch\_geometric.loader.DataLoader} using \texttt{batch\_size=100}, with shuffling enabled for the training loader and disabled otherwise.
    
    \subsection{Graph-neural-network architecture and optimization}
    \label{app:hc-gnn}
    The coercivity model instantiates \texttt{GNNModel} from \texttt{GNN\_Uncertainty/models.py}.
    The model uses
    \begin{itemize}
        \item an input embedding \texttt{Linear(orig\_node\_fea\_len, 18)}
        \item one message-passing layer \texttt{CustomConv} with mean aggregation and \texttt{edge\_fea\_len=1}
        \item a post-convolution block \texttt{Linear(18,16)}, dropout with rate 0.6, and \texttt{Softplus}
        \item graph pooling by \texttt{global\_mean\_pool}
        \item no additional hidden block beyond pooling because \texttt{n\_h=0}
        \item a final multilayer perceptron \texttt{Linear(16,8)} with \texttt{ReLU}, followed by \texttt{Linear(8,4)}
        \item two scalar output heads, one for the predictive mean and one for \texttt{log\_var}
    \end{itemize}
    The message function concatenates receiver-node features, sender-node features, and edge attributes, applies a linear layer, and then applies \texttt{ReLU}.
    Thus, the network predicts both a mean response and a log-variance term suitable for heteroscedastic regression.
    
    Optimization uses \texttt{AdamW} with \texttt{learning\_rate=0.009} and \texttt{weight\_decay=0.0005}.
    The training loss is \texttt{torch.nn.GaussianNLLLoss}, evaluated with the variance term \texttt{exp(log\_var)}.
    Learning-rate adaptation uses \texttt{ReduceLROnPlateau} with mode \texttt{min}, factor 0.5, patience 5, and threshold $10^{-4}$.
    The configuration file sets \texttt{epochs=150} and \texttt{patience=40} for the validation-based epoch search.
    The selected best epoch count is cached in \texttt{best\_epoch.txt} when available and reused for the repeated train-test runs.
    
    \subsection{Predictive uncertainty and evaluation outputs}
    \label{app:hc-uncertainty}
    Uncertainty is estimated by Monte Carlo dropout with \texttt{mc\_dropout\_samples=50}.
    During inference, dropout layers are explicitly switched back to training mode while all forward passes remain inside a \texttt{torch.no\_grad()} block.
    For each batch, the study computes the predictive mean as the average of the sampled means, epistemic variance as the variance across sampled means, and aleatoric variance as $\exp(\overline{\log \sigma^2})$, where the bar denotes the mean across stochastic passes.
    After inverse-scaling the target predictions, epistemic and aleatoric standard deviations are rescaled to physical units with the fitted label scaler.
    Total uncertainty is then reported as $\sigma = \sqrt{\sigma_\mathrm{e}^2 + \sigma_\mathrm{a}^2}$.
    
    For each seed, the script records train- and test set MAE, MSE, RMSE, and $R^2$.
    Across the ten seeds, it writes both per-seed metrics and summary statistics, and all reported scalar metrics in the manuscript are obtained by averaging the held-out test-split values across these ten runs.
    The representative seed for exported parity plots is again \texttt{123}.
    Confidence curves are computed on the held-out test predictions with MAE-based ranking, retained-sample step size 5, and \texttt{min\_remaining=10}.
    
    \subsection{Python packages specific to the graph-neural-network workflow}
    \label{app:hc-packages}
    Relative to the software already shared across all three studies, the \hc graph-neural-network workflow additionally depends on PyTorch 2.0.0 and PyTorch Geometric 1.15.2.
    PyTorch provides the tensor backend, \texttt{GaussianNLLLoss}, \texttt{AdamW}, CUDA-or-CPU device handling, checkpoint serialization, and Monte Carlo dropout execution.
    PyTorch Geometric provides the graph \texttt{Data} containers, batched \texttt{DataLoader}, the \texttt{MessagePassing} interface, and \texttt{global\_mean\_pool} for graph-level aggregation.
    The workflow also reads \texttt{config.yaml} through PyYAML, but that package is used only for configuration parsing rather than for the graph model itself.

%% file: 999_post.tex
\section{Data availability}
The dataset for the \tc study is publicly available on GitHub \href{https://github.com/msg-byu/ML-for-CurieTemp-Predictions}{msg-byu/ML-for-CurieTemp-Predictions}. 

The dataset for the \ms and \ha study are considered proprietary information and are the intellectual property of Toyota Motor Corporation. 

The dataset for the graph neural network study for prediction of \hc is available alongside the source code on GitHub \href{https://github.com/heisammoustafa/GNN_Uncertainty}{heisammoustafa/GNN\_Uncertainty}.

\section{Code availability}
The code generated during the present study is available on GitHub: \\ \href{https://github.com/wagerc97/magnetic-uncertainty}{wagerc97/magnetic-uncertainty}.

The graph neural network study for prediction of \hc is available on GitHub: \href{https://github.com/heisammoustafa/GNN_Uncertainty}{heisammoustafa/GNN\_Uncertainty}.

\section{Acknowledgment}
The financial support by the Austrian Federal Ministry of Energy, Economy and Tourism, the National Foundation for Research, Technology and Development and the Christian Doppler Research Association is gratefully acknowledged. 

This research was funded in whole or in part by the Austrian Science Fund (FWF) [\href{https://doi.org/10.55776/I6159}{10.55776/I6159}]. For open access purposes, the author has applied a CC BY public copyright license to any author accepted manuscript version arising from this submission.

\section{Author Contributions}
Clemens Wager: conceptualization, methodology, software, validation, investigation, writing - original draft, visualization.
Heisam Moustafa: conceptualization, methodology, software, validation, investigation, writing - original draft.
Thomas Schrefl: conceptualization, software, supervision, project administration, funding acquisition.
Harald Oezelt: funding acquisition, conceptualization, writing - review \& editing.
Qais Ali and Alexander Kovacs: writing - review \& editing.
All other authors contributed to manuscript revision, read, and approved the submitted version.

\section{Declaration of generative AI and AI-assisted technologies in the manuscript preparation process}

During the preparation of this work, the author(s) used Codex (OpenAI) and Gemini (Google) for code generation. DeepL and ChatGPT (OpenAI) were used for language editing and reformulation.
The author(s) reviewed and edited the output as needed and take full responsibility for the content of the published article.